\documentclass{aa}  
\DeclareUnicodeCharacter{02BC}{'}
\usepackage{afterpage}
\usepackage{graphicx}
\usepackage{txfonts}
\usepackage{amsmath} 
\usepackage{url}    
\usepackage{longtable}   
\usepackage{supertabular}
\usepackage{color}

\usepackage{float} 
\usepackage{caption}
\usepackage{subcaption}
\usepackage[]{hyperref}

\begin{document}

   \title{Exploring the central engines of gamma-ray bursts from prompt light curves}


   \author{Xue Zhang\inst{1}
          \and Shuang-Xi Yi\inst{1}
          \and Wei-Hua Lei\inst{2}
          \and Tong Liu\inst{3}
          \and Yu-Peng Yang\inst{1}
          \and Ying Qin\inst{4}
          \and Yan-Kun Qu\inst{1}
          \and Qing-Wen Tang\inst{5}
          \and Fa-Yin Wang\inst{6}
          }

   \institute{School of Physics and Physical Engineering, Qufu Normal University, Qufu 273165, China\\
              \email{yisx2015@qfnu.edu.cn}
         \and
             Department of Physics, Huazhong University of Science and Technology, Wuhan, 430074, China\\
             \email{leiwh@hust.edu.cn}
         \and
             Department of Astronomy, Xiamen University, Xiamen, Fujian 361005, China
         \and
             Department of Physics, Anhui Normal University, Wuhu, 241002, Anhui, China
         \and
             Department of Physics, Nanchang University, Nanchang 330031, Peopleʼs Republic of China
         \and
             School of Astronomy and Space Science, Nanjing University, Nanjing 210023, China
             }

   \date{Accepted February 2, 2026}

 
 \abstract
{Hyperaccreting stellar-mass black hole systems are leading candidates for the central engines of gamma-ray bursts (GRBs). Their jets are thought to be powered by either the Blandford–Znajek (BZ) process or neutrino-dominated accretion flows (NDAFs), but discriminating between these mechanisms remains challenging. To address this, we propose using the luminosity decay slope ($d$) of GRB light curves to distinguish between the BZ and NDAF mechanisms, thereby linking the light-curve morphology to the central engine physics. By analysing 85 single-peaked GRBs with fast-rise, exponential-decay (FRED) profiles observed by Swift/BAT using 64 ms background-subtracted light curves, we fit the decay slope ($d$) with the empirical Kocevski–Ryde–Liang (KRL) function and compare the results with theoretical predictions for the BZ ($d \approx 1.67$) and the NDAF ($d \approx 3.7\text{--}7.8$) mechanisms. We find that the decay slope ($d$) can differentiate central engine mechanisms, with 15 GRBs consistent with the BZ mechanism and 22 supporting the NDAF mechanism. However, most events exhibit slopes within the range \(2 < d < 4\), suggesting a hybrid of mechanisms, with NDAF being dominant.}

   \keywords{black hole physics --
                     accretion --
            gamma ray burst: general
               }
   \maketitle
%
\section{Introduction}
Gamma-ray bursts (GRBs) are among the most energetic explosions in the universe, and understanding their central engine mechanisms is crucial to probing jet dynamics and radiation processes. Currently, the central engines of GRBs are commonly attributed to two models: a magnetar \citep{1992ApJ...392L...9D,1998PhRvL..81.4301D,1998A&A...333L..87D,2010MNRAS.409..284M,2015ApJS..218...12L} and a hyperaccreting stellar-mass black hole \citep{1977MNRAS.179..433B,1991AcA....41..257P,1993ApJ...405..273W,1999ApJ...518..356P,2001ApJ...557..949N,2013ApJ...765..125L,2016ApJ...833..129X,2017ApJ...849...47L}. In the magnetar scenario, a rapidly rotating, strongly magnetised neutron star powers relativistic jets through accretion; continued energy injection may produce an extended X-ray plateau---a feature typically observed in bursts with lower energetics and longer plateau durations. Meanwhile, hyperaccreting black holes originate from either massive stellar collapse or compact binary mergers. These systems generate relativistic jets via the Blandford–Znajek (BZ) mechanism \citep{1977MNRAS.179..433B,2000ApJ...536..416L,2000PhR...325...83L,2005ApJ...630L...5M,2013ApJ...765..125L,2015ApJS..218...12L,2017ApJ...849...47L}, driven by magnetic fields or through the annihilation of neutrino–antineutrino pairs (NDAF) \citep{1997A&A...319..122R,1999ApJ...518..356P,2002ApJ...579..706D,2007ApJ...661.1025L,10.1111/j.1365-2966.2010.17600.x,2013ApJS..207...23X,2016MNRAS.458.1921S,2016ApJ...833..129X}. 

Some GRBs are attributed to the magnetar model, whereas the hyperaccreting black hole model can explain the majority of GRBs, particularly high-luminosity and ultra-long GRBs. Here, we focus on the hyperaccreting black hole system. The true jet luminosity, inferred by excluding X-ray emission from the central engine, can be linked to the NDAF and BZ mechanisms using parameters such as the isotropic radiated energy \( E_{\gamma, \text{iso}} \), the isotropic kinetic energy \( E_{k, \text{iso}} \), redshift \( z \), and the jet half-opening angles \( \theta_{j} \). \cite{2011ApJ...739...47F} estimated the accretion disc mass for short GRBs to range from \( 0.01 M_{\odot} \) to \( 0.1 M_{\odot} \), suggesting that the binary neutron star merger scenario can plausibly account for their observed characteristics. \cite{2015ApJS..218...12L} and \cite{2015ApJ...815...54S,2016MNRAS.458.1921S} further pointed out that the NDAF mechanism faces difficulties in explaining high-luminosity and ultra-long GRBs, particularly those requiring large accretion disc masses. \cite{2017JHEAp..13....1Y} found that the BZ mechanism matches the selected observational data better than the NDAF mechanism, a result further supported by \cite{2013ApJS..207...23X}. \cite{2021ApJ...908..242D} refined the jet-accretion rate constraints for high-luminosity GRBs within the BZ scenario.
  
Although previous studies have focused primarily on energy estimates or statistical analyses, some investigations have explored how light-curve variability reflects the central engine mechanism. The morphology of the light curve provides crucial insights into this distinction. \cite{2007A&A...468..563L} linked complex light-curve structures to BZ-driven jet precession and nutation, and successfully fitted the light curves of five GRBs.
\cite{2008MNRAS.388.1729K} explained the prompt emission phase and afterglow plateau through fallback accretion's impact on jet luminosity. \cite{2017ApJ...849...47L} provided a general luminosity fitting formula for both NDAF and BZ mechanisms, analysing light-curve evolution across prompt and afterglow phases. Thus, distinguishing the central engine based on the instantaneous light-curve features holds significant research value.

In this work, we simulate light curves during the prompt radiation phase for different initial parameters of hyperaccreting black hole systems to compare the light curves for BZ and NDAF scenarios.  
To validate our theoretical models, we select GRBs exhibiting single-peaked FRED profiles observed by Swift and compare their theoretical light curves. Through this comparison, we establish a new link between central engine physics and light-curve morphology. Section~\ref{section2} describes the hyperaccreting black hole and prompt emission mechanisms. Section~\ref{section3} details the sample selection and processing methods. Section~\ref{section4} presents the results, with Sections~\ref{section5} and~\ref{section6} discussing and concluding the study.

\section{Hyperaccreting black hole}
\label{section2}
Stellar-mass black hole hyperaccretion systems are widely considered as prominent candidate models for the central engine of GRBs \citep{1992ApJ...395L..83N,1999ApJ...524..262M}. These systems primarily involve two candidate mechanisms: the magnetically driven BZ mechanism and the NDAF mechanism. This section reviews both mechanisms and compares their effects on the light curves of GRBs.

\begin{figure*}[t]
\centering
\captionsetup[subfigure]{aboveskip=-0.5pt, belowskip=0pt, margin=0pt}
\setlength{\belowcaptionskip}{-3pt} 
\setlength{\tabcolsep}{-8pt} 

\begin{subfigure}[t]{0.5 \textwidth}
\includegraphics[width=\linewidth, trim=2 15 8 5, clip]{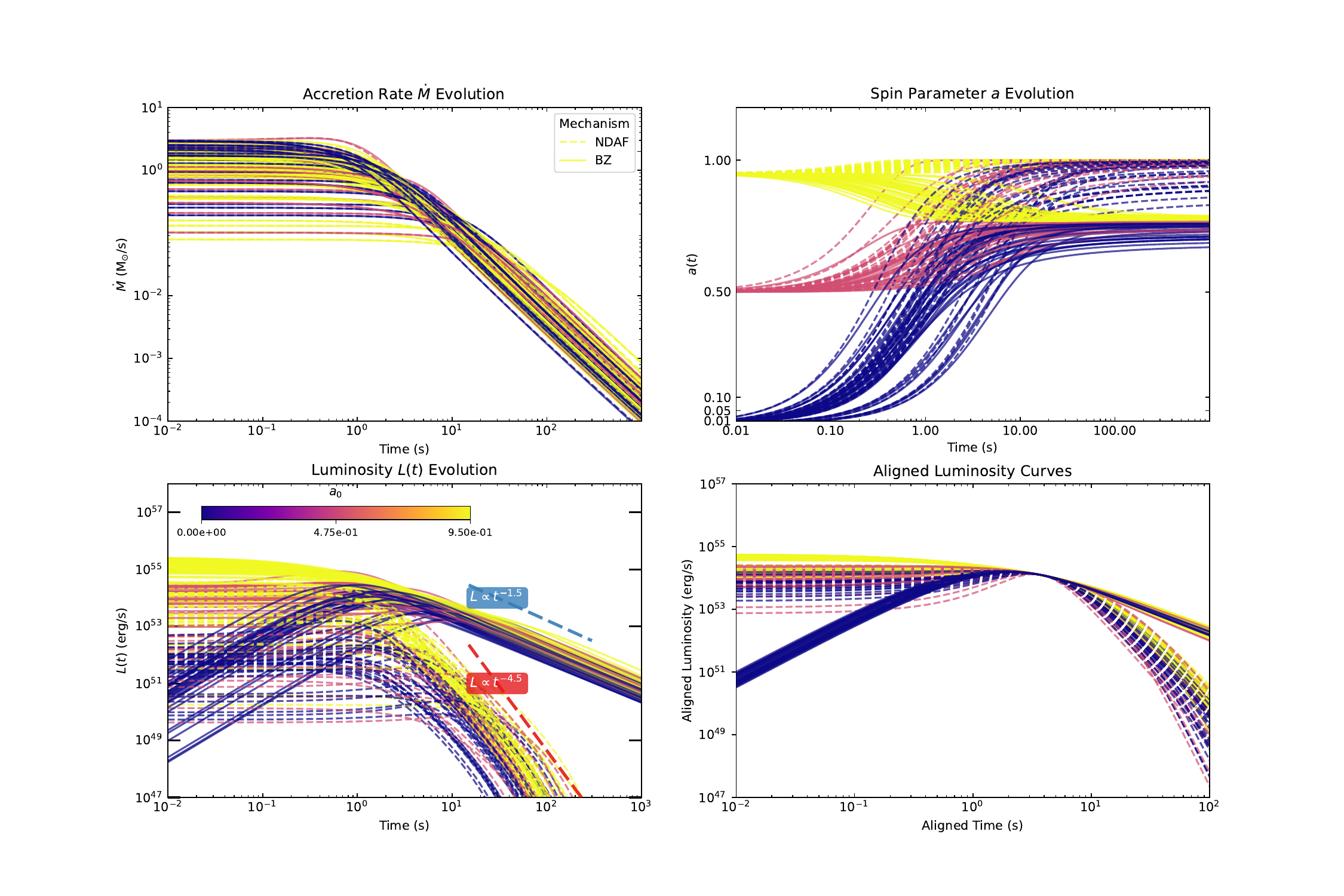}
\caption{$a_{\bullet 0}$}

\end{subfigure}\hspace{-0.8cm} 
\begin{subfigure}[t]{0.5\textwidth}
\includegraphics[width=\linewidth, trim=8 15 2 5, clip]{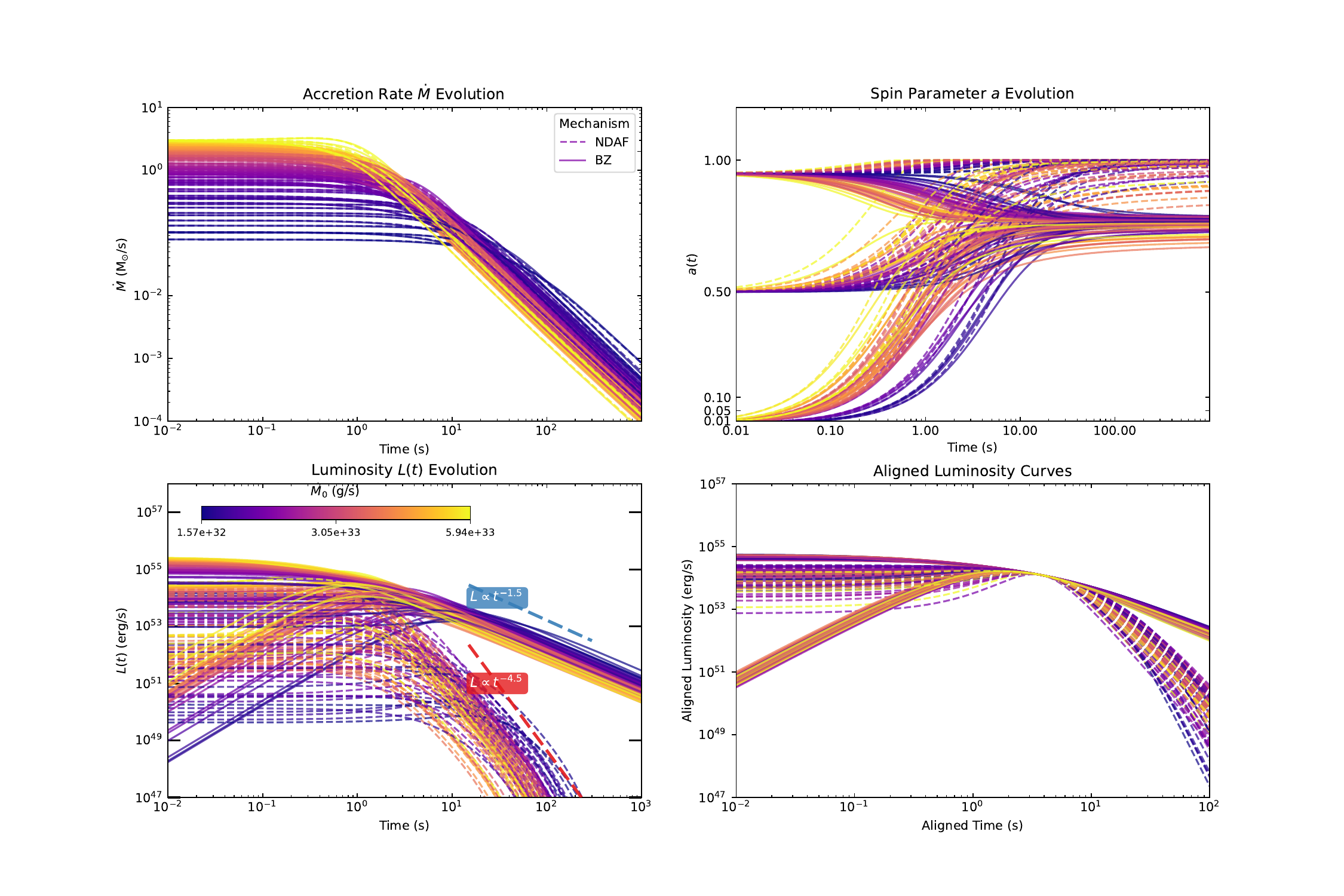}
\caption{$\dot{M}_0$}
\end{subfigure}

\vspace{-0.1cm}  

\begin{subfigure}[t]{0.5\textwidth}
\includegraphics[width=\linewidth, trim=2 15 8 5, clip]{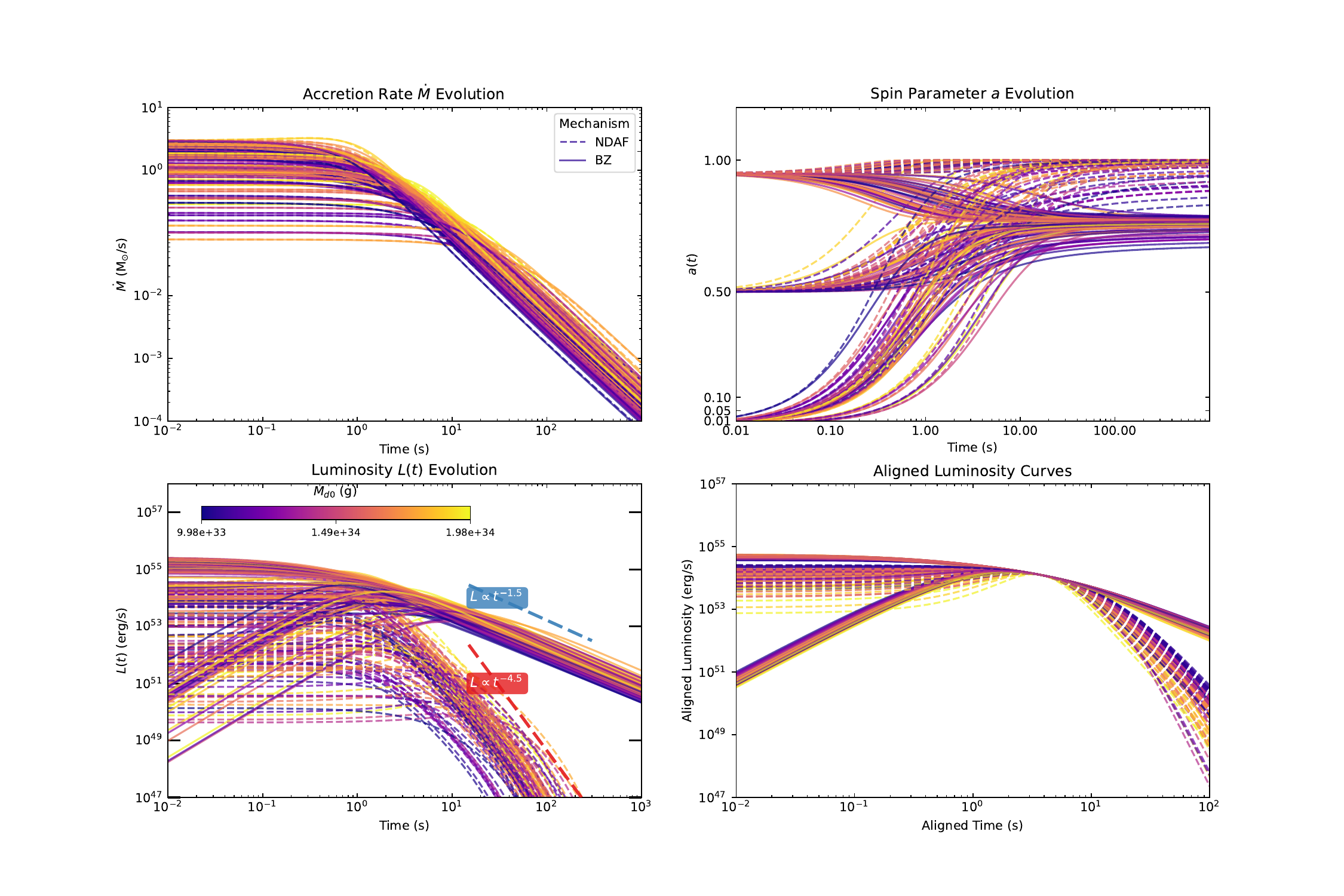}
\caption{$M_{d0}$}
\end{subfigure}\hspace{-0.8cm}
\begin{subfigure}[t]{0.5\textwidth}
\includegraphics[width=\linewidth, trim=8 15 2 5, clip]{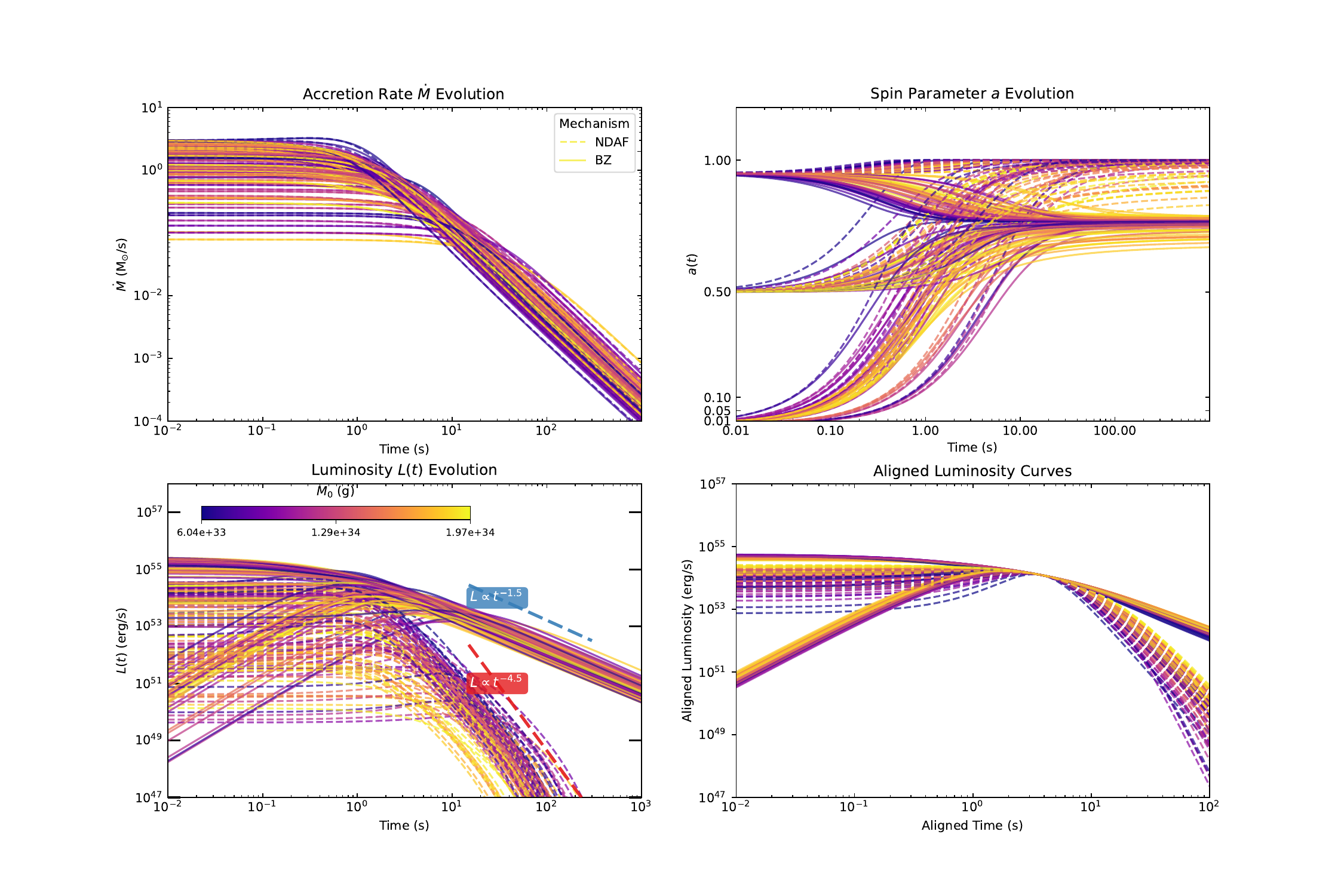}
\caption{$M_0$}
\end{subfigure}

\caption{Simulated long-GRB light curves for two central engine mechanisms: NDAF and BZ. The NDAF model exhibits a steep decay, whereas the BZ model shows a more gradual one. Colours represent variations in initial parameters—(a) black hole spin, (b) accretion rate, (c) accretion disc mass, and (d) black hole mass—illustrated by the colour bars. For each parameter set, the top left panel shows the evolution of the accretion rate over time; the top right panel shows the evolution of the spin parameter; the bottom left panel shows the luminosity evolution, where the NDAF model is plotted as red dashed lines (typical slope $-4.5$) and the BZ model as blue dashed lines (typical slope $-1.5$); and the bottom right panel compares the two mechanisms after alignment.}

\label{fig1}
\end{figure*}

\subsection{BZ mechanism}
\label{section2.1}
The BZ mechanism, first proposed by \cite{1977MNRAS.179..433B}, was initially introduced to explain the energy generation in active galactic nuclei. Later, \cite{2000ApJ...536..416L} and \cite{2000PhR...325...83L} extended this mechanism to explain the central engine of GRBs. They suggested that when a rotating Kerr black hole is threaded by a large-scale multipolar magnetic field from the accretion disc, the magnetic field efficiently extracts the black hole's rotational energy, driving relativistic jets. In this process, the magnetic field lines, dragged by the black hole's spin, induce an electric field and current near the event horizon, forming a stable magnetosphere. The current generated exerts an Ampère force on the black hole, causing it to decelerate. Consequently, the black hole's rotational energy is transmitted outward as Poynting flux, which drives the relativistic jets.

In the BZ mechanism, the energy extraction rate from the rotating black hole's magnetic field can be approximated as \citep{2017ApJ...849...47L}
\begin{equation}
L_{\rm BZ} = 1.7 \times 10^{20} a_{\bullet}^2 m_{\bullet} ^2 B_{\bullet, G} ^2  F(a_\bullet) \, \text{erg s}^{-1},
\end{equation}
where \(m_{\bullet} = M_{\bullet} / M_{\odot}\) is the dimensionless black hole mass; \(B_{\bullet}\) refers to the magnetic field strength at the black hole horizon, \(B_{\bullet, G} = B_{\bullet }  / 1 \, \text{G}\); and \( F(a_{\bullet}) \) is a dimensionless function of the black hole's spin parameter \( a_\bullet\), given by \( F(a_{\bullet}) = \left[ (1 + q^2) / q \right] \left[ \left( q + 1 / q \right) \arctan(q) - 1 \right] \), where \( q \) is defined as \(q = a_{\bullet} / \left( 1 + \sqrt{1 - a_{\bullet}^2} \right)\).

By assuming equilibrium between the accretion disc's gas pressure and the magnetic pressure at the disc's inner radius \citep{1997ApJ...482L..29M}, we obtain
\begin{equation}
\frac{B_{\bullet} ^2}{8 \pi} = P_{\text{in}} \sim \frac{\dot{M}  c}{4 \pi r_{\rm H}^2},
\end{equation}
where \( \dot{M}\) represents the accretion rate at the inner edge of the accretion disc, and \( r_{\rm H} \) is the radius of the black hole's event horizon, given by \( r_{\rm H} = \left(1 + \sqrt{1 - a_{\bullet}^2} \right) r_{\rm g} \). We then obtain 
\begin{equation}
B_{\bullet}  \simeq 7.4 \times 10^{16} \dot{m}^{1/2} m_{\bullet}^{-1} \left(1 + \sqrt{1 - a_{\bullet}^2}\right)^{-1} \, \text{G}.
\end{equation}

Finally, the luminosity of the BZ mechanism can be approximated as
\begin{equation}
\begin{aligned}
L_{\rm BZ} &= 9 \times 10^{53} a_{\bullet}^2 \dot{m} X(a_*) \, \text{erg s}^{-1} \\
&\simeq 1.5 \times 10^{53} a_{\bullet}^2 \dot{m}\, \text{erg s}^{-1},
\end{aligned}
\end{equation}
where \( \dot{m} = \dot{M} / (M_\odot \, \text{s}^{-1}) \) is the dimensionless mass-accretion rate of a black hole. 

To establish a connection between the BZ power and the observed gamma-ray luminosity, we use the following relation:
\begin{equation}
\eta L_{\rm BZ} = f_{\rm b} L_{\gamma, \text{iso}} = L_{\gamma},
\end{equation}
where \( f_{\rm b} \) is the beaming factor of the jet, and \( \eta \) is the efficiency of converting BZ power into gamma-ray radiation. For this study, we adopt typical values of \( f_{\rm b} = 0.01 \) and \( \eta = 0.2 \).

In the BZ mechanism, considering the energy and angular momentum conservation \citep{2000ApJ...536..416L,2000PhR...325...83L}, the evolution of the black hole mass and angular momentum can be expressed as
\begin{equation}
\frac{dM_{\bullet}c^2}{dt} = \dot{M}_{\bullet}c^2 E_{\text{ms}} - L_{\rm BZ},
\end{equation}
\begin{equation}
\frac{dJ_{\bullet}}{dt} = \dot{M}_{\bullet} L_{\text{ms}} - T_{\rm B},
\end{equation}
where \( T_{\rm B} \) represents the torque produced by the interaction of the black hole with the accretion disc, and can be written as
\begin{equation}
\begin{aligned}
T_{\rm B} &= \frac{L_{\rm BZ}}{\Omega_{\rm F}} = 1.8 \times 10^{49} a_{\bullet} \dot{m} m_{\bullet} F(a_{\ast}) \, \text{g} \, \text{cm}^{2} \, \text{s}^{-2} \\
&\simeq 1.2 \times 10^{49} a_{\bullet} \dot{m} m_{\bullet} \, \text{g} \, \text{cm}^{2} \, \text{s}^{-2},
\end{aligned}
\end{equation}
where \(\Omega_{\rm F} = 0.5 \Omega_{\bullet}\) \citep{2000JKPS...36..188L,2000PhR...325...83L}, and \(\Omega_{\bullet} = \frac{a_{\bullet} c}{2r} = \frac{a_{\bullet} c^3}{2GM_{\bullet}\left(1 + \sqrt{1 - a_{\bullet}^2}\right)}\).

Here, \( E_{\text{ms}} \) and \( L_{\text{ms}} \) correspond to the specific energy and specific angular momentum at the maximum stable circular orbit radius \( r_{\text{ms}} \) \citep{novikov1973black}, and are given by
\begin{equation}
E_{\text{ms}} = \frac{4 \sqrt{R_{\text{ms}}} - 3a_{\bullet}}{\sqrt{3 R_{\text{ms}}}},
\end{equation}
\begin{equation}
L_{\text{ms}} = \frac{GM_{\bullet}}{c} \frac{2(3 \sqrt{R_{\text{ms}}} - 2a_{\bullet})}{\sqrt{3 R_{\text{ms}}}},
\end{equation}
where \( R_{\text{ms}} = r_{\text{ms}} / r_{\rm g} \), and the expression for \( r_{\text{ms}} \) is given by
\begin{equation}
r_{\text{ms}} = r_{\rm g} \left[ 3 + Z_2 - \text{sgn}(a_{\bullet}) \left( [3 - Z_1](3 + Z_1 + 2Z_2) \right)^{1/2} \right].
\end{equation}
For \( 0 \leq a_{\bullet} \leq 1 \), the quantity \( Z_1 \) is given by \( Z_1 \equiv 1 + (1 - a_{\bullet}^2)^{1/3} \left[ (1 + a_{\bullet})^{1/3} + (1 - a_{\bullet})^{1/3} \right] \), while \( Z_2 \) is defined as \( Z_2 \equiv (3a_{\bullet}^2 + Z_1^2)^{1/2} \) \citep{1972ApJ...178..347B,novikov1998astrophysics,kato2008black}.

Building upon this foundation and incorporating the definition of the intrinsic angular momentum, one can derive the expression for the spin as \( a_{\bullet} = \frac{J_{\bullet} c}{G M_{\bullet}^2} \). The corresponding evolution equation for the Kerr black hole can then be derived as
\begin{equation}
\frac{da_{\bullet}}{dt} = \frac{(\dot{M} L_{\text{ms}} - T_{\rm B}) c}{G M_{\bullet}^2} - 2a_{\bullet} \left( \frac{\dot{M} c^2 E_{\text{ms}} - L_{\rm BZ}}{M_{\bullet}c^2} \right).
\end{equation}
This equation governs the time evolution of the black hole's spin, derived by considering the BZ mechanism in the context of energy extraction from the black hole.

\subsection{NDAF model}
\label{section2.2}

The mechanism of neutrino annihilation is an important channel of energy release, first proposed by \cite{1999ApJ...518..356P} to explain GRBs. The typical accretion rate associated with GRBs is in the range of 0.01--10\,$M_{\odot}$\,s$^{-1}$, resulting in an accretion disc with extremely high temperature and density. Under such conditions, the optical depth increases, photons become trapped, and the viscously dissipated energy can be carried away only via neutrino emission; this constitutes the NDAF mechanism. In this mechanism, a fraction of neutrinos annihilate into electron--positron pairs ($\nu + \bar{\nu} \rightarrow e^{-} + e^{+}$), which subsequently annihilate into gamma-ray photons that help launch relativistic jets. Because this process does not require strong magnetic fields, it is regarded as a complementary GRB launching mechanism to the BZ process.

To accurately describe the relativistic accretion system, we adopt a steady-state disc model around a Kerr black hole, incorporating neutrino radiation and rotation within the framework of general relativity, as proposed by \cite{2007ApJ...657..383C} and \cite{2017ApJ...849...47L}. For simplicity, we assume that the NDAF accretion disc is a standard thin disc and that the rotation of the accretion disc follows the Keplerian disc model.

Under the Kerr metric, the basic equations of the NDAF model are as follows:
\begin{enumerate}
    \item Mass conservation   
 \begin{align}
    \dot{M} = -4 \pi r v_r \rho H.
\end{align}

    \item Angular momentum conservation
\begin{align}
    \dot{M} \sqrt{GM_{\bullet} r} \frac{D}{A} = 4 \pi r^2 H \alpha P \sqrt{\frac{A}{BC}},
\end{align}
where \( A \), \( B \), \( C \), and \( D \) are general relativistic correction factors, given as
\begin{equation*}
A = 1 - \frac{2GM_{\bullet}}{c^2 r} + \left( \frac{GM_{\bullet} \cdot a_{\bullet}}{c^2 r} \right)^2,
\end{equation*}
\begin{equation*}
B = 1 - \frac{3GM_{\bullet}}{c^2 r} + 2a_{\bullet} \left( \frac{GM_{\bullet}}{c^2 r} \right)^{3/2},
\end{equation*}
\begin{equation*}
C = 1 - 4a_{\bullet} \left( \frac{GM_{\bullet}}{c^2 r} \right)^{3/2} + 3 \left( \frac{GM_{\bullet} \cdot a_{\bullet}}{c^2 r} \right)^2,
\end{equation*}
\begin{equation*}
D = Bf.
\end{equation*}

    \item Energy conservation
\begin{align}
    Q^+ = Q^-,
\end{align}
where \( Q^+ = Q_{\text{vis}} \) and \( Q^- = Q_{\nu} + Q_{\text{photo}} + Q_{\text{adv}} \). 
Here, \( Q_{\text{vis}} \) is the viscous heating rate, and the cooling terms (\( Q^- \)) include neutrino cooling \( Q_{\nu} \), photon emission \( Q_{\text{photo}} \), and advection energy \( Q_{\text{adv}} \).

    \item The total pressure on the NDAF disc
\begin{equation}
\begin{aligned}
    P &= \frac{11}{12} a T^4 + \frac{\rho k T}{m_{\rm p}} \left( 1 + \frac{3 X_{\text{nuc}}}{4} \right) + \frac{2 \pi h c}{3} \left( \frac{3}{8 \pi m_{\rm p}} \right)^{4/3} \\
    &\quad \times \left( \frac{\rho}{\mu_e} \right)^{4/3} + \frac{u_{\nu}}{3}.
\end{aligned}
\end{equation}
\end{enumerate}

The total energy emitted by neutrinos from the NDAF accretion disc per unit time represents the neutrino luminosity. The neutrino radiation luminosity of the NDAF is given by
\begin{equation}
    L_{\nu} = 4\pi \int_{r_{\text{ms}}}^{r_{\text{out}}} Q_{\nu} r \, dr,
\end{equation}
where \( r_{\text{out}} \) represents the outer radius of the accretion disc and \( Q_{\nu} \) denotes the neutrino cooling rate. The neutrino radiation luminosity is obtained by integrating \( Q_{\nu} \) over the area.

For the neutrino annihilation luminosity, \cite{1997A&A...319..122R}, \cite{1999ApJ...518..356P}, \cite{2003MNRAS.345.1077R} and others proposed dividing the NDAF disc into many grid points. Neutrinos are denoted as \( k \), while antineutrinos are labelled as \( k' \). By summing and integrating over all the grid points, the neutrino annihilation luminosity at each point is obtained:
\begin{equation}
\begin{aligned}
    \bar{E}_{\nu \bar{\nu}} &= A_1 \sum_k \frac{l_{\nu_i}^{k}}{d_k^2} \sum_{k'} \frac{l_{\nu_i}^{k'}}{d_{k'}^2} \left( \epsilon_{\nu_i}^{k} + \epsilon_{\bar{\nu}_i}^{k'} \right) \left( 1 - \cos \theta_{kk'} \right)^2 \\
    &\quad + A_2 \sum_k \frac{l_{\nu_i}^{k}}{d_k^2} \sum_{k'} \frac{l_{\nu_i}^{k'}}{d_{k'}^2} \left( \frac{\epsilon_{\nu_i}^{k} + \epsilon_{\bar{\nu}_i}^{k'}}{\epsilon_{\nu_i}^{k} \epsilon_{\bar{\nu}_i}^{k'}} \left( 1 - \cos \theta_{kk'} \right) \right),
\end{aligned}
\end{equation}
where \( A_1 \approx 1.7 \times 10^{-44} \, \text{cm} \, \text{erg}^{-2} \, \text{s}^{-1} \) and \( A_2 \approx 1.6 \times 10^{-56} \, \text{cm} \, \text{erg}^{-2} \, \text{s}^{-1} \).

By performing an integration over the black hole and the outer region of the accretion disc, the total neutrino annihilation luminosity can be written as
\begin{equation}
L_{\nu \bar{\nu}} = 4\pi \int \int l_{\nu \bar{\nu}} r \, dr \, dz.
\end{equation}

This method is physically well founded. However, due to the dependence on the accretion rate, black hole mass and spin parameters, an analytical derivation is not feasible. Consequently, many empirical fitting formulae have been proposed for the neutrino annihilation luminosity. For example, \citet{1999ApJ...518..356P} and \citet{2013ApJS..207...23X} have suggested various parameterised expressions, while some studies have provided corresponding analytical approximations (e.g., \cite{1977MNRAS.179..433B}). The choice of a specific formula should be based on the range of accretion rates and the research objectives to determine the most suitable fitting function.

The results of \cite{2017ApJ...849...47L} are consistent with the findings of \cite{2013ApJS..207...23X} and \cite{2011MNRAS.410.2302Z} with low and moderate accretion rates. This study adopts the segmented power-law fitting formula proposed by \cite{2017ApJ...849...47L}, which better reflects the true trend of annihilation luminosity in the range of \( \dot{M} = 0.01 - 10\, \text{M}_{\odot} \, \text{s}^{-1} \), with the black hole mass ranging from \( 3\, M_{\odot} \) to \( 10\, M_{\odot} \). This fitting is particularly suited for simulating the evolution of the central engine energy output during the prompt phase of GRBs.

\begin{equation}
\begin{aligned}
L_{\nu \bar{\nu}} &\simeq L_{\nu \bar{\nu},\text{ign}} \left[ \left( \frac{\dot{m}}{\dot{m}_{\text{ign}}} \right)^{-\alpha_{\nu \bar{\nu}}} + \left( \frac{\dot{m}}{\dot{m}_{\text{ign}}} \right)^{-\beta_{\nu \bar{\nu}}} \right]^{-1} \\
&\quad \times \left[ 1 + \left( \frac{\dot{m}}{\dot{m}_{\text{trap}}} \right)^{\beta_{\nu \bar{\nu}} - \gamma_{\nu \bar{\nu}}} \right]^{-1},
\end{aligned}
\end{equation}
where
\begin{equation}
L_{\nu\bar{\nu},\text{ign}} = 10^{\left(48.0 + 0.15 a_{\bullet}\right)} \left( \frac{m_{\bullet}}{3} \right)^{ \log \left( \frac{\dot{m}}{\dot{m}_{\text{ign}}} \right) - 3.3} \, \text{erg s}^{-1},
\end{equation}
with \( \alpha_{\nu\bar{\nu}} = 4.7 \), \( \beta_{\nu \bar{\nu}} = 2.23 \), \( \gamma_{\nu \bar{\nu}} = 0.3 \), \( \dot{m}_{\text{ign}} = 0.07 - 0.063 a_{\bullet} \), and \( \dot{m}_{\text{trap}} = 6.0 - 4.0 a_{\bullet}^3 \).

Similarly, the efficiency of the NDAF and its relation to the luminosity of the observed GRBs can be written as \( \eta_{\nu \bar{\nu}} L_{\nu \bar{\nu}} = f_{\rm b} L_{\gamma,\text{iso}} \).

It should be noted that the fitting parameters include the luminosity normalisation terms \( \dot{E}_{\nu,\text{ign}} \), \( \dot{E}_{\nu \bar{\nu}, \text{ign}} \), the power-law evolution indices \( \alpha \), \( \beta \), \( \gamma \), as well as the critical accretion rates \( \dot{m}_{\text{ign}} \), \( \dot{m}_{\text{trap}} \). These values are derived from the fits under the conditions of black hole mass \( M = 3 M_{\odot} \) and spin \( a_{\bullet} = 0.1 - 0.95 \).

\begin{figure*}[t]

\centering
\captionsetup[subfigure]{aboveskip=-0.5pt, belowskip=0pt, margin=0pt}
\setlength{\belowcaptionskip}{-3pt} 
\setlength{\tabcolsep}{-8pt} 

\begin{subfigure}[t]{0.5 \textwidth}
\includegraphics[width=\linewidth, trim=2 15 8 5, clip]{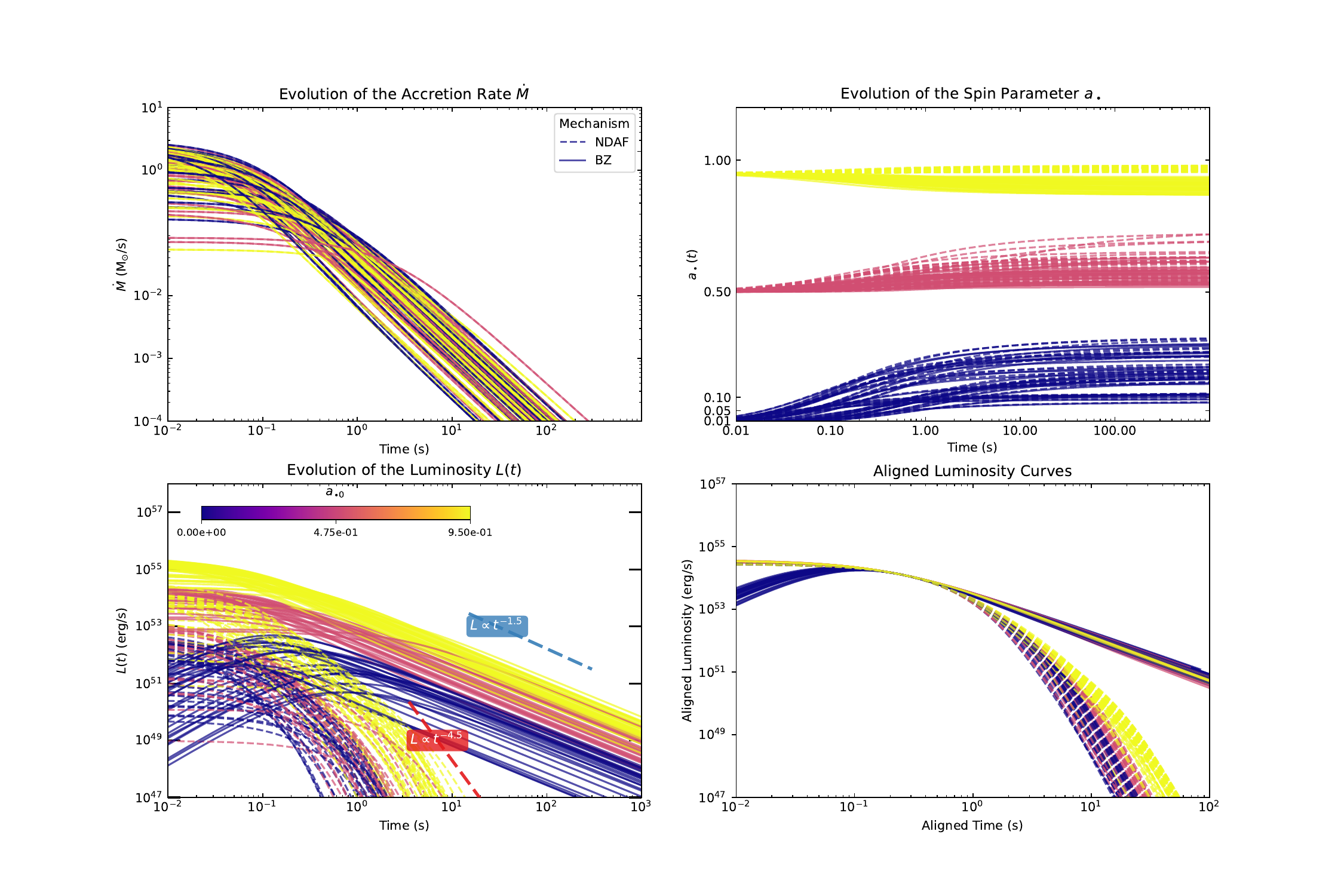}
\caption{$a_{\bullet 0}$}

\end{subfigure}\hspace{-0.8cm} 
\begin{subfigure}[t]{0.5\textwidth}
\includegraphics[width=\linewidth, trim=8 15 2 5, clip]{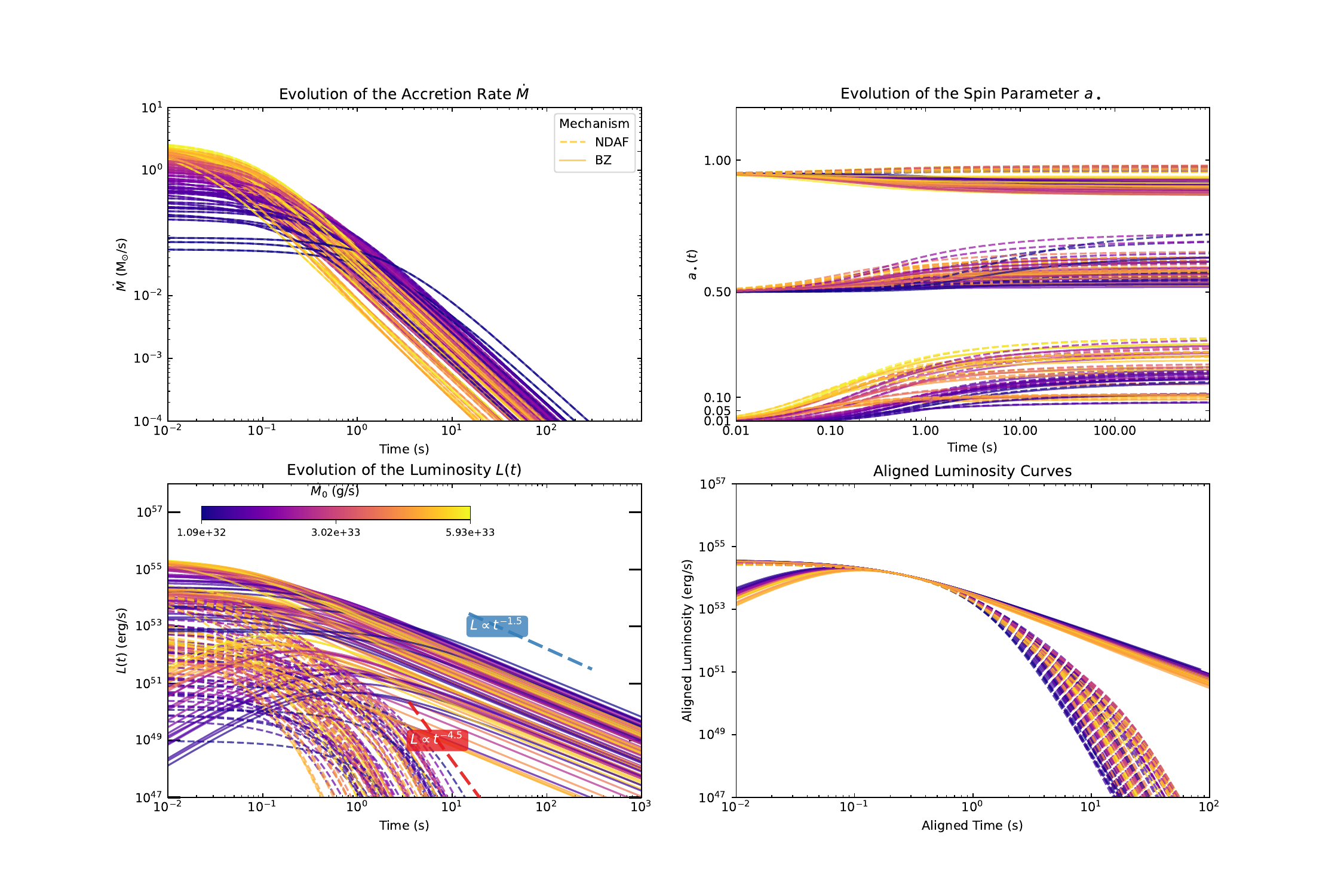}
\caption{$\dot{M}_0$}
\end{subfigure}

\vspace{-0.1cm}  

\begin{subfigure}[t]{0.5\textwidth}
\includegraphics[width=\linewidth, trim=2 15 8 5, clip]{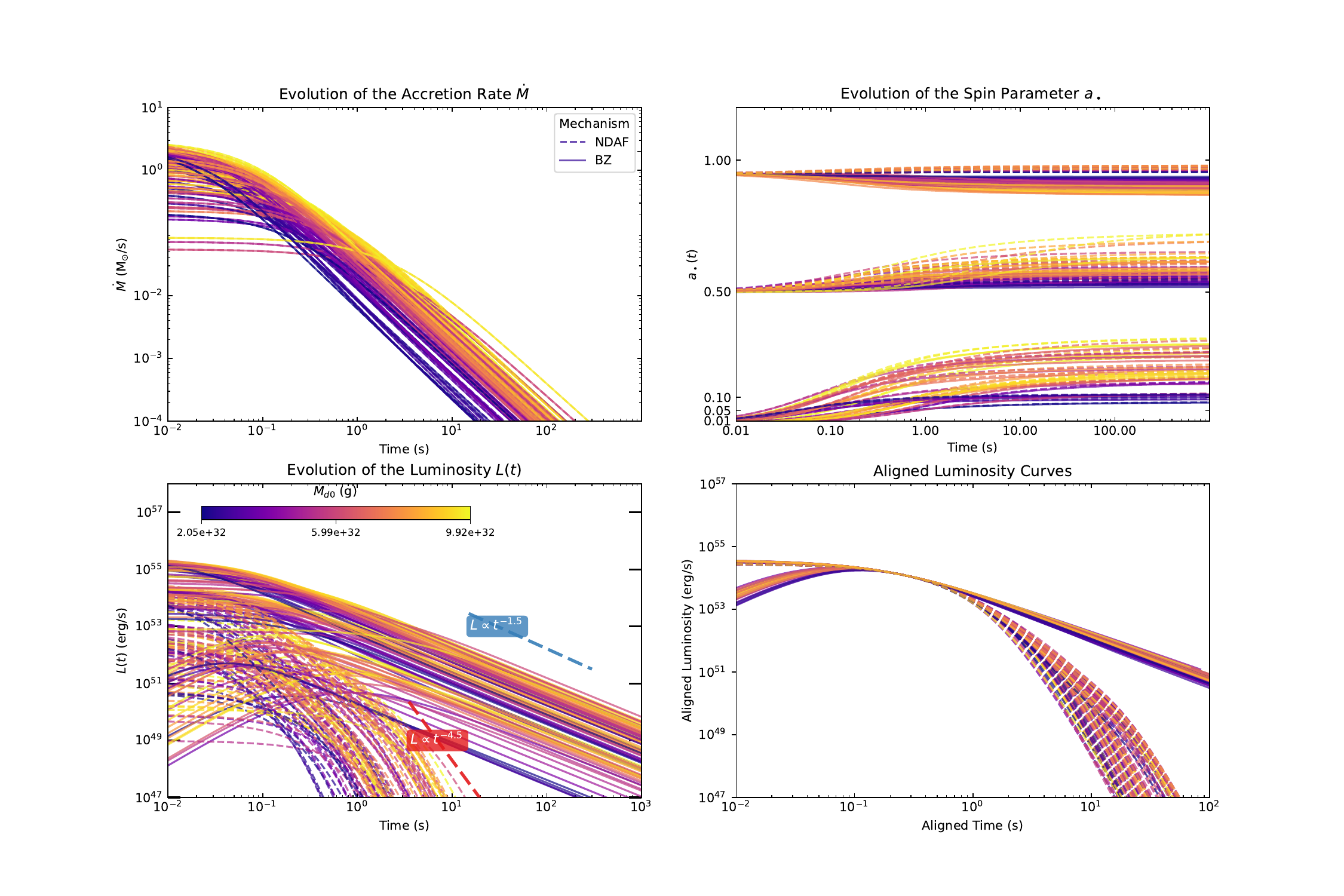}
\caption{$M_{d0}$}
\end{subfigure}\hspace{-0.8cm}
\begin{subfigure}[t]{0.5\textwidth}
\includegraphics[width=\linewidth, trim=8 15 2 5, clip]{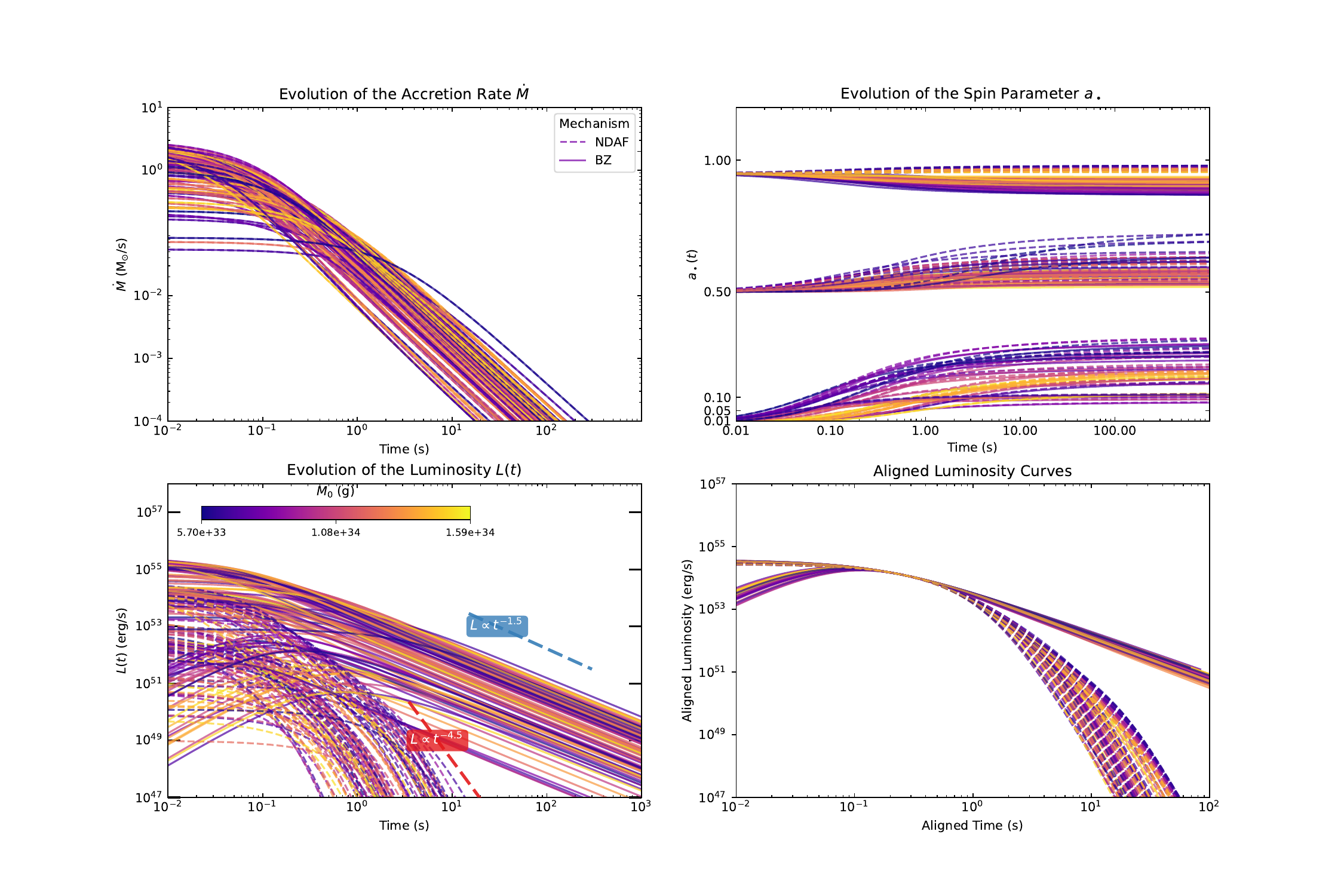}
\caption{$M_0$}
\end{subfigure}

\caption{Same as Fig. 1 but for simulated short-GRB light curves}
\label{fig2}
\end{figure*}

In the absence of the BZ mechanism, the evolution equation of the Kerr black hole can be derived from the conservation of energy and angular momentum:
\begin{equation}
\frac{dM_{\bullet} c^2}{dt} = \dot{M} c^2 E_{\text{ms}},
\end{equation}
\begin{equation}
\frac{dJ_{\bullet}}{dt} = \dot{M} L_{\text{ms}}.
\end{equation}

Using these physical quantities, the evolution equation of the black hole spin parameter, driven by the neutrino annihilation mechanism, can be derived as
\begin{equation}
\frac{da_{\bullet}}{dt} =  \frac{\dot{M} L_{\text{ms}} c}{G M_{\bullet}^2} -  2a_{\bullet}\frac{\dot{M} E_{\text{ms}}}{M_{\bullet}}.
\end{equation}

\subsection{Comparison of theoretical light curves from two models}
\label{section2.3}

To investigate the influence of different central engine mechanisms on the optical behaviour of GRBs during the prompt phase, we adopt the simplified annular accretion disc model proposed by \cite{2008MNRAS.388.1729K,2008Sci...321..376K,2008MNRAS.390..781M,2017ApJ...849...47L}, which is regarded as a ring structure concentrated at a representative radius \( r_{\rm d} \). Under this assumption, the unit mass angular momentum of the disc is calculated using classical orbital mechanics, while the total mass \( M_{\rm d} \) and angular momentum \( J_{\rm d} \) of the disc are equivalently distributed across the ring. The following expression can be derived from the conservation of angular momentum:
\begin{equation}
j(r_{\rm d}) = \left( G M_{\bullet} r_{\rm d} \right)^{1/2} = \frac{J_{\rm d}}{M_{\rm d}}.
\end{equation}
To further obtain the instantaneous accretion rate \( \dot{M} \), we relate it to the viscous accretion timescale as
\begin{equation}
\dot{M} = \frac{M_{\rm d}}{t_{\text{acc}}}.
\end{equation}
Here, \( \dot{M} \) denotes the mass accretion rate, and the accretion timescale \( t_{\text{acc}} \) is given by \( t_{\text{acc}} = \frac{r_{\rm d}^2}{\nu} \sim \frac{2}{\alpha \Omega_{\rm K}} \) \citep{1973A&A....24..337S}, where \( r_{\rm d} \) is the radial distance, \( \nu \) is the kinematic viscosity, and \( \alpha = 0.1 \) is the viscosity coefficient. The angular velocity \( \Omega_{\rm K} \), based on the Keplerian velocity under the pseudo-Newtonian potential, is expressed as \( \Omega_{\rm K} = \left( \frac{GM}{R} \right)^{1/2} \frac{1}{R - R_{\rm g}} \), where \( G \) is the gravitational constant, \( M \) is the mass of the central object, \( R \) is the radius, and \( R_{\rm g} \) is the Schwarzschild radius.

During the accretion process, the mass and angular momentum of the disc decay and evolve over time, which can be expressed as
\begin{equation}
\dot{M}_{\rm d} = -\dot{M},
\end{equation}
\begin{equation}
\dot{J}_{\rm d} = -L_{\text{ms}} \dot{M}.
\end{equation}
By combining the luminosity functions \( L_{\nu\bar{\nu}}(t) \) and \( L_{\rm BZ}(t) \) derived in Sections~\ref{section2.1} and~\ref{section2.2}, we establish the coupled evolution equations for the black hole and the accretion disc based on the conservation of angular momentum. Solving for the time-dependent evolution of the spin parameter \( a_{\bullet}(t) \), black hole mass \( M(t) \) and accretion rate \( \dot{M} \), we can obtain the theoretical evolution of the luminosity light curve driven by the central engine for different initial parameters.

Figures~\ref{fig1} and~\ref{fig2} present the simulated GRB light curves for varying initial parameters, including black hole mass \( M \), spin parameter \( a_{\bullet} \), accretion disc mass \( m_{\rm d} \), accretion rate \( \dot{m} \), radiation efficiency \( \epsilon \) and beaming factor \( f_{\rm b} \). The initial conditions are based on different types of compact object mergers or collisions. In merger scenarios, black holes formed from double neutron star mergers typically have masses less than \( 4 M_{\odot} \), while black holes merging with neutron stars undergo tidal disruption, leading to the formation of an accretion disc, with resulting black hole masses between \( 4\)--\(8 M_{\odot} \) \citep{2009PhRvD..79d4030S}. For black holes formed via the collapse of massive stars, the initial mass is typically within the range of \( 3\)--\(10 M_{\odot} \) \citep{2016MNRAS.458.1921S}. In these cases, the initial black hole mass has a minimal effect on GRB production, with the accretion disc mass and accretion process being the main contributors \citep{2011ApJ...740L..27L,2022ApJ...929...83Q}. The mass of accretion discs formed in double compact object mergers ranges from \( 0.1\)--\(0.5 M_{\odot} \) \citep{2001MNRAS.328..583L,2011ApJ...739...47F,2012ApJ...760...63L,2015ApJ...805...89L}, while for massive star collapses, it ranges from \( 1\)--\(5 M_{\odot} \) \citep{2003ApJ...586..356Z,2016MNRAS.458.1921S}. Thus, we set the initial parameter ranges as follows: for long GRBs, \( m:(3, 10) \); \( \dot{m}:(0.01, 3) \); \( m_{\rm d}:(1, 5) \); and for short GRBs, \( m:(2.3, 8) \); \( \dot{m}:(0.01, 3) \); \( m_{\rm d}:(0.1, 0.5) \). For typical cases, we adopt a radiation efficiency of \( \epsilon = 0.2 \), a beaming factor of \( f_{\rm b} = 0.01 \), and the black hole spin parameter is taken as \( a_{\bullet} = (0, 0.5, 0.95) \). The study of spin equilibrium by \cite{2017ApJ...849...47L} found \( a_{\mathrm{eq}} \sim 0.87 \), while our choice of \( a_{\bullet} = 0.95 \) is motivated by a rapidly rotating progenitor, which is necessary for powering a luminous jet at the GRB onset, before the black hole evolves towards equilibrium.

By fitting the decay slope of the theoretical light curve, we find that on the logarithmic time scale, the light curves dominated by the two mechanisms exhibit distinct differences in the slope of the decay phase for both long and short GRBs. The details are as follows: 
\begin{enumerate}
    \item When the luminosity varies by several orders of magnitude, the slope of the decay phase effectively differentiates between the two mechanisms.
    
    \item A steeper luminosity decay in the NDAF mechanism, with slopes from \(-4.05 \pm 0.08\) to \(-7.01 \pm 0.40\), where the typical luminosity decays as \( L \propto t^{-4.5} \). Theoretically, the expected evolution of the NDAF luminosity depends on the accretion regime. For high accretion rates (\( \dot{m} \gg \dot{m}_{\rm ign} \)), the luminosity scales as \( L_{\nu\bar{\nu}} \propto \dot{M}^{2.23} \propto t^{-3.72} \). For low accretion rates (\( \dot{m} \ll \dot{m}_{\rm ign} \)), it scales as \( L_{\nu\bar{\nu}} \propto \dot{M}^{4.7} \propto t^{-7.83} \). The decay slopes range approximately from \( t^{-3.7} \) to \( t^{-7.8} \), consistent with the fitted slopes.

    \item The luminosity decay for the BZ mechanism is relatively shallow, with slopes typically ranging from  \(-1.08 \pm 0.04\) to \(-1.49 \pm 0.05\), consistent with \( L \propto t^{-1.5} \). Theoretically, the BZ luminosity is linearly proportional to the accretion rate, thus predicting a decay slope of \( L_{\rm BZ} \propto \dot{M} \propto t^{-5/3} \approx t^{-1.67} \), closely matching the fitted results.
\end{enumerate}

This phenomenon suggests that the slope of the light curve during the decay phase serves as a critical parameter for distinguishing between the different mechanisms of the central engine of GRBs. We next integrate actual data from single-pulse GRBs exhibiting FRED structures with the theoretical curves outlined above to perform a fitting analysis, with the aim of evaluating the model's potential for differentiating these mechanisms.

\section{Samples and methods}
\label{section3}
To ensure clear and well-defined light curves with reliable fitting, this study focuses on single-peaked GRB events that exhibit a typical fast rise exponential decay (FRED) profile. Here, a ``single-peaked GRB'' refers to the emergent overall profile that results from the superposition of multiple local dissipation episodes throughout the GRB duration, with each episode contributing to the overall pulse. We assume that the peak flux of this variation is proportional to the instantaneous energy release of the central engine. In this case, the overall pulse profile primarily traces the temporal evolution of the central engine. This provides a physical basis for our statistical study of the GRB central engine based on the pulse profile. This simple structure improves fitting accuracy and facilitates mechanism identification, while also aiding the identification and interpretation of the underlying physical mechanisms.

To connect jet power with observed pulse shapes, we derive the observed variability timescale \( t_{\text{obs}} \) based on radius \( R \), Lorentz factor \( \Gamma \), and redshift \( z \) of the emitting region. The relationship is given by \( t_{\text{obs}} = \frac{R(1+z)}{2c\Gamma^2} \). For typical GRB parameters, such as \( \Gamma \sim 100-300 \) and \( R \sim 10^{13} \, \text{cm} \), the timescale is approximately 0.01 s. If the radiation region extends to \( R \sim 10^{14} \, \text{cm} \), \( t_{\text{obs}} \) would be about 0.17 s. This suggests that even a single-peaked pulse contains multiple dissipation events, with the overall decay trend linked to energy injection from the central engine. Thus, the decay provides a reliable indicator of the central engine's temporal power evolution.

The light curve data used in this study are from the Burst Alert Telescope (BAT) on the \textit{Swift} satellite from December 2004 to May 2025 \citep{2016ApJ...829....7L}. This instrument provides multi-channel count rate data that have been processed with mask-weighting, offering high time resolution. The four energy bands of BAT are as follows: 15--25 keV, 25--50 keV, 50--100 keV and 100--350 keV. To reflect the total luminosity variation of the GRB, we combined the data from all four energy bands and analysed the broadband light curves. Additionally, \textit{Swift} offers light curve data with varying time resolutions. By comparing these light curves, we found that the 2 ms and 8 ms resolution light curves are suitable for short GRBs. However, the 1 s resolution light curves, with fewer data points in the pulse regions, result in rougher curves and may exhibit complex pulse overlaps. Therefore, we selected 64 ms resolution data to identify and analyse the FRED single-pulse structure.

We selected GRB events with single-pulse structures and 64 ms resolution from a specific database.\footnote{\url{https://swift.gsfc.nasa.gov/archive/grbtable/}} A total of 85 single-pulse samples were collected, including 39 with redshift. For GRBs lacking redshift, we estimate the pseudo-redshift using the standard $\Lambda$CDM cosmological parameters (\( H_0 = 70 \, \text{km} \, \text{s}^{-1} \, \text{Mpc}^{-1} \), \( \Omega_m = 0.27 \), \( \Omega_\Lambda = 0.73 \) \citep{2003ApJS..148..175S}) in combination with the Yonetoku relation. This approach ensures that the calculated luminosity evolution of the light curves can be compared with the theoretical light curves.

\begin{figure*}[t]
\centering
\captionsetup[subfigure]{aboveskip=-8pt, belowskip=-1pt, margin=0pt}
\setlength{\belowcaptionskip}{-10pt} 
\setlength{\tabcolsep}{-2pt} 
\begin{subfigure}{0.32\textwidth} 
    \centering
    \includegraphics[width=\linewidth]{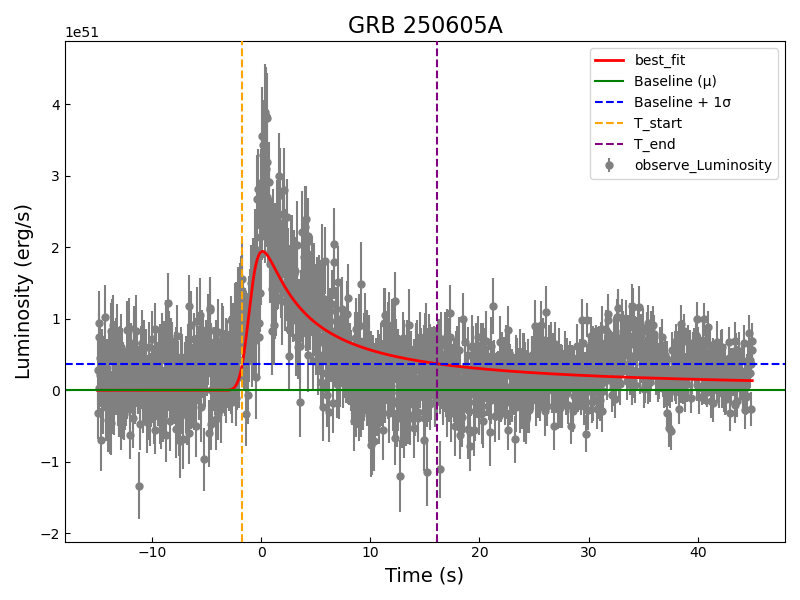}
    \label{fig:sub1}
\end{subfigure}\hfill
\begin{subfigure}{0.32\textwidth}
    \centering
    \includegraphics[width=\linewidth]{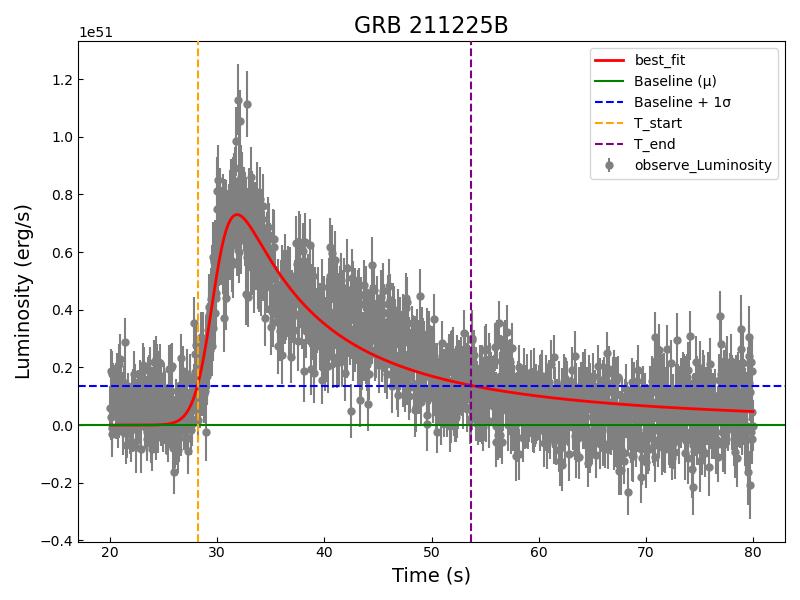}
    \label{fig:sub2}
\end{subfigure}\hfill
\begin{subfigure}{0.32\textwidth}
    \centering
    \includegraphics[width=\linewidth]{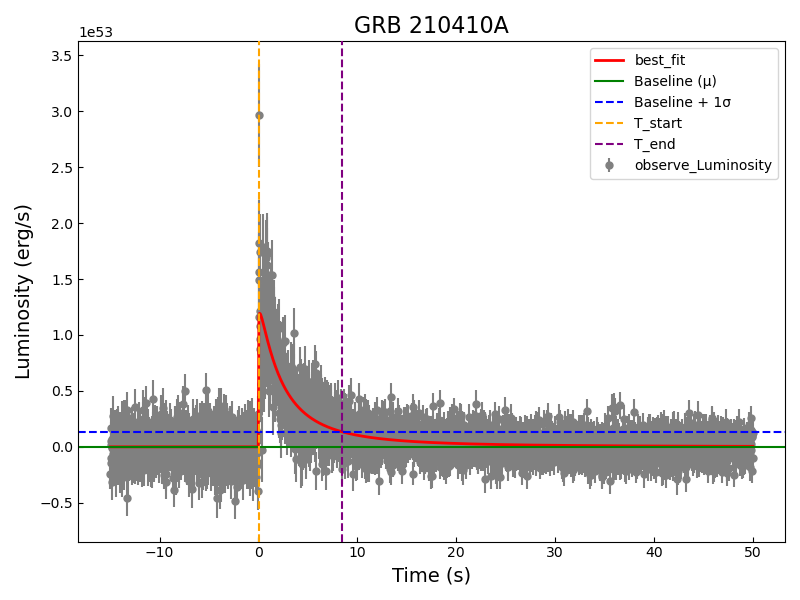}
    \label{fig:sub3}
\end{subfigure}

\vspace{5pt}
\begin{subfigure}{0.32\textwidth}
    \centering
    \includegraphics[width=\linewidth]{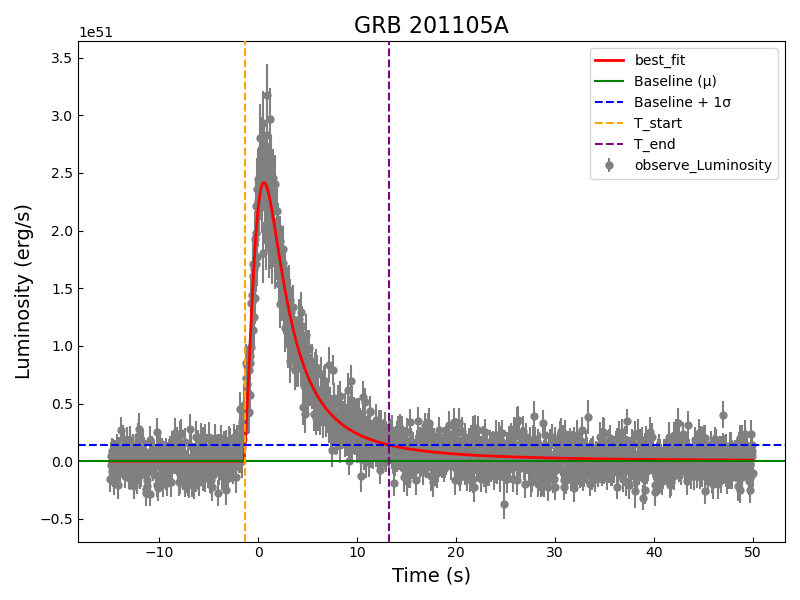}
    \label{fig:sub4}
\end{subfigure}\hfill
\begin{subfigure}{0.32\textwidth}
    \centering
    \includegraphics[width=\linewidth]{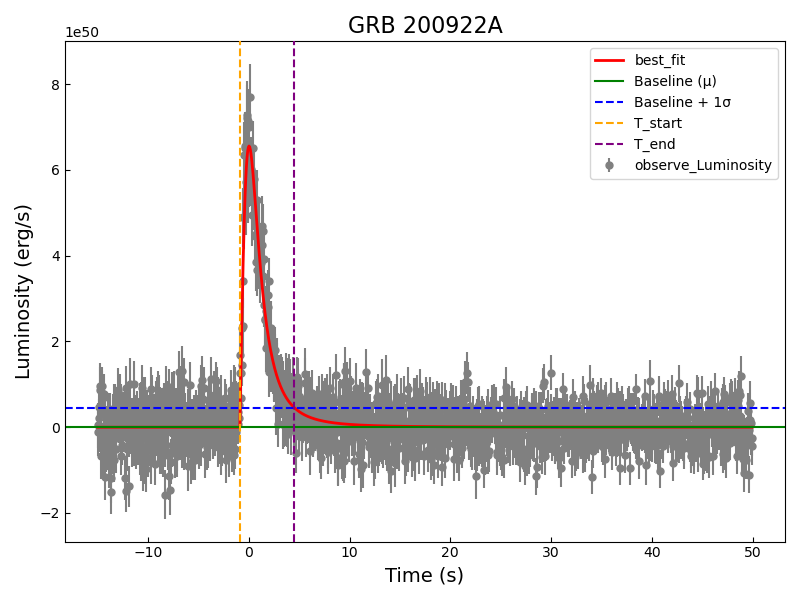}
    \label{fig:sub5}
\end{subfigure}\hfill
\begin{subfigure}{0.32\textwidth}
    \centering
    \includegraphics[width=\linewidth]{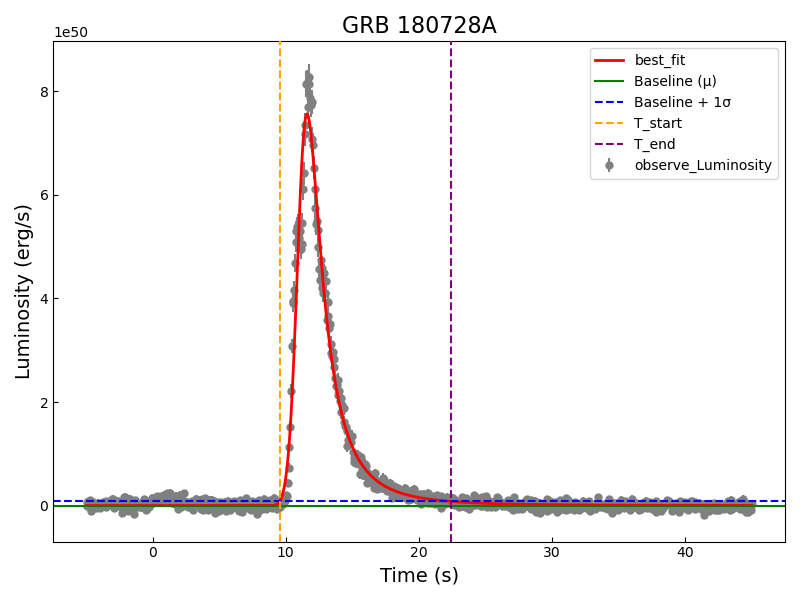}
    \label{fig:sub6}
\end{subfigure}

\vspace{5pt}
\begin{subfigure}{0.32\textwidth}
    \centering
    \includegraphics[width=\linewidth]{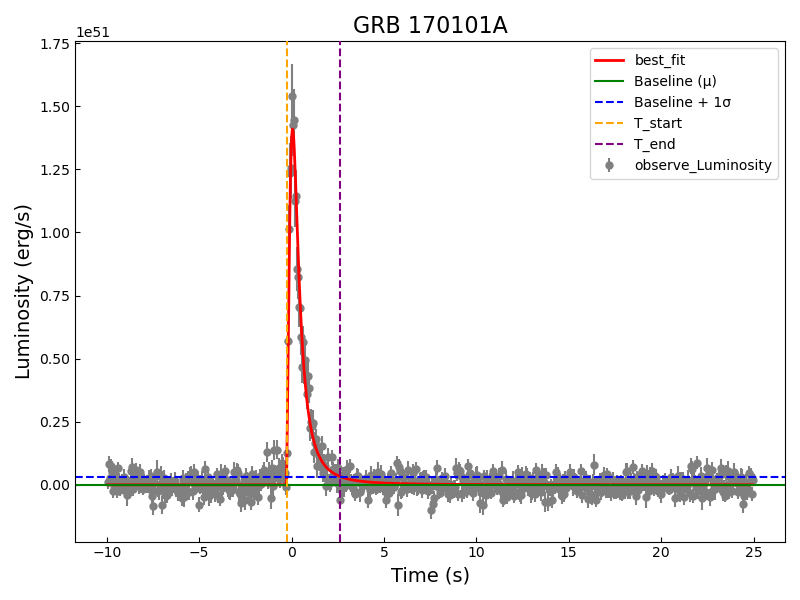}
    \label{fig:sub7}
\end{subfigure}\hfill
\begin{subfigure}{0.32\textwidth}
    \centering
    \includegraphics[width=\linewidth]{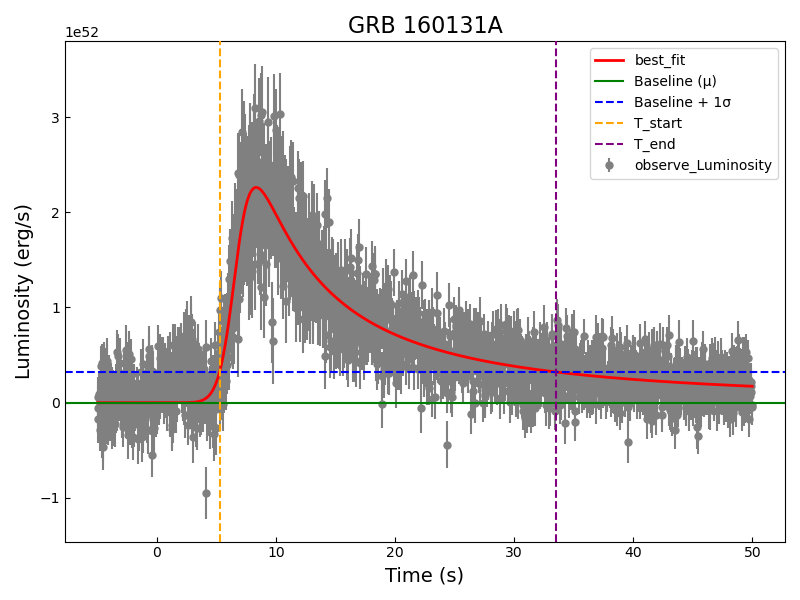}
    \label{fig:sub8}
\end{subfigure}\hfill
\begin{subfigure}{0.32\textwidth}
    \centering
    \includegraphics[width=\linewidth]{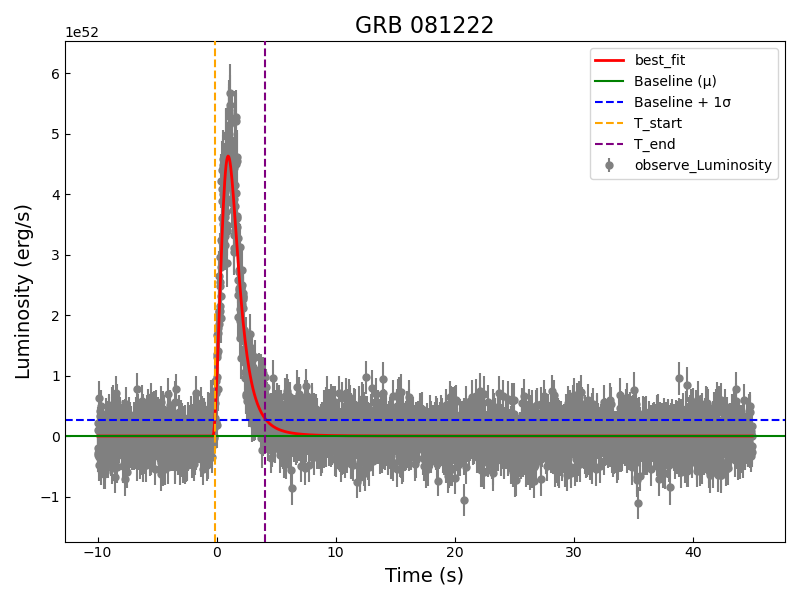}
    \label{fig:sub9}
\end{subfigure}

\caption{Examples of fitting light curves (1--10\,000 keV) to the KRL model (red lines) for GRB 081222, GRB 161218A, GRB 170101A, GRB 180728A, GRB 200922A, GRB 201105A, GRB 210410A, GRB 211225B and GRB 250605A. The dashed vertical lines indicate the pulse start and end times. The observed data (shown in grey) have been corrected for redshift, and the corresponding fitting parameters are listed in Table~\ref{table1}.}
\label{fig3}
\end{figure*}


To analyse the structural characteristics of GRB light curves in the prompt phase, we use the MCMC method with the empirical Kocevski–Ryde–Liang (KRL) function to fit the pulses and evaluate the goodness of fit for the FRED profile using the minimum chi-square method \citep{2003ApJ...596..389K,2010ApJ...725.2209L,2013PASP..125..306F}.
\begin{equation}
F(t) = F_{\rm p} \left( \frac{t + t_0}{t_{\rm p} + t_0} \right)^r \left[ \frac{d}{r + d} + \frac{r}{r + d} \left( \frac{t + t_0}{t_{\rm p} + t_0} \right)^{r+1} \right]^{-\frac{r + d}{r + 1}}.
\end{equation}

The KRL function consists of five adjustable parameters that allow for flexible fitting of light curves with various shapes. Specifically, \( F_{\rm p} \) represents the peak luminosity, \( t_{\rm p} \) is the peak time of the pulse, and \( r \) (\( d \)) are the power-law indices for the rise (decay) phase, respectively. This equation is valid for \( t > -t_0 \), where \( t_0 \) represents the offset of the pulse start time relative to the trigger time. The temporal boundaries of the pulse (\( t_{\text{start}} \) and \( t_{\text{end}} \)) are determined by the signal-to-noise ratio threshold (\( \mathrm{S/N} = 1 \)), corresponding to the initial and final epochs when the pulse profile intersects the detection threshold.

To determine \( t_0 \), we adopt an empirical method similar to that of \citet{2003ApJ...596..389K}, in which 10\% of the standard deviation of the background noise is added to its mean, i.e., \( t_0 = \mu + 0.1 \times \sigma \), where \( \mu \) is the mean and \( \sigma \) is the standard deviation of the background noise. The signal onset is defined as the point where the signal first exceeds this threshold. This method simplifies the KRL function to four adjustable parameters. By applying the \( t_0 \)-corrected KRL function, we can directly compare its parameters with those of theoretical models, such as the BZ mechanism or the NDAF models, which assume that the light curve evolves from \( t = 0 \) without accounting for the trigger delay.

The data processing procedure is as follows:
\begin{enumerate}
    \item The count rates (\( \text{counts} \, \text{s}^{-1} \, \text{det}^{-1} \))\footnote{``det'' refers to a detector area of \( 0.4 \times 0.4 = 0.16 \, \text{cm}^2 \).} for each energy channel are converted to energy flux (\( \text{erg} \, \text{cm}^{-2} \, \text{s}^{-1} \)) using the best-fitting model (power law or broken power law).

    \item For samples lacking redshift, we estimate the pseudo-redshift using the Yonetoku relation. The flux values from the four energy bands are then combined, and \( K \)-correction is applied to obtain the light curve showing the luminosity evolution over time.
    
    \item We determine the pulse start time for each sample and apply MCMC with the KRL model (with \( t_0 \) fixed) to fit the pulses. The observed time \( T_{\rm obs} \) is converted to rest-frame time \( T_{\rm rest} = \frac{T_{\rm obs}}{1+z} \) (with \( z \) being the source redshift), and the decay slope parameter \( d \) is derived.
\end{enumerate}

\setlength{\unitlength}{1pt} 
\begin{figure}[ht]
  \centering
  \begin{minipage}[b]{0.45\textwidth}\centering
    \includegraphics[width=\textwidth]{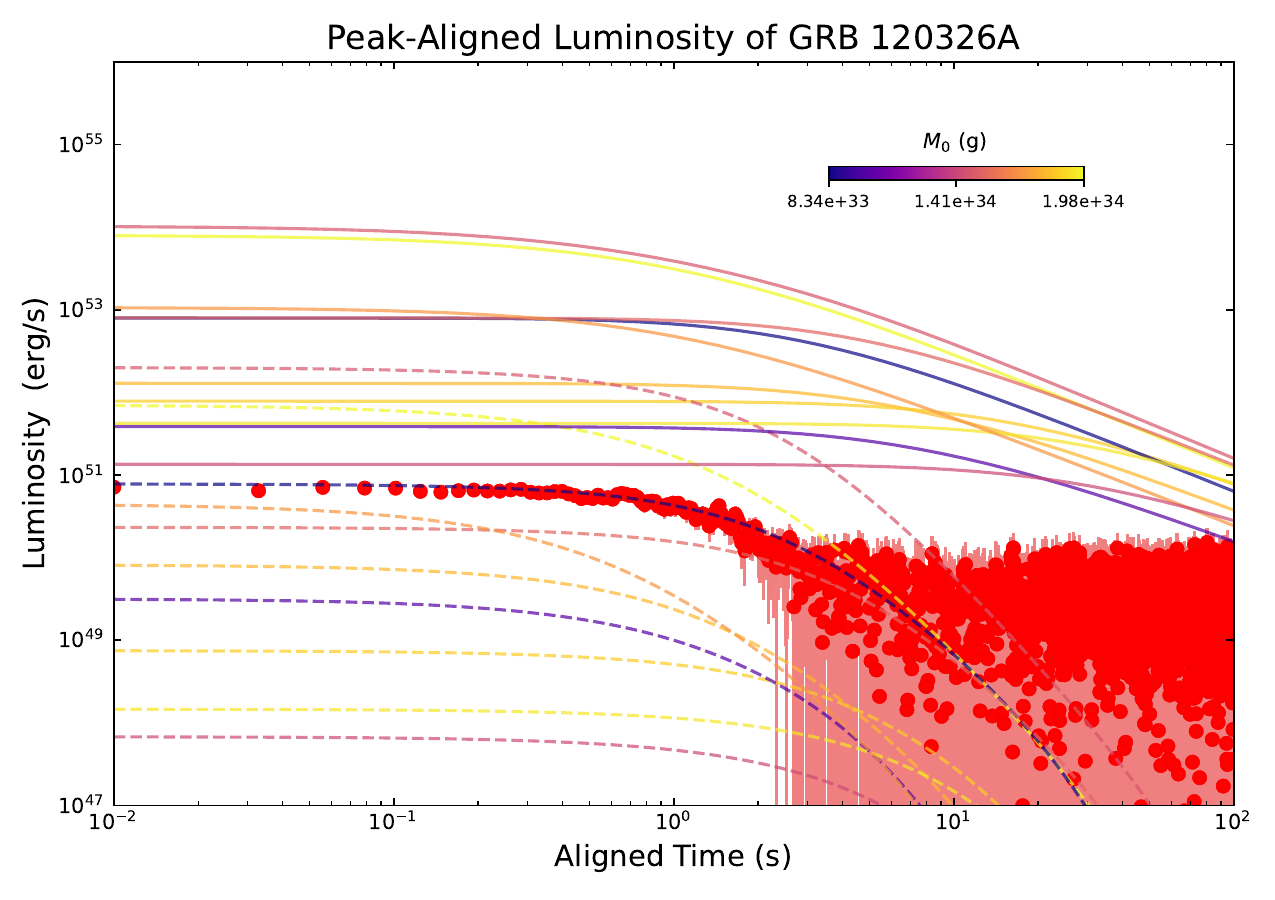}%
    \begin{picture}(0,0)
      \put(-30,140){\textbf{(a)}} 
    \end{picture}
  \end{minipage}\hspace{0.05\textwidth}
  \begin{minipage}[b]{0.45\textwidth}\centering
    \includegraphics[width=\textwidth]{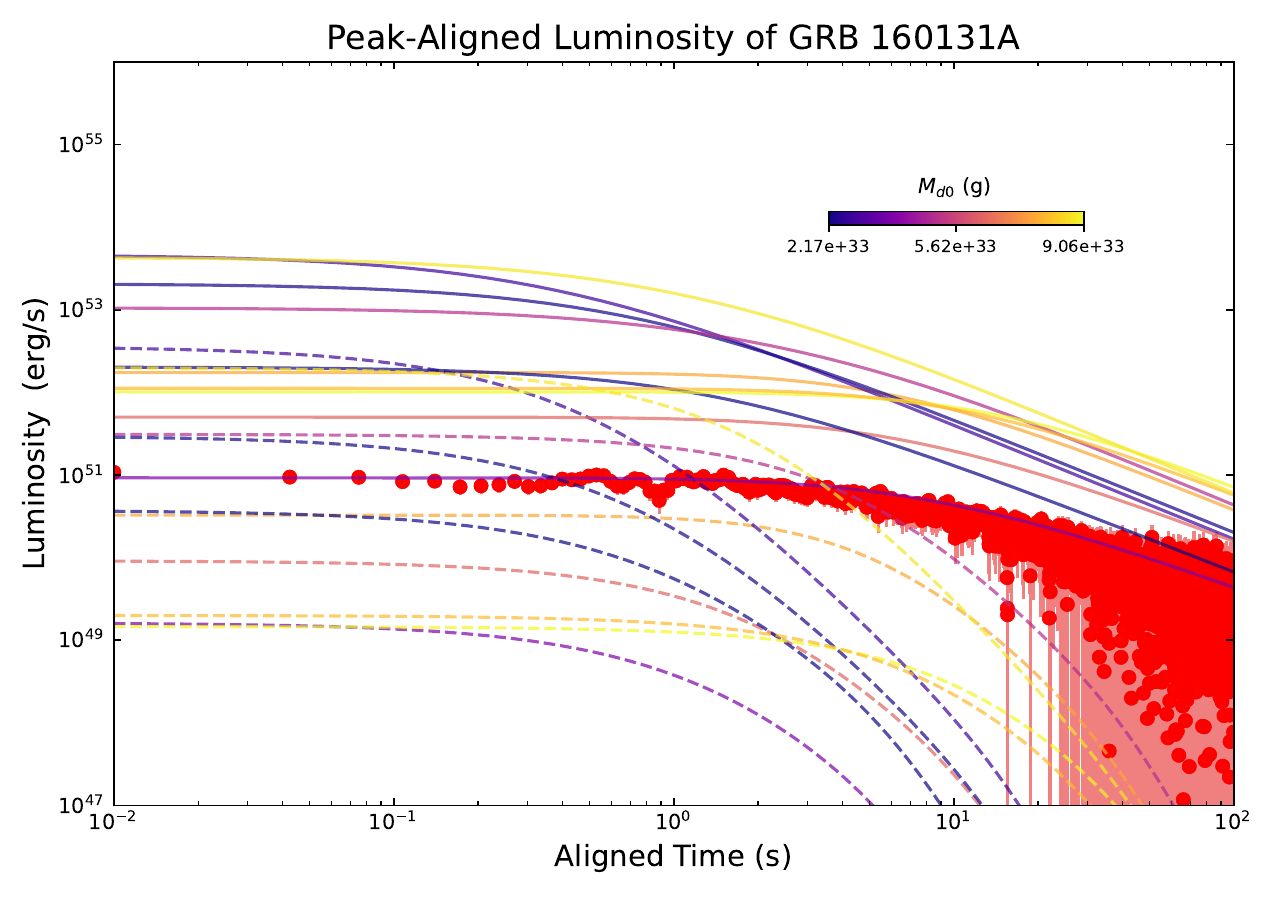}%
    \begin{picture}(0,0)
      \put(-30,140){\textbf{(b)}}
    \end{picture}
  \end{minipage}

  \caption{Comparison between theoretical models and observational data for (a) GRB~120326A and (b) GRB~160131A. Theoretical light curves are shown for the Blandford–Znajek (BZ) mechanism (solid lines) and the neutrino-dominated accretion flow (NDAF) mechanism (dashed lines), following the same convention as in Figures~\ref{fig1} and~\ref{fig2}. Observational data are indicated by red circles, with lighter shading showing the error bars. The upper panel exhibits a steep luminosity decay (\( L_{\nu\bar{\nu}} \propto t^{-3.72}\)--\(t^{-7.83} \)), consistent with NDAF predictions, whereas the lower panel shows a shallower decay (\( L \propto t^{-1.67} \)), matching BZ model expectations.}
  \label{fig4}
\end{figure}
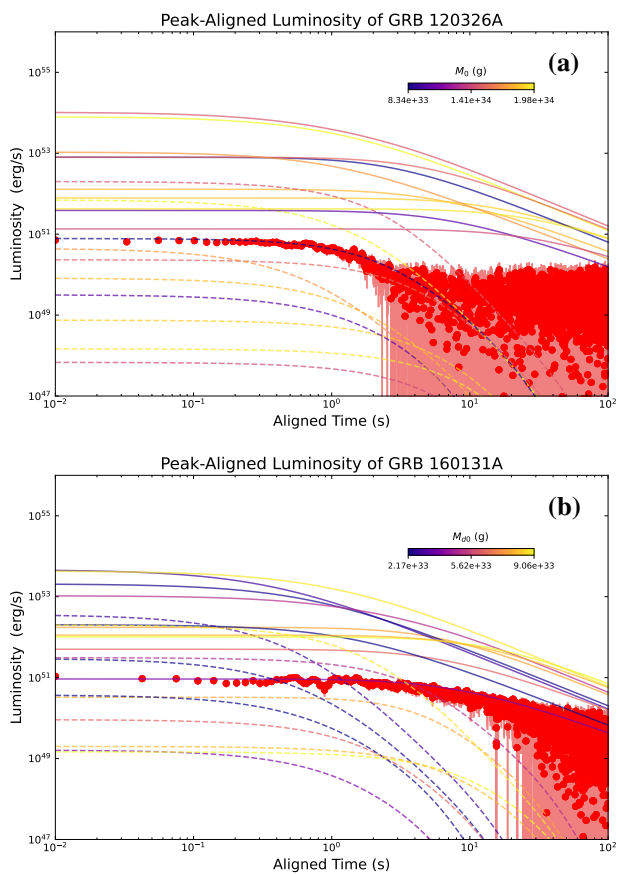

The spectral fitting in this study uses both the power law (PL) and broken power law (CPL) models. The specific forms of these models are as follows:
\begin{equation}
N_{E,\mathrm{CPL}}(E) = N_{0,\mathrm{CPL}} \left( \frac{E}{100~\mathrm{keV}} \right)^{\alpha_{\mathrm{CPL}}} \exp\left( -\frac{E}{E_{\rm p}} \right),
\end{equation}
\begin{equation}
N_{E,\mathrm{PL}}(E) = N_{0,\mathrm{PL}} \left( \frac{E}{100~\mathrm{keV}} \right)^{\alpha_{\mathrm{PL}}},
\end{equation}
where \( N_{E,\mathrm{CPL}}(E) \) and \( N_{E,\mathrm{PL}}(E) \) represent the differential photon spectra for the CPL and PL models, respectively \citep{2007ApJ...660...16S}. Here, \( \alpha_{\mathrm{CPL}} \) and \( \alpha_{\mathrm{PL}} \) are the spectral indices, and \( E_{\rm p} \) is the break energy for the CPL model.

The count rate is converted into flux using the following equation:
\begin{equation}
F_{\rm p} = P_{\rm p} \frac{\int_{E_{\text{min}}}^{E_{\text{max}}} E N(E) \, dE}{\int_{E_{\text{min}}}^{E_{\text{max}}} N(E) \, dE},
\end{equation}
where \( F_{\rm p} \) is the flux (\( \text{erg} \, \text{cm}^{-2} \, \text{s}^{-1} \)), \( P_{\rm p} \) is the photon peak flux (\( \text{photons} \, \text{cm}^{-2} \, \text{s}^{-1} \)), and the integrals are performed over the energy range \( [E_{\text{min}}, E_{\text{max}}] \). These integrals are evaluated over the energy intervals \( 15\)--\(25 \, \text{keV} \), \( 25\)--\(50 \, \text{keV} \), \( 50\)--\(100 \, \text{keV} \) and \( 100\)--\(350 \, \text{keV} \). The total flux is then computed by summing the contributions from each range.

The isotropic luminosity \( L_{\text{iso}} \) is then calculated as
\begin{equation}
L_{\text{iso}} = 4\pi D_{\rm L}^2 F_{\rm p} k,
\end{equation}
where \( D_{\rm L} \) is the luminosity distance and \( k \) is the \( K \)-correction factor, which accounts for the cosmological redshift and is given by
\begin{equation}
k = \frac{\int_{1/(1+z)}^{10^4/(1+z)} E N(E) \, dE}{\int_{E_{\text{min}}}^{E_{\text{max}}} E N(E) \, dE},
\end{equation}
where \( z \) is the redshift, \( [E_{\text{min}}, E_{\text{max}}] = 15\)--\(350 \, \text{keV} \), and both integrals are evaluated in the source frame.

We estimate the spectral peak energy \( E_{\rm p} \) using the empirical \( E_{\rm p} - L \) correlation \citep{2004ApJ...609..935Y,2012MNRAS.421.1256N}
\begin{equation}
\log \left[ E_{\rm p} (1 + z) \right] = -25.33 + 0.53 \log L,
\end{equation}
where \( L \) is the isotropic luminosity and \( z \) is the redshift.

\section{Results}
\label{section4}

To investigate the differences in light-curve morphology between NDAF and BZ mechanisms, we performed spectral fitting using the KRL function and analysed 85 single-pulse GRBs. The results of the light-curve fits, along with the isotropic gamma-ray energy (\( E_{\gamma,\mathrm{iso}} \)) and duration (\( T_{90} \)), are listed in Table~\ref{table1}.

Figure~\ref{fig3} compares the observed data (black) with the fitted KRL models (red lines), highlighting the decay behaviours characteristic of each mechanism. The dotted lines indicate the pulse start and end times, and the data have been corrected for redshift. The energy range for the fitting was initially 15--350 keV and was then extended to the full energy range of 1--10\,000 keV through \( K \)-correction. Each panel represents a unique GRB, illustrating the applicability of the KRL model across varying attenuation slopes and mechanisms, and providing insights into the physical processes governing the light-curve evolution.

To further assess the accuracy of the model, Figure~\ref{fig4} compares the theoretical light curves of typical samples with the observed data after aligning their peak values. The \( x \)-axis represents the observation time corrected for redshift, i.e., \( T/(1+z) \). In the upper panel of Figure~\ref{fig4}, the red curve represents GRB~120326A, which exhibits an observed decay slope of \( d \approx 4.49 \), matching the steep luminosity decline predicted for the NDAF mechanism (\( d \approx 3.7\)--\(7.8 \)). Conversely, the lower panel shows GRB~160131A, with a shallower decay slope of \( d \approx 1.59 \), in close agreement with the BZ mechanism prediction (\( d \approx 1.67 \)). Theoretical and observational results are in strong agreement, providing robust evidence for these mechanisms in explaining the observed GRB light curves.

\begin{figure}[t]
\centering
\includegraphics[width=0.5\textwidth]{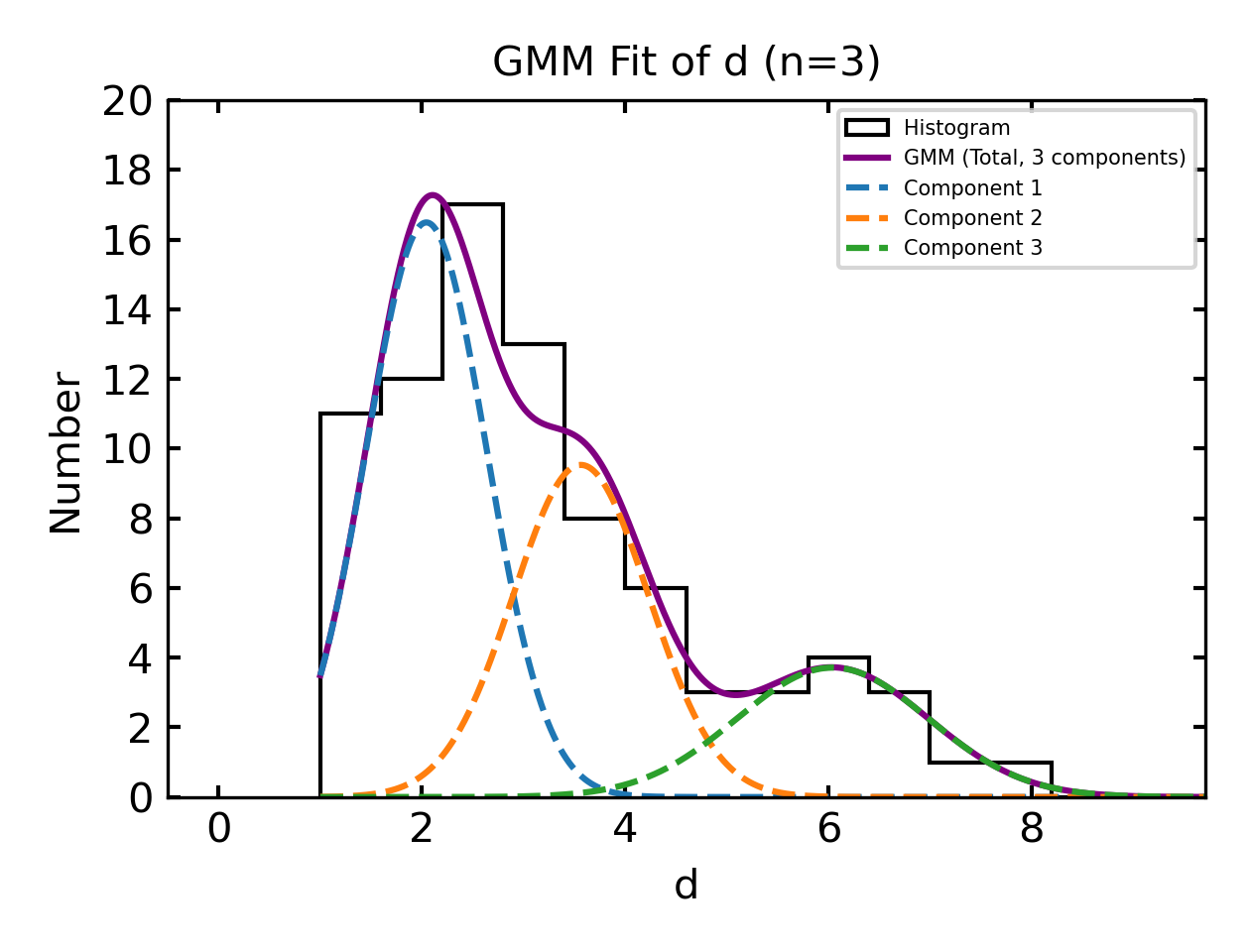}
\caption{Distribution of decay slopes \( d \) for 85 GRBs, fitted with a three-component Gaussian mixture model (GMM). Three extreme cases (GRB~070306, GRB~140209A and GRB~161004B) with \( d > 7.8 \) were excluded to avoid skewing the high-\( d \) tail due to potential observational or physical anomalies, and their removal does not affect the main results. The best-fit components peak at \( d \approx 2.04 \), \( d \approx 3.56 \) and \( d \approx 6.04 \) with relative weights of 0.50, 0.32 and 0.18, respectively. Most events fall in the range \( 2 < d < 4 \), suggesting hybrid central engines, differing from idealised BZ (\( d \approx 1.67 \)) or NDAF (\( d \approx 3.7\)--\(7.8 \)) predictions.}
\label{fig5}
\end{figure}

In a sample of 85 GRBs, we find that the decay slopes of 15 bursts (e.g., GRB~050717, GRB~070318, GRB~070808, GRB~080805, GRB~090530) are consistent with the BZ mechanism, while 22 bursts (e.g., GRB~081222, GRB~090129, GRB~091018, GRB~110318A, GRB~120213A, GRB~120326A) are more in line with the NDAF mechanism (see Table~\ref{table1}). However, based on the distribution of decay slopes across all samples (Figure~\ref{fig5}), most bursts cluster in the range \( 2 < d < 4 \). This range lies between the typical values predicted by the NDAF (\( d \approx 3.7\)--\(7.8 \)) and BZ (\( d \approx 1.67 \)) mechanisms shown in Figures~\ref{fig1} and~\ref{fig2}, indicating that many events cannot be classified into a single mechanism solely from their decay slopes. For the statistical analysis in Figure~\ref{fig5}, we excluded three extreme cases (GRB~070306, GRB~140209A and GRB~161004B) with \( d > 7.8 \), as such steep slopes are likely influenced by observational uncertainties or atypical physical conditions, and would disproportionately skew the high-\( d \) tail of the distribution. A Gaussian mixture model (GMM) fit to the distribution yields three peaks at \( d \approx 2.04 \) (weight 0.50), \( d \approx 3.56 \) (weight 0.32) and \( d \approx 6.04 \) (weight 0.18). The first two peaks fall within \( 2 < d < 4 \), while the third represents the high-\( d \) tail of the distribution.

\section{Discussion}
\label{section5}

The interpretation of GRB pulse profiles should account for both local and global physical mechanisms. Models involving high-latitude emission from a spherical shell or localized magnetic reconnection events \citep{2009MNRAS.399.1328G,2015ApJ...808...33U,2016ApJ...824L..16U,2024ApJ...963L..30U} can produce a decaying trend without invoking a decaying jet power, and these processes likely govern the fine-scale temporal structure (e.g., micro-spikes) observed within pulses. However, in this work, we focus on the macroscopic envelope of the pulse, which is primarily influenced by the global energy injection history of the central engine. We suggest that the global decay trend, particularly over timescales longer than those governed by local dissipation processes, may trace the temporal evolution of the jet power. While local processes, such as high-latitude emission and magnetic reconnection, govern the rapid fluctuations within the pulse, the overall decay remains coupled to the global energy injection history of the central engine. We argue that the overall decay trend remains statistically indicative of the temporal evolution of the central engine.

This framework motivates the analysis of the temporal decay slope as a proxy for the engine's behavior. The observed concentration of intermediate slopes (\( 2 < d < 4 \)) may reflect hybrid central engine mechanisms or transitional accretion regimes operating in a significant fraction of bursts. Assuming that the evolution of the light curve is primarily driven by accretion processes, this slope can be used to differentiate between the two mechanisms (see Figures~\ref{fig1} and~\ref{fig2}). When the photometric attenuation does not reach a full order of magnitude, the measured slope is often intermediate between the two canonical values. In such cases, the trend in our sample leans toward NDAF-like behaviour. Instrumental factors such as orientation effects or calibration uncertainties can also introduce data gaps that obscure the intrinsic decay pattern. For bright, long-duration events with high-quality light curves, however, the slope remains a useful diagnostic for distinguishing between the two mechanisms.

A Gaussian mixture model (GMM) fit to the slope distribution supports this interpretation: while a two-component model is favoured by the Bayesian information criterion (BIC), a three-component fit reveals peaks near \( d \approx 2.04 \), \( 3.56 \) and \( 6.04 \), with the high-\( d \) tail likely tracing rare NDAF-dominated bursts or specific geometric and radiative configurations (Figure~\ref{fig5}). The concentration of slopes in the \( 2 < d < 4 \) range persists regardless of whether two- or three-component fits are adopted, reinforcing the prevalence of intermediate decay behaviours. Physical processes beyond the idealised models may further shape the observed slopes. Strong disc outflows, for instance, can remove a substantial fraction of the accreting mass before it reaches the black hole \citep{2018ApJ...852...20L}, lowering the late-time accretion rate and steepening the decay.

The light-curve slope may be influenced by magnetic field variations during the BZ process, such as attenuation or sudden disappearance of the magnetic field, which could increase the slope steepness. While magnetic flux is assumed constant for simplicity, variations may affect the slope. We identify GRBs such as GRB~170317A, GRB~141004A, GRB~120521C, GRB~120326A, GRB~100814A, GRB~100704A, GRB~080430, GRB~071003, GRB~070306, GRB~061202 and GRB~060904B, which exhibit plateaus and a steep decay for afterglows, indicative of the magnetar model \citep{2004ApJ...606.1000D,2006ApJ...642..354Z}. Magnetic field evolution could alter the prompt light-curve structure, influencing intensity and spectral features. The correlation between the prompt and afterglow phases requires further investigation through multi-band data (optical, X-ray, radio) \citep{2016ApJS..224...20Y,2021MNRAS.507.1047Y,2022ApJ...924...69Y} and numerical simulations. Our analysis focuses on the prompt phase; if the afterglow phase involves continued accretion or fallback, the actual accreted mass could be higher, consistent with conclusions drawn from X-ray afterglow analyses (e.g., \citealt{2013ApJ...767L..36W,2014ApJ...781L..19H,2016ApJ...826..141G}). Additionally, the complexity of GRB pulse structures should be considered. Some GRB pulses may result from the superposition of multiple components, complicating light-curve analysis and making mechanism differentiation more difficult. Small peak structures, typically generated by internal shock wave collisions, can subtly influence the overall light-curve behaviour, despite being unrelated to the primary interaction. The choice of binning, fitting interval range and initial time point \( t_0 \) may also influence the slope calculation, though their effects are generally minimal.

In summary, our study provides new insights into GRB central engine mechanisms through light-curve slope analysis. Under low photometric attenuation, slopes in the transition zone between the BZ and NDAF mechanisms are harder to distinguish, yet well-sampled, bright events retain clear diagnostic value. Targeted multi-wavelength campaigns, combined with prompt--afterglow correlations, will be essential to connect slope evolution with magnetic field dynamics and the underlying engine activity.


\section{Conclusions}
\label{section6}
We propose a method to distinguish between the BZ and NDAF mechanisms through analysis of GRB light-curve decay slopes. Our results show that the BZ mechanism typically follows \( L_{\mathrm{BZ}} \propto t^{-1.67} \), while the NDAF mechanism exhibits slopes ranging from \( L_{\nu\bar{\nu}} \propto t^{-3.7} \) to \( t^{-7.8} \), highlighting the decay slope as a key diagnostic. From 85 \textit{Swift}/FRED single-pulse GRB samples at 64 ms resolution, 15 events are consistent with the BZ mechanism and 22 with the NDAF mechanism. These results suggest that the NDAF mechanism explains the majority of GRBs in our sample, complementing previous studies that indicate high-luminosity GRBs may require the BZ mechanism \citep{2015ApJS..218...12L,2017JHEAp..13....1Y}. The contrasting decay behaviours align with differences in accretion processes: the BZ mechanism produces smoother decays via gradual energy extraction, whereas the NDAF mechanism yields steeper decays due to rapid accretion rate decline in the present model framework.

While these findings establish a potential link between decay slope and central engine mechanism, they also motivate further investigation. Several limitations remain, including the thin-disc approximation and the assumption of purely single-mechanism models. Future studies should relax these assumptions and incorporate multi-band observations, together with correlations between X-ray afterglows, gravitational wave signals and central engine activity. Model refinements that account for hybrid mechanisms, disc precession and magnetic reconnection, supported by new telescopes, improved calibration and multi-band data, combined with numerical simulations, will be essential for a more complete understanding of GRB light-curve formation, enhancing the accuracy of mechanism identification and deepening our understanding of the underlying GRB physics.

\begin{acknowledgements}
     \textcolor{black}{ We thank Xiu-Juan Li, Wen-Long Zhang and Cui-Ying Song for the helpful discussions. This work is supported by Shandong Provincial Natural Science Foundation (ZR2025MS47, ZR2021MA021), the National Natural Science Foundation of China (Grant Nos. 12494575, 12473012 and 12494572), the Natural Science Foundation of Jiangxi Province of China (grant No. 20242BAB26012), the National Natural Science Foundation of China under grants 12473012 and 12533005, and the National Key R\&D Program of China (No. 2023YFC2205901) and Manned Spaced Project (CMS-CSST-2021-A12).}

\end{acknowledgements}

\bibliography{aa631}{} 
\bibliographystyle{aa}


\begin{appendix} 

\begin{sidewaystable*}
\tiny
\setlength{\abovecaptionskip}{0pt}
\setlength{\belowcaptionskip}{-5pt}
\caption{Fitting parameters of 2004-2025 FRED-individual pulses in Swift.}
\centering
\tabcolsep 2pt
\begin{tabular}{l@{\hspace{3pt}}c@{\hspace{3pt}}c@{\hspace{3pt}}ccc@{\hspace{3pt}}cc@{\hspace{3pt}}c@{\hspace{3pt}}c@{\hspace{3pt}}cccc@{\hspace{3pt}}c@{\hspace{5pt}}c@{\hspace{5pt}}cc@{\hspace{10pt}}c@{\hspace{3pt}}cc}

\hline\hline
GRB & $T_{90}$ & z & Model & $\alpha$ & $E_{\text{peak}}$ & Fluence & $E_{\gamma,\text{iso}}$ & $T_{0}$ & $T_{start}$ & $T_{end}$ & $T_{p}$ & $L_{p}$ & r & d & $\chi^2$ & dof & Red-$\chi^2$ & BZ & & NDAF \\
\hline
& (s) & & & & (keV) & ($10^{-7}$erg/cm$^2$) & ($10^{52}$erg) &  & & & & ($10^{50}$erg/s) & & & & & & & \\
(1)&(2)&(3)&(4)&(5)&(6)&(7)&(8)&(9)&(10)&(11)&(12)&(13)&(14)&(15)&(16)&(17)&(18)&(19)&&(20)\\
\hline

$050717^{\ddagger}$ & $85$ & $2$ & PL & $1.3$ & \textit{--} & $129^{+6.53}_{-6.26}$ & $67.634^{+3.42}_{-3.3}$ & $1.34$ & $0.62$ & $10.23$ & $1.52^{+0.06}_{-0.06}$ & $498.9^{+25.37}_{-22.56}$ & $1.18^{+0.17}_{-0.13}$ & $1.3^{+0.06}_{-0.06}$ & $4353.62$ & $3512$ & $1.24$ & $\checkmark$ & \\ 
\hline
050724 & $96$ & $0.26$ & PL & $1.89$ & \textit{--} & $14.2^{+2.84}_{-2.44}$ & $702^{+0.014}_{-1.2}$ & $-0.08$ & $-0.03$ & $0.55$ & $0.04^{+0.01}_{-0.01}$ & $5.67^{+0.74}_{-0.7}$ & $3.75^{+1.64}_{-0.99}$ & $1.96^{+0.6}_{-0.34}$ & $186.4$ & $153$ & $1.22$  & & \textit{--}  & \\
\hline
051111 & $46.1$ & $1.55$ & PL & $1.32$ & \textit{--} & $84.6^{+5.48}_{-5.21}$ & $29.978^{+1.94}_{-1.8}$ & $-6.84$ & $-2.24$ & $6.99$ & $-0.26^{+0.08}_{-0.08}$ & $174.84^{+5.57}_{-5.51}$ & $1.18^{+0.13}_{-0.12}$ & $2.07^{+0.19}_{-0.16}$ & $2893.88$ & $2983$ & $0.97$  & & \textit{--}  & \\
\hline
$060110^{\dagger}$ & $26$ & $0.59$ & PL & $1.64$ & \textit{--} & $24.6^{+1.75}_{-1.65}$ & $0.88^{+0.0626}_{-5.9}$ & $-1.35$ & $-0.64$ & $9.43$ & $1.47^{+0.14}_{-0.13}$ & $6.58^{+0.32}_{-0.32}$ & $0.86^{+0.15}_{-0.13}$ & $2.53^{+0.66}_{-0.42}$ & $1353.92$ & $1348$ & $1$  & & \textit{--}  & \\
\hline
060904B & $171.5$ & $0.7$ & PL & $1.64$ & \textit{--} & $27.7^{+4.22}_{-3.78}$ & $1.402^{+0.214}_{-1.9}$ & $-1.2$ & $-0.34$ & $4.31$ & $0.89^{+0.08}_{-0.08}$ & $15.47^{+0.91}_{-0.9}$ & $1.44^{+0.31}_{-0.25}$ & $3.19^{+1.18}_{-0.59}$ & $1952.75$ & $1991$ & $0.98$  & & \textit{--}  & \\
\hline
060912A & $5$ & $0.94$ & PL & $1.74$ & \textit{--} & $21.2^{+1.61}_{-1.51}$ & $1.596^{+0.121}_{-1.1}$ & $-0.49$ & $-0.19$ & $2.29$ & $0.15^{+0.02}_{-0.02}$ & $76.76^{+4.26}_{-4.14}$ & $1.38^{+0.26}_{-0.21}$ & $2^{+0.24}_{-0.19}$ & $2165.75$ & $2265$ & $0.96$  & & \textit{--}  & \\
\hline
$061202^{\dagger}$ & $91.2$ & $0.69$ & PL & $1.58$ & \textit{--} & $59.4^{+4.05}_{-3.83}$ & $3.271^{+0.223}_{-2.1}$ & $37.21$ & $42.42$ & $57.2$ & $45.04^{+0.13}_{-0.13}$ & $13.93^{+0.37}_{-0.37}$ & $29.05^{+2.77}_{-2.68}$ & $3.58^{+0.13}_{-0.12}$ & $2179.31$ & $1982$ & $1.1$  & & \textit{--}  & \\
\hline
070306 & $209.5$ & $1.5$ & PL & $1.66$ & \textit{--} & $89.8^{+8.2}_{-7.62}$ & $17.009^{+1.55}_{-1.4}$ & $40$ & $36.64$ & $43.96$ & $39.17^{+0.07}_{-0.06}$ & $95.97^{+2.51}_{-2.5}$ & $26.16^{+2.09}_{-1.91}$ & $10.7^{+0.76}_{-0.7}$ & $2345.79$ & $2142$ & $1.1$  & & \textit{--}  & \\
\hline
070318 & $74.6$ & $0.84$ & PL & $1.42$ & \textit{--} & $50.2^{+4.91}_{-4.55}$ & $5.358^{+0.524}_{-4.9}$ & $-1.02$ & $-0.36$ & $7.97$ & $1.01^{+0.15}_{-0.16}$ & $21.11^{+1.25}_{-1.3}$ & $0.56^{+0.16}_{-0.15}$ & $1.36^{+0.23}_{-0.15}$ & $2219.2$ & $2148$ & $1.03$  & $\checkmark$ & \\ 
\hline
$070808^{\dagger}$ & $32$ & $0.97$ & PL & $1.47$ & \textit{--} & $23.7^{+3.7}_{-3.31}$ & $2.931^{+0.458}_{-4.1}$ & $-1.05$ & $-0.43$ & $3.17$ & $0.17^{+0.09}_{-0.1}$ & $34.77^{+3.09}_{-2.81}$ & $0.94^{+0.39}_{-0.28}$ & $1.34^{+0.19}_{-0.14}$ & $2359.07$ & $2301$ & $1.03$  & $\checkmark$ & \\ 
\hline
$070917^{\dagger}$ & $7.3$ & $0.41$ & PL & $1.52$ & \textit{--} & $37.6^{+1.86}_{-1.79}$ & $0.84^{+0.0416}_{-4.0}$ & $-0.13$ & $-0.06$ & $4.75$ & $0.41^{+0.02}_{-0.02}$ & $21.61^{+0.62}_{-0.64}$ & $0.66^{+0.07}_{-0.06}$ & $2.65^{+0.32}_{-0.25}$ & $1710.13$ & $1643$ & $1.04$  & & \textit{--}  & \\
\hline
071003 & $150$ & $1.1$ & PL & $1.36$ & \textit{--} & $164^{+10.6}_{-9.91}$ & $31.636^{+2.04}_{-1.9}$ & $-2.02$ & $-0.94$ & $9.87$ & $0.83^{+0.09}_{-0.09}$ & $163.23^{+5.87}_{-5.71}$ & $0.48^{+0.07}_{-0.06}$ & $3.7^{+1.21}_{-0.7}$ & $2869.64$ & $2457$ & $1.17$ & & \textit{--} & \\
\hline
071010B & $>35.7$  & $0.95$ & CPL & $1.52$ & $54.78$ & $64.1^{+2.04}_{-1.99}$ & $2.991^{+0.0952}_{-9.3}$ & $-1.54$ & $-0.58$ & $7.04$ & $1.06^{+0.03}_{-0.03}$ & $38.76^{+0.69}_{-0.65}$ & $1.27^{+0.06}_{-0.06}$ & $3.19^{+0.24}_{-0.2}$ & $1872.4$ & $1822$ & $1.03$  & & \textit{--}  & \\
\hline
080413B & $8$ & $1.1$ & PL & $1.74$ & \textit{--} & $51.9^{+2.74}_{-2.64}$ & $5.234^{+0.276}_{-2.7}$ & $-1.29$ & $-0.48$ & $1.78$ & $0.07^{+0.01}_{-0.01}$ & $271.94^{+9.31}_{-9.28}$ & $2.33^{+0.21}_{-0.19}$ & $3.72^{+0.43}_{-0.34}$ & $1855.88$ & $1801$ & $1.03$  & & \textit{--}  & \\
\hline
080430 & $16.2$ & $0.75$ & PL & $1.73$ & \textit{--} & $18.2^{+1.37}_{-1.28}$ & $0.911^{+0.0686}_{-6.4}$ & $-0.63$ & $-0.24$ & $4.38$ & $0.84^{+0.09}_{-0.08}$ & $14.03^{+0.75}_{-0.71}$ & $0.75^{+0.16}_{-0.13}$ & $3.04^{+1.29}_{-0.67}$ & $1500.42$ & $1500$ & $1$  & & \textit{--}  & \\
\hline
$080714^{\dagger}$  & $33$ & $0.6$ & PL & $1.52$ & \textit{--} & $46.9^{+3.75}_{-3.53}$ & $2.212^{+0.177}_{-1.7}$ & $-4.2$ & $-1.58$ & $4.89$ & $-0.08^{+0.06}_{-0.06}$ & $22.84^{+1.16}_{-1.13}$ & $3.3^{+0.6}_{-0.51}$ & $2.36^{+0.24}_{-0.19}$ & $1629.34$ & $1495$ & $1.09$ & & \textit{--}  & \\
\hline
080805 & $78$ & $1.5$ & PL & $1.53$ & \textit{--} & $46.5^{+3.92}_{-3.66}$ & $10.706^{+0.903}_{-8.4}$ & $-4.32$ & $-0.8$ & $2.94$ & $0.48^{+0.15}_{-0.15}$ & $51.27^{+2.32}_{-2.26}$ & $0.7^{+0.13}_{-0.11}$ & $1.72^{+0.25}_{-0.19}$ & $2236.73$ & $2149$ & $1.04$  & $\checkmark$ & \\ 
\hline
081222 & $24$ & $2.7$ & CPL & $1.2$ & $217.88$ & $83.9^{+9.64}_{-7.8}$ & $18.477^{+2.12}_{-1.7}$ & $-1.12$ & $-0.15$ & $4.09$ & $0.97^{+0.01}_{-0.01}$ & $462.88^{+6.39}_{-6.46}$ & $1.64^{+0.06}_{-0.06}$ & $4.6^{+0.41}_{-0.32}$ & $3561.89$ & $3176$ & $1.12$ & & &$\checkmark$\\
\hline
$090129^{\dagger}$& $17.5$ & $0.63$ & PL & $1.88$ & \textit{--} & $3.27^{+1.75}_{-1.67}$ & $0.102^{+0.0544}_{-5.2}$ & $-0.27$ & $-0.14$ & $6.52$ & $0.69^{+0.07}_{-0.08}$ & $12^{+0.48}_{-0.46}$ & $0.31^{+0.07}_{-0.06}$ & $4.34^{+2.62}_{-1.31}$ & $1425.17$ & $1398$ & $1.02$  & &  & $\checkmark$\\
\hline
090530 & $48$ & $1.27$ & PL & $1.61$ & \textit{--} & $18.7^{+3.23}_{-2.86}$ & $2.86^{+0.494}_{-4.4}$ & $-0.31$ & $-0.09$ & $1.4$ & $0.05^{+0.02}_{-0.02}$ & $109.94^{+15.15}_{-12.75}$ & $1.96^{+1.1}_{-0.65}$ & $1.11^{+0.14}_{-0.1}$ & $1921.62$ & $1944$ & $0.99$  & $\checkmark$ & \\ 
\hline
091018 & $4.4$ & $0.97$ & CPL & $1.77$ & $19.43$ & $18.6^{+0.71}_{-0.69}$ & $2.756^{+0.106}_{-1.0}$ & $-0.54$ & $-0.08$ & $2.24$ & $0.59^{+0.02}_{-0.02}$ & $141.61^{+4.59}_{-4.29}$ & $2.05^{+0.22}_{-0.2}$ & $4.07^{+0.74}_{-0.53}$ & $1784.4$ & $1690$ & $1.06$ & &  &$\checkmark$\\
\hline
091020 & $34.6$ & $1.71$ & PL & $1.53$ & \textit{--} & $66^{+4.36}_{-4.14}$ & $18.544^{+1.23}_{-1.2}$ & $-2.05$ & $-0.58$ & $4.97$ & $0.48^{+0.05}_{-0.05}$ & $239.73^{+8.9}_{-8.7}$ & $1.06^{+0.12}_{-0.12}$ & $2.37^{+0.3}_{-0.22}$ & $2319.34$ & $2325$ & $1$  & & \textit{--}  & \\
\hline
$091208A^{\dagger}$ & $29.1$ & $1.38$ & CPL & $0.29$ & $115.83$ & $43.2^{+10.1}_{-4.34}$ & $2.496^{+0.584}_{-2.5}$ & $-0.34$ & $-0.13$ & $4.71$ & $0.12^{+0.04}_{-0.04}$ & $35.38^{+2.38}_{-2.3}$ & $0.45^{+0.16}_{-0.13}$ & $1.23^{+0.14}_{-0.11}$ & $2234.75$ & $2037$ & $1.1$  & $\checkmark$ & \\ 
\hline
100704A & $197.5$ & $3.6$ & PL & $1.73$ & \textit{--} & $92.3^{+5.27}_{-5.02}$ & $63.045^{+3.6}_{-3.4}$ & $-3.98$ & $-0.8$ & $3$ & $-0.13^{+0.02}_{-0.02}$ & $975.43^{+26.31}_{-25.02}$ & $1.17^{+0.1}_{-0.09}$ & $2.23^{+0.21}_{-0.16}$ & $4360.84$ & $3949$ & $1.1$  & & \textit{--}  & \\
\hline
100814A & $174.5$ & $1.44$ & PL & $1.47$ & \textit{--} & $161^{+6.37}_{-6.16}$ & $38.329^{+1.52}_{-1.5}$ & $-3.41$ & $-1.33$ & $59.98$ & $0.06^{+0.05}_{-0.05}$ & $154.39^{+3.88}_{-3.68}$ & $1.18^{+0.1}_{-0.09}$ & $1.04^{+0.03}_{-0.03}$ & $3585.78$ & $2855$ & $1.26$  & $\checkmark$ & \\ 
\hline
$110318A^{\dagger}$ & $16$ & $0.56$ & PL & $1.58$ & \textit{--} & $63.2^{+4.62}_{-4.44}$ & $2.332^{+0.17}_{-1.6}$ & $-0.55$ & $-0.28$ & $5.95$ & $1.18^{+0.07}_{-0.07}$ & $32.37^{+1.23}_{-1.13}$ & $0.67^{+0.08}_{-0.07}$ & $6.23^{+2.25}_{-1.65}$ & $2023.31$ & $1826$ & $1.11$ & & &$\checkmark$\\
\hline
$110319B^{\dagger}$  & $14.5$ & $0.67$ & PL & $1.39$ & \textit{--} & $25.9^{+2.76}_{-2.53}$ & $2.014^{+0.215}_{-2.0}$ & $-0.22$ & $0.17$ & $6.02$ & $1.68^{+0.2}_{-0.2}$ & $12.67^{+0.84}_{-0.84}$ & $0.53^{+0.17}_{-0.13}$ & $2.16^{+0.89}_{-0.45}$ & $1606.53$ & $1554$ & $1.03$  & & \textit{--}  & \\
\hline
110503A & $10$ & $1.61$ & CPL & $1.02$ & $229.44$ & $192^{+39.4}_{-33.5}$ & $17.342^{+3.56}_{-3.0}$ & $-2.1$ & $-0.65$ & $2.29$ & $0.03^{+0.02}_{-0.02}$ & $554.19^{+19.98}_{-19.33}$ & $1.94^{+0.21}_{-0.2}$ & $3.38^{+0.43}_{-0.33}$ & $2465.25$ & $2446$ & $1.01$  & & \textit{--}  & \\
\hline
$110519A^{\dagger}$  & $27.2$ & $0.6$ & PL & $2.09$ & \textit{--} & $52.4^{+2.41}_{-2.32}$ & $1.465^{+0.0674}_{-6.5}$ & $-4.95$ & $-2.74$ & $11.13$ & $0.31^{+0.12}_{-0.12}$ & $10.5^{+0.34}_{-0.33}$ & $1.05^{+0.11}_{-0.11}$ & $3.03^{+0.48}_{-0.36}$ & $1595.26$ & $1492$ & $1.07$  & & \textit{--}  & \\
\hline
$120213A^{\dagger}$ & $48.9$ & $0.47$ & PL & $2.37$ & \textit{--} & $22.3^{+1.33}_{-1.26}$ & $0.553^{+0.033}_{-3.1}$ & $-7.45$ & $-3.74$ & $9.88$ & $0.73^{+0.27}_{-0.28}$ & $2.55^{+0.11}_{-0.1}$ & $0.85^{+0.14}_{-0.12}$ & $4.4^{+2.03}_{-1.12}$ & $1375.61$ & $1373$ & $1$ & & &$\checkmark$\\
\hline
120326A & $69.6$ & $1.8$ & CPL & $1.41$ & $46.84$ & $36.8^{+3.47}_{-2.46}$ & $5.614^{+0.529}_{-3.8}$ & $-0.38$ & $0.08$ & $4.46$ & $1.29^{+0.03}_{-0.03}$ & $108.04^{+2.78}_{-2.85}$ & $1.51^{+0.13}_{-0.12}$ & $4.49^{+0.93}_{-0.62}$ & $2609.29$ & $2620$ & $1$ & & &$\checkmark$\\
\hline
120521C & $26.7$ & $6$ & PL & $1.73$ & \textit{--} & $19.5^{+2.32}_{-2.12}$ & $26.034^{+3.1}_{-2.8}$ & $-1.57$ & $-0.17$ & $0.88$ & $-0^{+0.03}_{-0.02}$ & $1207.85^{+112.71}_{-111.47}$ & $1.17^{+0.59}_{-0.4}$ & $1.16^{+0.19}_{-0.13}$ & $876.03$ & $871$ & $1.01$  & $\checkmark$ & \\ 
\hline
120729A & $71.5$ & $0.8$ & PL & $1.62$ & \textit{--} & $41.6^{+3.59}_{-3.35}$ & $2.771^{+0.239}_{-2.2}$ & $-3.19$ & $-1.48$ & $10.23$ & $-0.16^{+0.08}_{-0.09}$ & $18.52^{+0.95}_{-1.03}$ & $1.53^{+0.32}_{-0.3}$ & $1.33^{+0.11}_{-0.09}$ & $1707.52$ & $1684$ & $1.01$  & $\checkmark$ & \\ 
\hline
$121031A^{\dagger}$ & $226$ & $0.68$ & PL & $1.5$ & \textit{--} & $140^{+7.39}_{-7.07}$ & $8.755^{+0.462}_{-4.4}$ & $-3.59$ & $-1.11$ & $9.28$ & $1.98^{+0.19}_{-0.18}$ & $13.07^{+0.51}_{-0.51}$ & $0.96^{+0.14}_{-0.13}$ & $2.43^{+0.44}_{-0.32}$ & $1616.19$ & $1574$ & $1.03$  & & \textit{--}  & \\
\hline
130420A & $123.5$ & $1.3$ & CPL & $1.52$ & $33.36$ & $96.7^{+3.41}_{-3.45}$ & $10.945^{+0.386}_{0.391}$ & $99.4$ & $44.74$ & $57.9$ & $48.45^{+0.13}_{-0.12}$ & $42.22^{+1.03}_{-1.04}$ & $1.46^{+0.14}_{-0.13}$ & $2.42^{+0.26}_{-0.2}$ & $1498.35$ & $1432$ & $1.05$  & & \textit{--}  & \\
\hline
130701A & $4.38$ & $1.16$ & PL & $1.58$ & \textit{--} & $54.4^{+3.29}_{-3.2}$ & $7.468^{+0.452}_{-4.4}$ & $-0.87$ & $-0.34$ & $2.72$ & $0.29^{+0.02}_{-0.02}$ & $324.99^{+11.88}_{-11.31}$ & $1^{+0.1}_{-0.09}$ & $5.4^{+1.68}_{-0.97}$ & $2280.07$ & $2016$ & $1.13$ & & &$\checkmark$\\
\hline
130831A & $32.5$ & $0.48$ & PL & $1.93$ & \textit{--} & $91.8^{+3.33}_{-3.24}$ & $1.592^{+0.0578}_{-5.6}$ & $-0.57$ & $-0.09$ & $10.46$ & $1.41^{+0.03}_{-0.03}$ & $19.08^{+0.47}_{-0.47}$ & $1.84^{+0.13}_{-0.12}$ & $2.08^{+0.09}_{-0.08}$ & $1525.5$ & $1382$ & $1.1$  & & \textit{--}  & \\
\hline
131004A & $1.54$ & $0.72$ & PL & $1.81$ & \textit{--} & $4.13^{+0.48}_{-0.44}$ & $0.174^{+0.0202}_{-1.9}$ & $-0.34$ & $-0.17$ & $0.57$ & $0.02^{+0.02}_{-0.02}$ & $20.82^{+1.7}_{-1.63}$ & $1.3^{+0.34}_{-0.26}$ & $3.92^{+2.75}_{-1.21}$ & $197.69$ & $211$ & $0.94$  & & \textit{--}  & \\
\hline
$140209A^{\dagger}$ & $21.3$ & $0.28$ & PL & $1.44$ & \textit{--} & $76.8^{+4.78}_{-4.57}$ & $1.034^{+0.0644}_{-6.2}$ & $1.61$ & $1.22$ & $2.67$ & $1.54^{+0}_{-0}$ & $102.6^{+2.75}_{-2.72}$ & $23.12^{+1.15}_{-1.09}$ & $8.17^{+0.29}_{-0.28}$ & $403.38$ & $297$ & $1.36$  & & \textit{--}  & \\
\hline
$140215A^{\dagger}$ & $84.2$ & $0.78$ & PL & $1.19$ & \textit{--} & $167^{+13}_{-12.2}$ & $27.379^{+2.13}_{-2.0}$ & $-1.54$ & $-0.81$ & $8.88$ & $0.85^{+0.14}_{-0.12}$ & $176.7^{+6.96}_{-7.34}$ & $0.48^{+0.07}_{-0.06}$ & $4.98^{+2.36}_{-1.22}$ & $2109.78$ & $2084$ & $1.01$ & & & $\checkmark$\\
\hline
$140529A^{\dagger}$ & $4.5$ & $0.42$ & CPL & $1.31$ & $60.04$ & $32.9^{+5.4}_{-3.28}$ & $0.222^{+0.0365}_{-2.2}$ & $-2.33$ & $-0.18$ & $2.85$ & $0.55^{+0.04}_{-0.04}$ & $7.92^{+0.37}_{-0.38}$ & $9.15^{+0.61}_{-0.93}$ & $3.65^{+0.35}_{-0.29}$ & $355.24$ & $373$ & $0.95$  & & \textit{--}  & \\
\hline
141004A & $3.92$ & $0.57$ & PL & $1.86$ & \textit{--} & $9.7^{+0.75}_{-0.7}$ & $0.249^{+0.0191}_{-1.8}$ & $-0.12$ & $-0.06$ & $1.51$ & $0.04^{+0.01}_{-0.01}$ & $29.85^{+1.66}_{-1.55}$ & $1.7^{+0.27}_{-0.23}$ & $1.47^{+0.11}_{-0.09}$ & $219.9$ & $193$ & $1.14$  & $\checkmark$ & \\ 
\hline
150204A & $12$ & $0.67$ & PL & $1.62$ & \textit{--} & $14.6^{+1.6}_{-1.47}$ & $0.704^{+0.0771}_{-7.1}$ & $1.96$ & $2.3$ & $6.29$ & $3.68^{+0.12}_{-0.11}$ & $8.8^{+0.6}_{-0.61}$ & $2.26^{+0.51}_{-0.43}$ & $3.21^{+1.09}_{-0.6}$ & $873.53$ & $852$ & $1.03$ & & \textit{--}  & \\
\hline
$150212A^{\dagger}$ & $11.4$ & $0.93$ & PL & $1.5$ & \textit{--} & $16^{+1.91}_{-1.75}$ & $1.728^{+0.206}_{-1.9}$ & $-0.72$ & $-0.06$ & $1.76$ & $0.55^{+0.08}_{-0.08}$ & $28.41^{+2.2}_{-2.12}$ & $0.88^{+0.27}_{-0.21}$ & $2.49^{+1.06}_{-0.53}$ & $1638.68$ & $1651$ & $0.99$  & & \textit{--}  & \\
\hline
$150222A^{\dagger}$ & $15.9$ & $0.65$ & PL & $1.61$ & \textit{--} & $36.1^{+2.74}_{-2.59}$ & $1.657^{+0.126}_{-1.2}$ & $1.5$ & $1.99$ & $7.21$ & $3.52^{+0.06}_{-0.06}$ & $21.44^{+1.09}_{-1.09}$ & $3.19^{+0.51}_{-0.45}$ & $3.32^{+0.54}_{-0.37}$ & $936.24$ & $886$ & $1.06$  & & \textit{--}  & \\
\hline

\label{table1}
\end{tabular}

\end{sidewaystable*}

\begin{sidewaystable*}
\tiny
\setlength{\abovecaptionskip}{0pt}
\setlength{\belowcaptionskip}{-5pt}
\vspace{1em}
{\Large \textbf{(continued)}} \\[0.5em]
\vspace{0.5em}

\centering
\begin{tabular}{l@{\hspace{3pt}}c@{\hspace{3pt}}c@{\hspace{3pt}}ccc@{\hspace{3pt}}cc@{\hspace{3pt}}c@{\hspace{3pt}}c@{\hspace{3pt}}cccc@{\hspace{3pt}}c@{\hspace{5pt}}c@{\hspace{5pt}}cc@{\hspace{10pt}}c@{\hspace{3pt}}cc}  

\hline\hline
GRB & $T_{90}$ & z & Model & $\alpha$ & $E_{\text{peak}}$ & Fluence & $E_{\gamma,\text{iso}}$ & $T_{0}$ & $T_{start}$ & $T_{end}$ & $T_{p}$ & $L_{p}$ & r & d & $\chi^2$ & dof & Red-$\chi^2$ & BZ & & NDAF \\
\hline
& (s) & & & & (keV) & ($10^{-7}$erg/cm$^2$) & ($10^{52}$erg) &  & & & & ($10^{50}$erg/s) & & & & & & & \\
(1)&(2)&(3)&(4)&(5)&(6)&(7)&(8)&(9)&(10)&(11)&(12)&(13)&(14)&(15)&(16)&(17)&(18)&(19)&&(20)\\
\hline

150301B & $12.44$ & $1.52$ & PL & $1.46$ & \textit{--} & $36^{+2.73}_{-2.58}$ & $9.46^{+0.717}_{-6.8}$ & $-1.1$ & $-0.28$ & $4.28$ & $0.86^{+0.07}_{-0.07}$ & $133.93^{+5.62}_{-5.23}$ & $0.84^{+0.12}_{-0.1}$ & $4.46^{+1.68}_{-0.99}$ & $2443.72$ & $2356$ & $1.04$ & & &$\checkmark$\\
\hline
150314A & $14.79$ & $1.76$ & CPL & $1.08$ & $764.68$ & $60.5^{+6.38}_{-6.29}$ & $12.099^{+1.28}_{-1.3}$ & $-2.27$ & $-0.8$ & $6.78$ & $0.08^{+0.01}_{-0.01}$ & $1732.76^{+24.48}_{-22.98}$ & $0.9^{+0.03}_{-0.03}$ & $3.77^{+0.17}_{-0.15}$ & $4068.48$ & $3228$ & $1.26$  & & &$\checkmark$\\
\hline
$151006A^{\dagger}$ & $203.9$ & $0.48$ & PL & $1.56$ & \textit{--} & $128^{+7.6}_{-7.23}$ & $3.628^{+0.215}_{-2.0}$ & $-4.83$ & $-3.03$ & $20.67$ & $0.48^{+0.21}_{-0.21}$ & $7.68^{+0.23}_{-0.22}$ & $0.53^{+0.08}_{-0.07}$ & $1.6^{+0.17}_{-0.14}$ & $1675.25$ & $1729$ & $0.97$ & $\checkmark$ & \\ 
\hline
$151229A ^{\ddagger}$ & $1.78$ & $0.5$ & PL & $1.84$ & \textit{--} & $8.6^{+0.85}_{-0.79}$ & $0.169^{+0.0167}_{-1.6}$ & $-0.19$ & $-0.12$ & $0.95$ & $0.11^{+0.02}_{-0.02}$ & $15.3^{+0.96}_{-0.91}$ & $0.7^{+0.12}_{-0.1}$ & $6.95^{+2.13}_{-2.42}$ & $200.79$ & $159$ & $1.26$ & & &$\checkmark$\\

\hline
160131A & $325$ & $0.97$ & PL & $1.24$ & \textit{--} & $430^{+19.8}_{-19}$ & $8.392^{+4.09}_{-3.9}$ & $-1.49$ & $5.3$ & $33.53$ & $8.32^{+0.08}_{-0.07}$ & $226.07^{+3.72}_{-3.99}$ & $9.49^{+0.35}_{-0.49}$ & $1.59^{+0.03}_{-0.02}$ & $1831.07$ & $1689$ & $1.08$  & $\checkmark$ & \\ 
\hline
$160424A^{\dagger}$ & $6.3$ & $0.68$ & PL & $1.4$ & \textit{--} & $23.5^{+2.63}_{-2.44}$ & $6.415^{+0.206}_{-1.9}$ & $-1.29$ & $-0.41$ & $1.48$ & $0.12^{+0.04}_{-0.04}$ & $62.31^{+4.36}_{-4.35}$ & $2.45^{+0.52}_{-0.43}$ & $3.34^{+0.93}_{-0.58}$ & $1737.06$ & $1706$ & $1.02$  & & \textit{--}  & \\
\hline
161004B & $15.9$ & $0.49$ & CPL & $1.21$ & $177.05$ & $160^{+10.9}_{-9.28}$ & $1.302^{+0.0856}_{-7.3}$ & $-5.85$ & $-3.28$ & $9.01$ & $0.84^{+0.05}_{-0.05}$ & $8.52^{+0.11}_{-0.11}$ & $1.48^{+0.05}_{-0.04}$ & $8.77^{+0.81}_{-0.94}$ & $1628.78$ & $1505$ & $1.08$  & & \textit{--}  & \\
\hline
$161218A^{\dagger}$ & $7.1$ & $0.51$ & CPL & $0.64$ & $148.02$ & $107^{+8.88}_{-7.52}$ & $1.244^{+0.0738}_{-6.2}$ & $-1.23$ & $-0.72$ & $5.52$ & $0.53^{+0.02}_{-0.02}$ & $19.36^{+0.43}_{-0.4}$ & $1.06^{+0.06}_{-0.06}$ & $5.42^{+0.88}_{-0.66}$ & $1811.1$ & $1529$ & $1.18$ & & &$\checkmark$\\
\hline
161219B & $6.94$ & $0.15$ & CPL & $1.29$ & $61.88$ & $22.6^{+1.65}_{-1.57}$ & $1.497^{+1.3}_{-1.2}$ & $-1.18$ & $-0.23$ & $5.36$ & $1.68^{+0.1}_{-0.12}$ & $0.26^{+0.02}_{-0.01}$ & $1.86^{+0.36}_{-0.31}$ & $5.18^{+2.21}_{-1.37}$ & $1221.58$ & $1162$ & $1.05$ & & &$\checkmark$\\
\hline
$170101A^{\dagger}$ & $2.43$ & $0.18$ & PL & $1.87$ & \textit{--} & $94.3^{+4.8}_{-4.62}$ & $3.144^{+0.0115}_{-1.1}$ & $-0.33$ & $-0.23$ & $2.63$ & $0.06^{+0.01}_{-0.01}$ & $14.12^{+0.57}_{-0.56}$ & $1.28^{+0.11}_{-0.1}$ & $2.6^{+0.18}_{-0.15}$ & $735.42$ & $640$ & $1.15$  & & \textit{--}  & \\
\hline
$170126A^{\dagger}$ & $9.5$ & $0.59$ & PL & $1.54$ & \textit{--} & $54.5^{+8.29}_{-7.51}$ & $4.858^{+0.368}_{-3.3}$ & $-1$ & $-0.4$ & $3.38$ & $0.52^{+0.12}_{-0.11}$ & $36.59^{+4.71}_{-4.23}$ & $1.26^{+0.62}_{-0.42}$ & $2.41^{+1.16}_{-0.44}$ & $1453.43$ & $1491$ & $0.97$& & \textit{--}  & \\
\hline
$170317A^{\dagger}$  & $11.94$ & $0.75$ & CPL & $0.81$ & $58.66$ & $14.6^{+2.29}_{-1.61}$ & $1.186^{+0.0404}_{-2.8}$ & $-0.53$ & $0.09$ & $3.69$ & $1.12^{+0.09}_{-0.08}$ & $6.98^{+0.62}_{-0.56}$ & $1.49^{+0.59}_{-0.4}$ & $2.87^{+1.41}_{-0.66}$ & $1670.12$ & $1640$ & $1.02$  & & \textit{--}  & \\
\hline
$170419B^{\dagger}$ & $77.2$ & $1.04$ & PL & $1.47$ & \textit{--} & $37.1^{+5.56}_{-5}$ & $4.957^{+0.774}_{-7.0}$ & $-0.46$ & $0.15$ & $2.86$ & $1.02^{+0.07}_{-0.06}$ & $64.32^{+3.94}_{-3.82}$ & $1.73^{+0.35}_{-0.29}$ & $3.34^{+1.35}_{-0.73}$ & $1945.18$ & $1905$ & $1.02$  & & \textit{--}  & \\
\hline
170607A & $23$ & $0.56$ & PL & $1.61$ & \textit{--} & $126^{+8.79}_{-8.32}$ & $4.277^{+0.301}_{-2.9}$ & $-5.21$ & $-2.81$ & $10.28$ & $0.54^{+0.14}_{-0.13}$ & $12.41^{+0.43}_{-0.42}$ & $0.98^{+0.11}_{-0.1}$ & $3.02^{+0.5}_{-0.35}$ & $1570.26$ & $1456$ & $1.08$ & & \textit{--}  & \\
\hline
$170711A^{\dagger}$ & $31.3$ & $0.6$ & PL & $1.75$ & \textit{--} & $21.6^{+2.09}_{-1.94}$ & $3.396^{+0.0659}_{-6.1}$ & $-0.72$ & $-0.34$ & $3.32$ & $0.23^{+0.03}_{-0.03}$ & $14.26^{+0.75}_{-0.7}$ & $1.11^{+0.23}_{-0.18}$ & $2^{+0.3}_{-0.21}$ & $1528.9$ & $1495$ & $1.02$  & & \textit{--}  &  \\ 
\hline
$180630A^{\dagger}$  & $18.85$ & $0.43$ & PL & $2.07$ & \textit{--} & $23.4^{+1.91}_{-1.79}$ & $2.976^{+0.0264}_{-2.5}$ & $0.32$ & $1.04$ & $5.78$ & $2.82^{+0.07}_{-0.07}$ & $5.04^{+0.26}_{-0.26}$ & $2.1^{+0.34}_{-0.25}$ & $6.29^{+1.97}_{-1.42}$ & $1274.9$ & $1335$ & $0.95$ & & &$\checkmark$\\
\hline
180728A & $8.68$ & $0.12$ & PL & $1.97$ & \textit{--} & $416^{+6.07}_{-6.03}$ & $2.951^{+5.7}_{-5.7}$ & $9.29$ & $9.54$ & $22.44$ & $11.6^{+0.01}_{-0.01}$ & $7.56^{+0.05}_{-0.05}$ & $5.94^{+0.07}_{-0.06}$ & $3.56^{+0.03}_{-0.03}$ & $2036.8$ & $869$ & $2.34$  & & \textit{--}  & \\
\hline
$200215A^{\dagger}$ & $11.7$ & $0.91$ & PL & $1.41$ & \textit{--} & $21.1^{+2.19}_{-2.03}$ & $5.828^{+0.274}_{-2.5}$ & $-2.13$ & $-0.64$ & $2$ & $0.15^{+0.05}_{-0.05}$ & $52.82^{+3.58}_{-3.52}$ & $2.57^{+0.57}_{-0.48}$ & $3.07^{+0.76}_{-0.49}$ & $1782.48$ & $1783$ & $1$  & & \textit{--}  & \\
\hline
200829A & $13.04$ & $1.25$ & CPL & $0.92$ & $694.21$ & $1120^{+83.6}_{-74.6}$ & $2.912^{+9.84}_{-8.8}$ & $-62.67$ & $-27.81$ & $-22.67$ & $-26.67^{+0.01}_{-0.01}$ & $2861.24^{+22.44}_{-22.13}$ & $1.47^{+0.03}_{-0.03}$ & $7.39^{+0.29}_{-0.27}$ & $3724.42$ & $1226$ & $3.04$ & & &$\checkmark$\\
\hline
$200922A^{\dagger}$ & $10.3$ & $0.37$ & CPL & $1.43$ & $39.14$ & $29.9^{+2.17}_{-1.76}$ & $2.054^{+0.0151}_{-1.2}$ & $-1.2$ & $-0.81$ & $4.52$ & $0.04^{+0.03}_{-0.03}$ & $6.55^{+0.19}_{-0.18}$ & $0.99^{+0.09}_{-0.08}$ & $2.94^{+0.38}_{-0.29}$ & $1341.33$ & $1384$ & $0.97$  & & \textit{--}  & \\
\hline
$201105A^{\dagger}$ & $33.82$ & $0.51$ & CPL & $0.9$ & $368.74$ & $280^{+10.1}_{-9.78}$ & $1.845^{+0.124}_{-1.2}$ & $-2.21$ & $-1.32$ & $13.18$ & $0.56^{+0.03}_{-0.03}$ & $24.17^{+0.44}_{-0.42}$ & $1.27^{+0.07}_{-0.06}$ & $2.2^{+0.08}_{-0.07}$ & $1636.4$ & $1528$ & $1.07$  & & \textit{--}  & \\
\hline
$210308A^{\dagger}$ & $5.3$ & $0.5$ & CPL & $0.84$ & $119.83$ & $45.9^{+7.3}_{-5.3}$ & $0.35^{+0.0557}_{-4.0}$ & $0.33$ & $0.63$ & $3.56$ & $1.68^{+0.03}_{-0.03}$ & $12.87^{+0.53}_{-0.52}$ & $2.72^{+0.25}_{-0.22}$ & $6.51^{+1.58}_{-1.22}$ & $1640.11$ & $1522$ & $1.08$ & & &$\checkmark$\\
\hline
$210410A^{\ddagger}$  & $52.88$ & $2$ & PL & $1.04$ & \textit{--} & $135^{+8.06}_{-7.66}$ & $119^{+7.11}_{-6.8}$ & $-0.03$ & $0.04$ & $8.5$ & $0.13^{+0.08}_{-0.07}$ & $1193.92^{+76.86}_{-62.47}$ & $0.06^{+0.04}_{-0.03}$ & $2.32^{+0.3}_{-0.24}$ & $3166.97$ & $3043$ & $1.04$  & & \textit{--}  & \\
\hline
$210723A^{\dagger}$  & $48.54$ & $0.77$ & CPL & $0.85$ & $94.89$ & $100^{+10.2}_{-9.41}$ & $1.791^{+0.183}_{-1.7}$ & $-4.89$ & $-2.56$ & $11.29$ & $0.52^{+0.18}_{-0.17}$ & $7.45^{+0.25}_{-0.24}$ & $0.56^{+0.07}_{-0.07}$ & $3.97^{+1.27}_{-0.78}$ & $1811.14$ & $1793$ & $1.01$  & & \textit{--}  & \\
\hline
$210730A^{\dagger}$ & $3.86$ & $0.67$ & PL & $1.41$ & \textit{--} & $35.6^{+11.7}_{-9.3}$ & $2.604^{+0.856}_{-6.8}$ & $-0.35$ & $-0.1$ & $1.85$ & $0.26^{+0.03}_{-0.03}$ & $72.7^{+5.86}_{-5.57}$ & $1.53^{+0.46}_{-0.35}$ & $2.44^{+0.58}_{-0.35}$ & $1766.57$ & $1689$ & $1.05$  & & \textit{--}  & \\
\hline
$211225B^{\dagger}$ & $121.54$ & $0.57$ & CPL & $1.32$ & $139.78$ & $220^{+17.9}_{-14.9}$ & $2.478^{+0.202}_{-1.7}$ & $28$ & $28.19$ & $53.69$ & $31.85^{+0.1}_{-0.11}$ & $7.3^{+0.12}_{-0.12}$ & $12.02^{+0.77}_{-0.71}$ & $1.95^{+0.04}_{-0.04}$ & $1919.68$ & $1468$ & $1.31$  & & \textit{--}  & \\
\hline
$221016A^{\dagger}$ & $21.86$ & $0.92$ & PL & $1.28$ & \textit{--} & $45.6^{+4.3}_{-4.01}$ & $7.825^{+0.738}_{-6.9}$ & $-3.18$ & $-1.34$ & $3.78$ & $0.3^{+0.09}_{-0.09}$ & $79.06^{+3.65}_{-3.52}$ & $1.09^{+0.14}_{-0.11}$ & $6.87^{+2.08}_{-1.88}$ & $1654.11$ & $1643$ & $1.01$ & & &$\checkmark$\\
\hline
$230405B^{\ddagger}$ & $14.14$ & $2$ & PL & $1.21$ & \textit{--} & $82.5^{+5.04}_{-4.8}$ & $51.761^{+3.16}_{-3.0}$ & $-0.28$ & $0.14$ & $1.91$ & $0.71^{+0.01}_{-0.01}$ & $2274.92^{+71.92}_{-69.39}$ & $2.93^{+0.19}_{-0.19}$ & $6.04^{+0.73}_{-0.55}$ & $2753.25$ & $2574$ & $1.07$ & & &$\checkmark$\\
\hline
$230723B^{\dagger}$ & $6.64$ & $0.63$ & CPL & $1.06$ & $63.91$ & $18.7^{+2.94}_{-2.01}$ & $0.248^{+0.039}_{-2.7}$ & $-1.17$ & $-0.35$ & $3$ & $0.85^{+0.08}_{-0.08}$ & $7.27^{+0.44}_{-0.42}$ & $1.38^{+0.24}_{-0.19}$ & $5.97^{+2.46}_{-1.91}$ & $1408.16$ & $1394$ & $1.01$ & & &$\checkmark$\\
\hline
$230802A^{\dagger}$  & $175.62$ & $0.87$ & PL & $2.3$ & $1.6$ & $81.6^{+5.82}_{-5.51}$ & $6.746^{+0.481}_{-4.6}$ & $-7.28$ & $-2.78$ & $10.82$ & $1.96^{+0.14}_{-0.15}$ & $24.37^{+0.59}_{-0.59}$ & $1^{+0.07}_{-0.06}$ & $7.68^{+1.52}_{-1.64}$ & $1730.22$ & $1604$ & $1.08$ & & &$\checkmark$\\
\hline
230818A & $9.82$ & $2.42$ & PL & $1.36$ & \textit{--} & $39.9^{+6.19}_{-5.56}$ & $24.058^{+3.73}_{-3.4}$ & $-0.8$ & $0.43$ & $2.31$ & $0.54^{+0.01}_{-0.01}$ & $1063.57^{+70.75}_{-81.25}$ & $28.43^{+1.18}_{-2.95}$ & $1.86^{+0.14}_{-0.11}$ & $3463.14$ & $3470$ & $1$  & $\checkmark$ & \\ 
\hline
$230826A^{\dagger}$ & $41.07$ & $0.8$ & CPL & $1.22$ & $136.97$ & $52.6^{+7.97}_{-5.89}$ & $1.142^{+0.173}_{-1.3}$ & $-6.58$ & $-2.69$ & $9.02$ & $0.97^{+0.13}_{-0.13}$ & $7.67^{+0.2}_{-0.2}$ & $1.28^{+0.11}_{-0.1}$ & $4.27^{+0.99}_{-0.65}$ & $2044.59$ & $1828$ & $1.12$ & & & $\checkmark$\\
\hline
$231110A^{\dagger}$ & $11.34$ & $0.57$ & PL & $1.35$ & \textit{--} & $28.3^{+3.75}_{-3.42}$ & $1.799^{+0.238}_{-2.2}$ & $-1.64$ & $-0.54$ & $3.16$ & $0.47^{+0.09}_{-0.1}$ & $24.47^{+3.09}_{-2.61}$ & $1.58^{+0.83}_{-0.53}$ & $2.4^{+0.9}_{-0.42}$ & $1471.94$ & $1591$ & $0.93$  & & \textit{--}  & \\
\hline
231117A & $0.67$ & $0.26$ & PL & $1.55$ & \textit{--} & $38.7^{+1.58}_{-1.54}$ & $0.331^{+0.0135}_{-1.3}$ & $-0.06$ & $-0.02$ & $1.09$ & $0.04^{+0}_{-0}$ & $91.88^{+2.9}_{-2.79}$ & $1.63^{+0.15}_{-0.14}$ & $2.65^{+0.12}_{-0.11}$ & $725.04$ & $585$ & $1.24$  & & \textit{--}  & \\
\hline
$231205B^{\dagger}$ & $46.94$ & $0.36$ & PL & $1.38$ & \textit{--} & $143^{+8.06}_{-7.72}$ & $3.649^{+0.206}_{-2.0}$ & $-8.33$ & $-5.27$ & $4.51$ & $-3.01^{+0.06}_{-0.06}$ & $19.23^{+0.6}_{-0.57}$ & $2.25^{+0.2}_{-0.18}$ & $2.6^{+0.2}_{-0.17}$ & $741.66$ & $636$ & $1.17$  & & \textit{--}  & \\
\hline
240825A & $57.2$ & $0.66$ & PL & $1.2$ & \textit{--} & $526^{+18.1}_{-17.6}$ & $63.022^{+2.17}_{-2.1}$ & $1.05$ & $0.67$ & $10.3$ & $1.09^{+0.01}_{-0.01}$ & $2054.04^{+30.2}_{-28.54}$ & $1.7^{+0.04}_{-0.04}$ & $2.11^{+0.03}_{-0.03}$ & $2870.48$ & $774$ & $3.71$  & & \textit{--}  & \\
\hline
$250330A^{\dagger}$ & $11.27$ & $0.68$ & PL & $1.72$ & \textit{--} & $14.4^{+1.43}_{-1.33}$ & $0.605^{+0.0601}_{-5.6}$ & $-0.31$ & $-0.12$ & $2.67$ & $0.41^{+0.04}_{-0.04}$ & $18.08^{+1.15}_{-1.08}$ & $1.08^{+0.24}_{-0.2}$ & $2.47^{+0.65}_{-0.41}$ & $857.35$ & $784$ & $1.09$  & & \textit{--}  & \\
\hline
250424A & $19.03$ & $0.31$ & CPL & $1.54$ & $120.42$ & $441^{+21.8}_{-19.4}$ & $1.64^{+0.0811}_{-7.2}$ & $-22.71$ & $-16.02$ & $-1.37$ & $-11.47^{+0.02}_{-0.02}$ & $13.88^{+0.12}_{-0.11}$ & $2.53^{+0.05}_{-0.04}$ & $5.33^{+0.15}_{-0.15}$ & $1398.51$ & $1019$ & $1.37$ & & &$\checkmark$\\
\hline
$250509A^{\dagger}$ & $55.43$ & $0.86$ & PL & $1.54$ & \textit{--} & $42.1^{+3.36}_{-3.16}$ & $3.674^{+0.293}_{-2.8}$ & $-3.84$ & $-1.68$ & $9.87$ & $0.43^{+0.15}_{-0.14}$ & $21.5^{+1.07}_{-1.07}$ & $1.06^{+0.23}_{-0.2}$ & $1.64^{+0.32}_{-0.22}$ & $888.49$ & $867$ & $1.02$  & $\checkmark$ & \\ 
\hline
$250605A^{\dagger}$ & $176.05$ & $0.53$ & PL & $1.5$ & \textit{--} & $173^{+7.25}_{-7.01}$ & $6.865^{+0.288}_{-2.8}$ & $-5.54$ & $-1.7$ & $16.12$ & $0.13^{+0.09}_{-0.09}$ & $19.42^{+1.01}_{-0.98}$ & $4.79^{+1.1}_{-0.84}$ & $1.12^{+0.04}_{-0.04}$ & $2138.02$ & $1435$ & $1.49$  & $\checkmark$ & \\ 
\hline
\end{tabular}
\begin{list}{}{}
\item \textbf{Notes:} Column (1) lists the GRB name; a ``\( ^{\dagger} \)'' symbol indicates an estimated redshift, while a ``\( ^{\ddagger} \)'' symbol denotes an assumed redshift, the typical redshifts are \( z = 2 \) for long GRBs and \( z = 0.5 \) for short GRBs \citep{2015ApJ...815...54S}. Column (2) represents the duration \( T \). Column (3) gives the cosmological redshift. Columns (4) to (8) contain the respective observed parameters, including \( E_{\rm p} \), fluence, \( E_{\gamma, \text{iso}} \) and other derived quantities. Columns (9) to (15) list the results of the KRL function fitting, including \( T_{\rm p} \), \( r \) and \( d \). Columns (19) and (20) show the results for the BZ and NDAF fitted models, respectively. A hyphen (``---'') indicates uncertainty or absence of data for a particular model.

\end{list}
\end{sidewaystable*}

\begin{figure*}[ht]
    \centering
    \captionsetup[subfigure]{aboveskip=-8pt, belowskip=-1pt, margin=0pt}
    \setlength{\belowcaptionskip}{-10pt}
    \setlength{\tabcolsep}{-2pt}
    
    \begin{subfigure}{0.19\textwidth}
        \includegraphics[width=\linewidth]{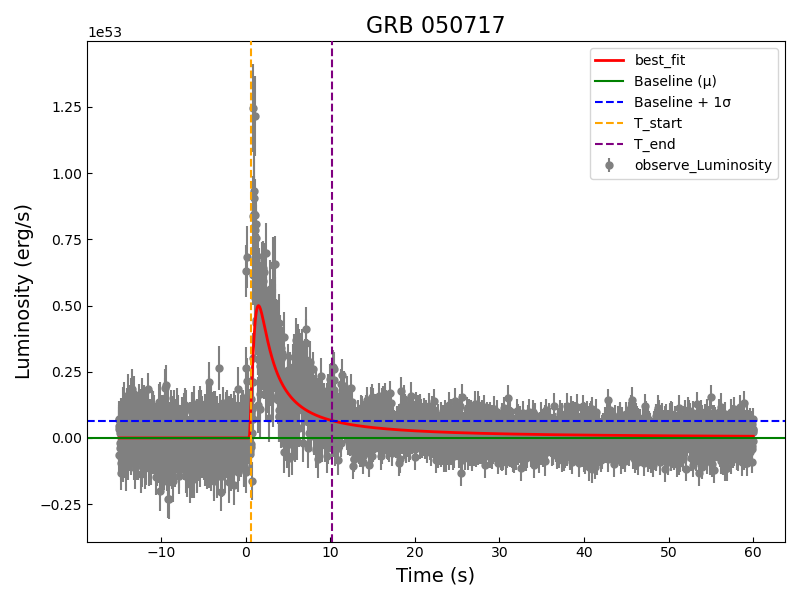}
    \end{subfigure}\hfill
    \begin{subfigure}{0.19\textwidth}
        \includegraphics[width=\linewidth]{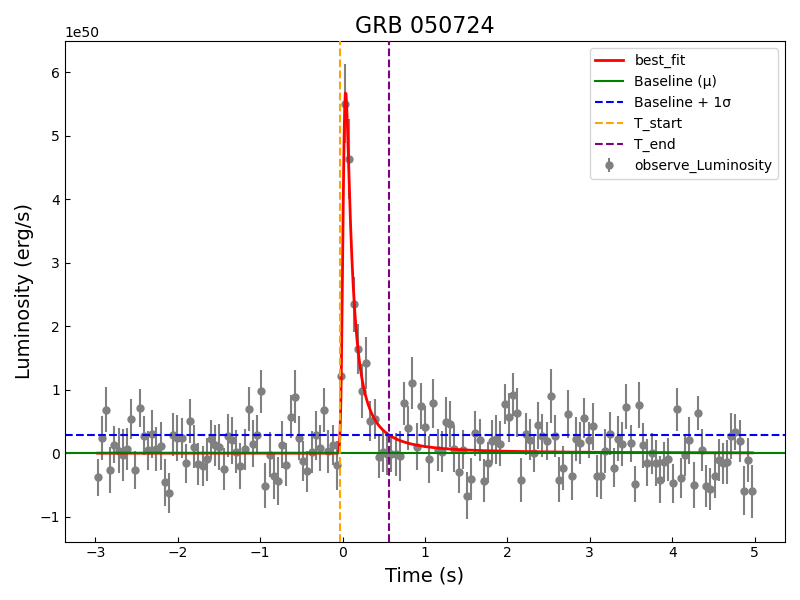}
    \end{subfigure}\hfill   
    \begin{subfigure}{0.19\textwidth}
        \includegraphics[width=\linewidth]{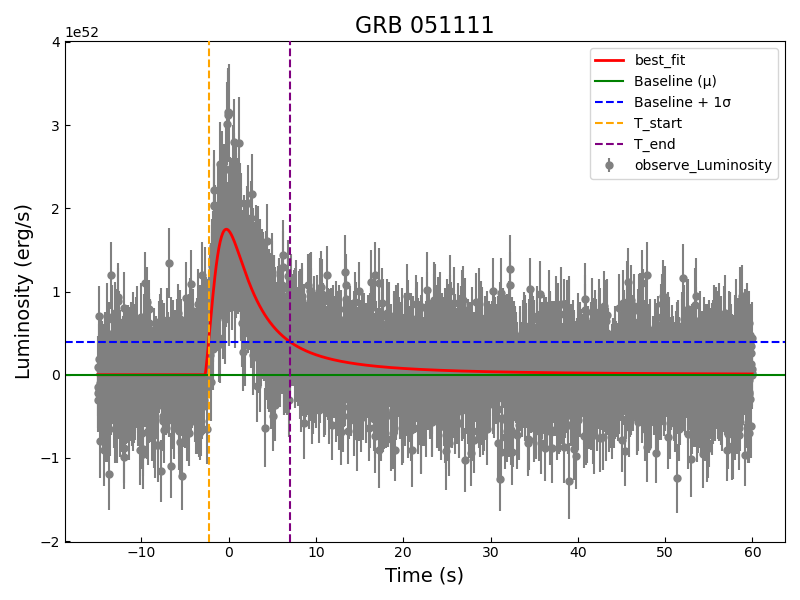}
    \end{subfigure}\hfill
    \begin{subfigure}{0.19\textwidth}
        \includegraphics[width=\linewidth]{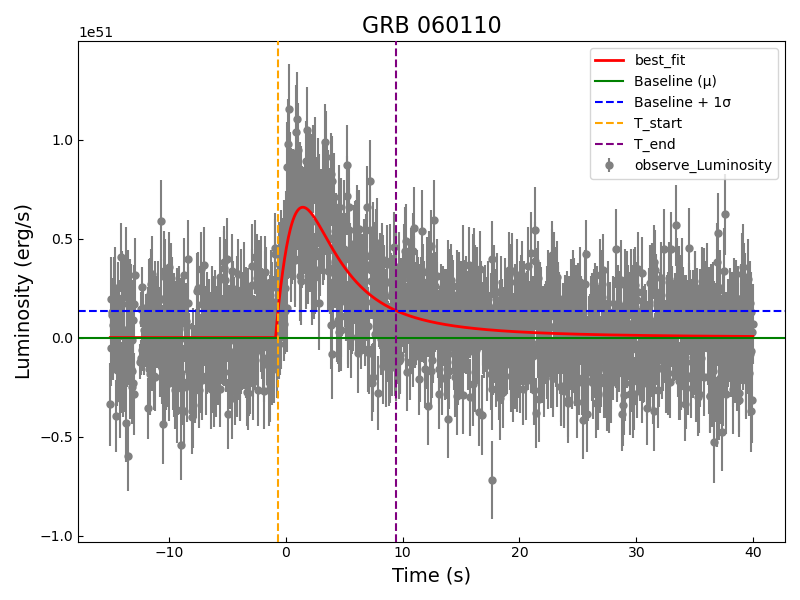}
    \end{subfigure}\hfill
    \begin{subfigure}{0.19\textwidth}
        \includegraphics[width=\linewidth]{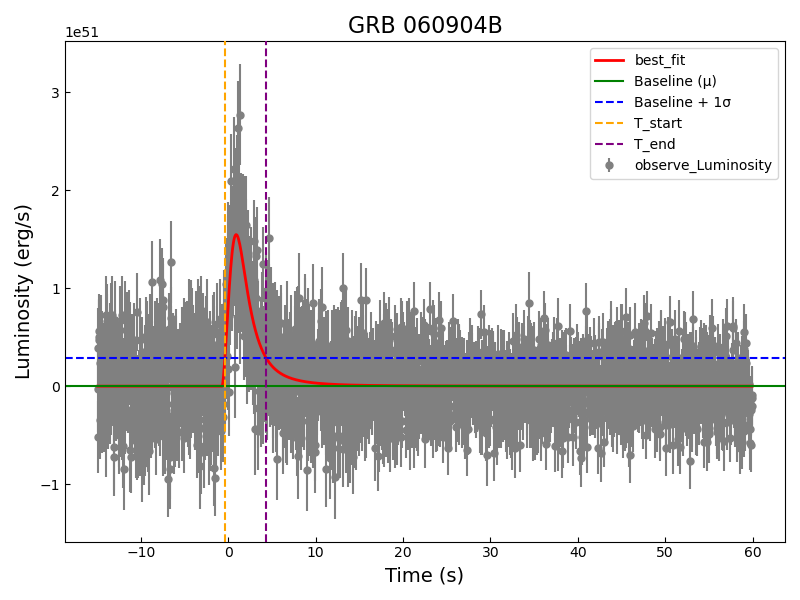}
    \end{subfigure}

    \begin{subfigure}{0.19\textwidth}
        \includegraphics[width=\linewidth]{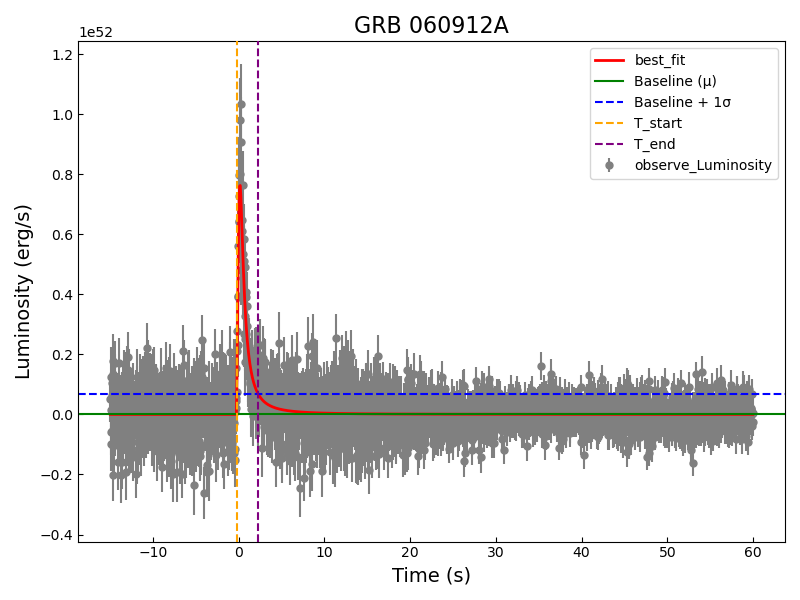}
    \end{subfigure}\hfill
    \begin{subfigure}{0.19\textwidth}
        \includegraphics[width=\linewidth]{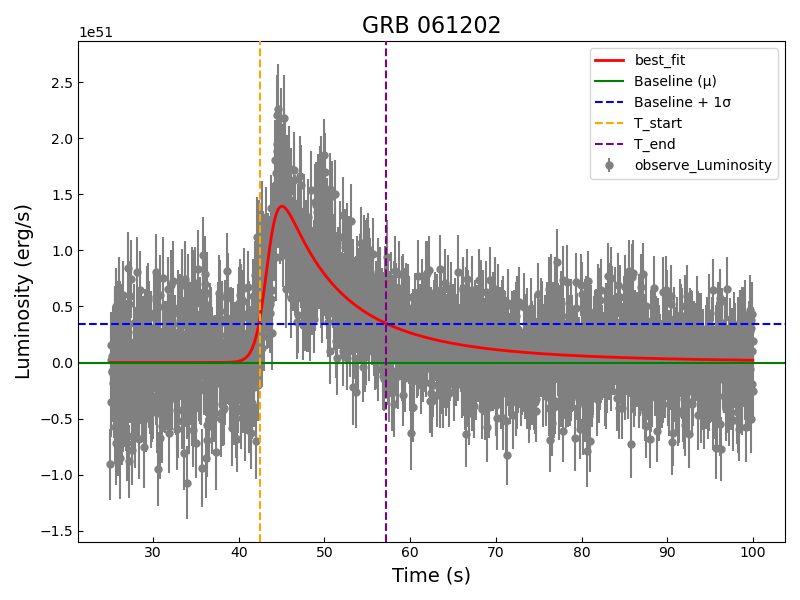}
    \end{subfigure}\hfill
    \begin{subfigure}{0.19\textwidth}
        \includegraphics[width=\linewidth]{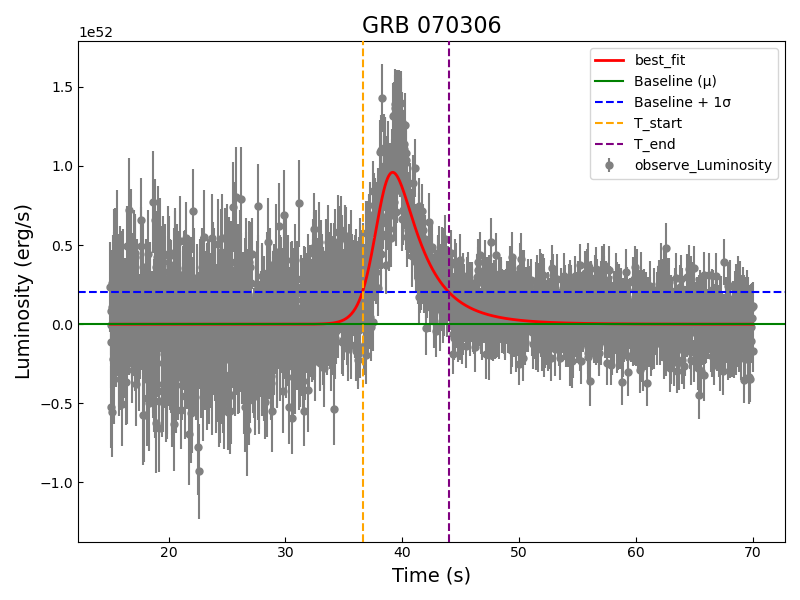}
    \end{subfigure}\hfill
    \begin{subfigure}{0.19\textwidth}
        \includegraphics[width=\linewidth]{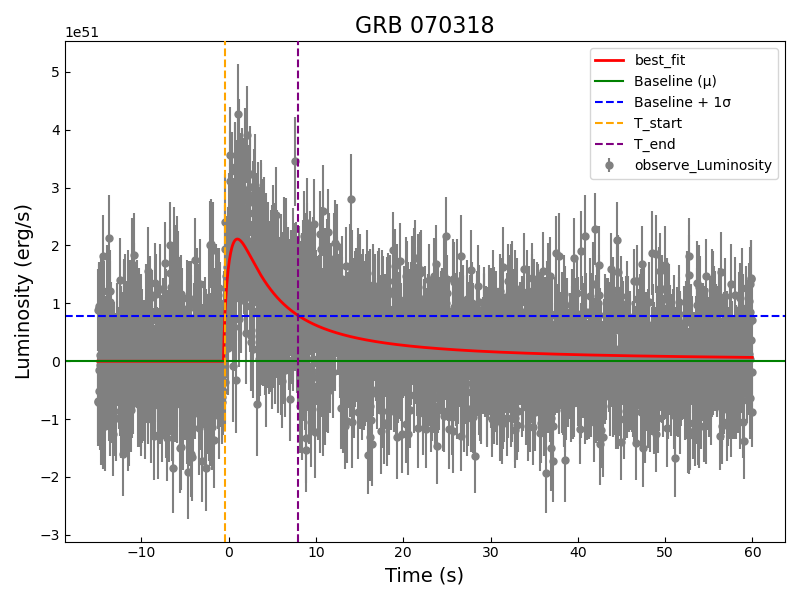}
    \end{subfigure}\hfill
    \begin{subfigure}{0.19\textwidth}
        \includegraphics[width=\linewidth]{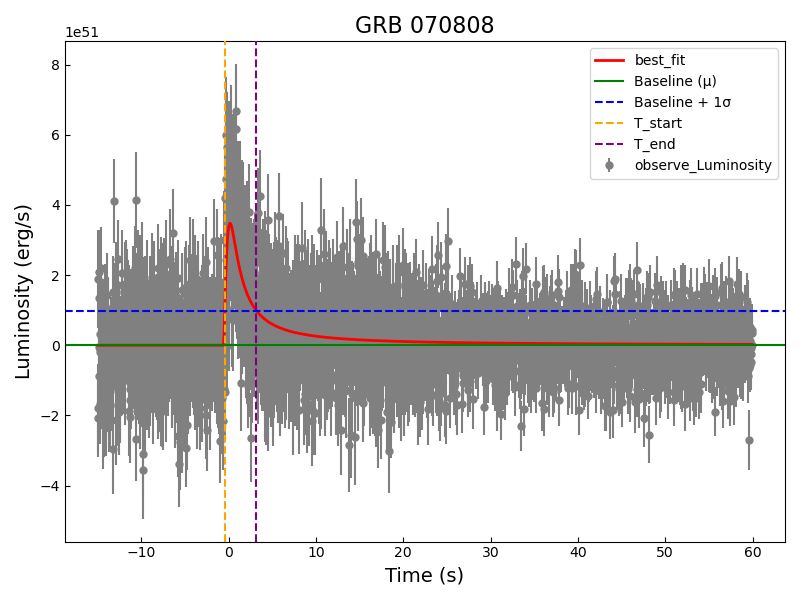}
    \end{subfigure}

    \begin{subfigure}{0.19\textwidth}
        \includegraphics[width=\linewidth]{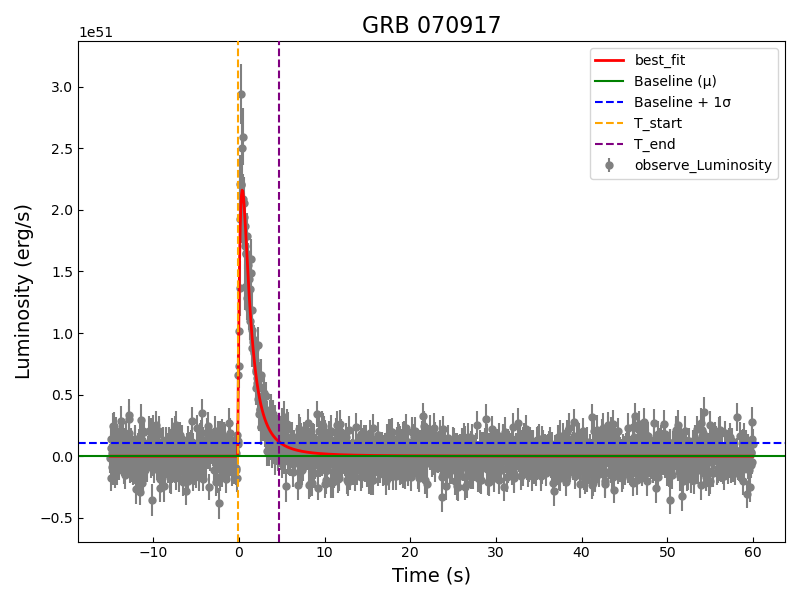}
    \end{subfigure}\hfill
    \begin{subfigure}{0.19\textwidth}
        \includegraphics[width=\linewidth]{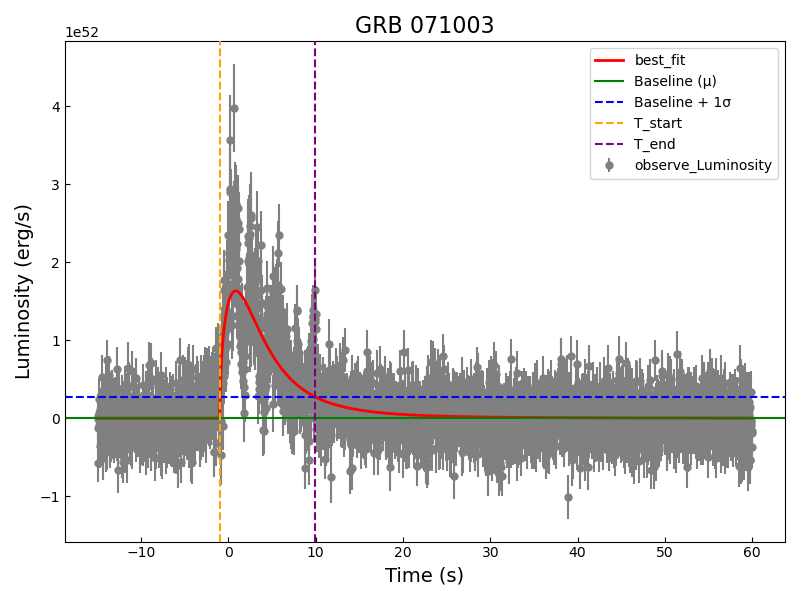}
    \end{subfigure}\hfill
    \begin{subfigure}{0.19\textwidth}
        \includegraphics[width=\linewidth]{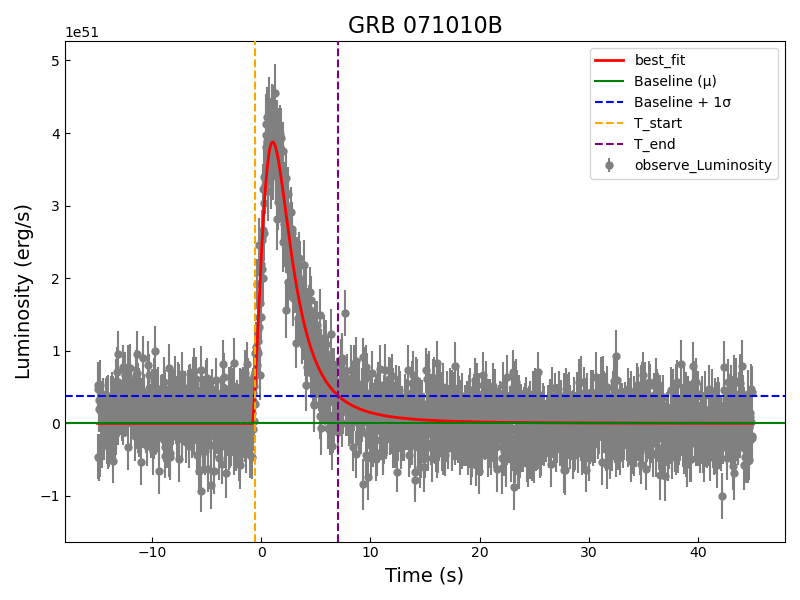}
    \end{subfigure}\hfill
    \begin{subfigure}{0.19\textwidth}
        \includegraphics[width=\linewidth]{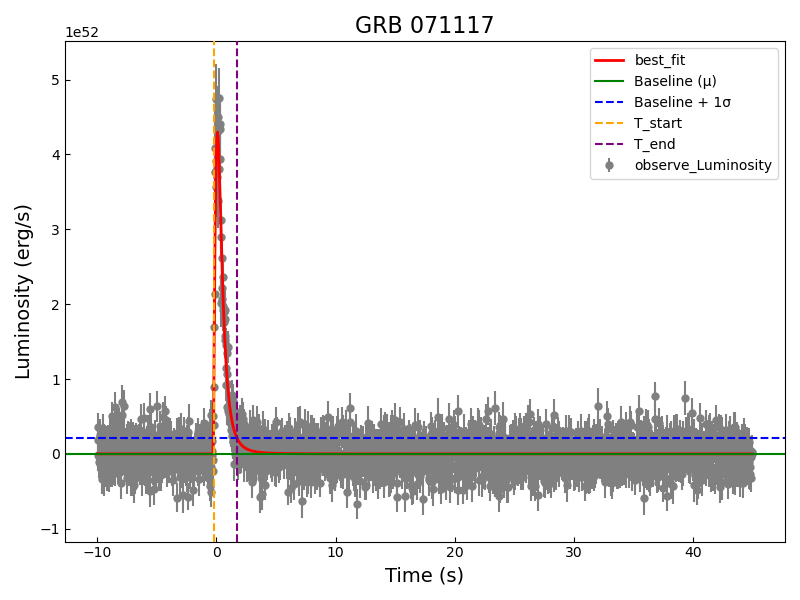}
    \end{subfigure}\hfill
    \begin{subfigure}{0.19\textwidth}
        \includegraphics[width=\linewidth]{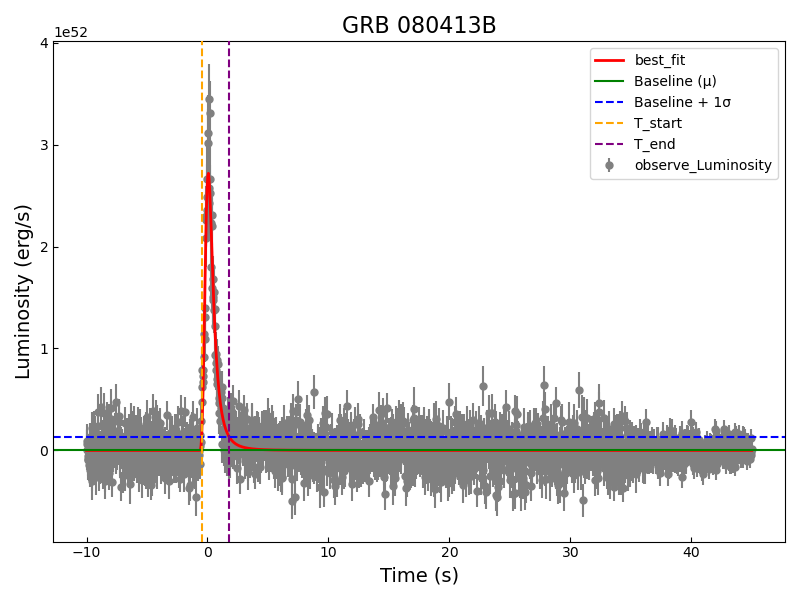}
    \end{subfigure}

    \begin{subfigure}{0.19\textwidth}
        \includegraphics[width=\linewidth]{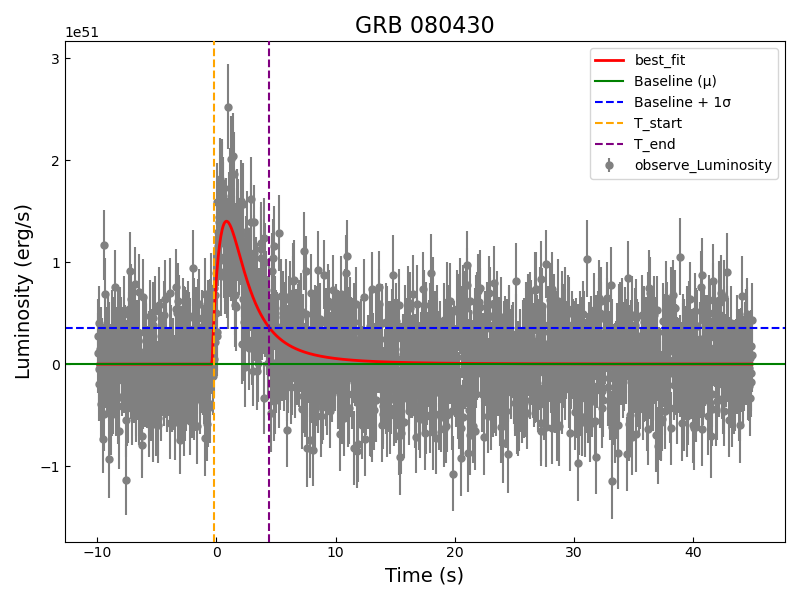}
    \end{subfigure}\hfill
    \begin{subfigure}{0.19\textwidth}
        \includegraphics[width=\linewidth]{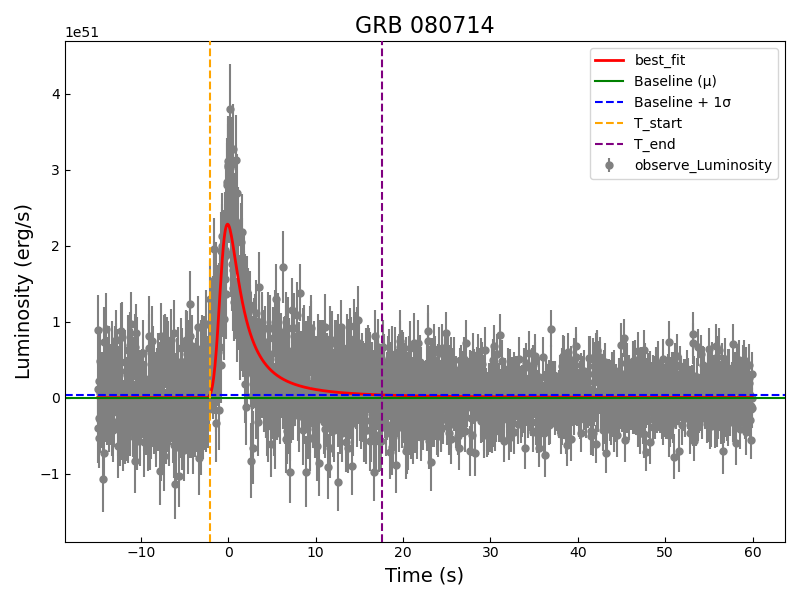}
    \end{subfigure}\hfill
    \begin{subfigure}{0.19\textwidth}
        \includegraphics[width=\linewidth]{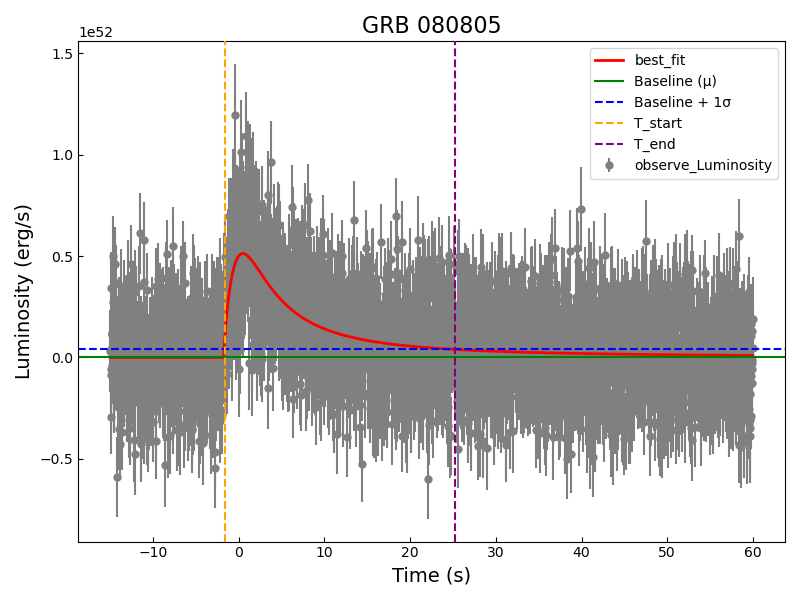}
    \end{subfigure}\hfill
    \begin{subfigure}{0.19\textwidth}
        \includegraphics[width=\linewidth]{GRB081222_luminosity_fit.png}
    \end{subfigure}\hfill
    \begin{subfigure}{0.19\textwidth}
        \includegraphics[width=\linewidth]{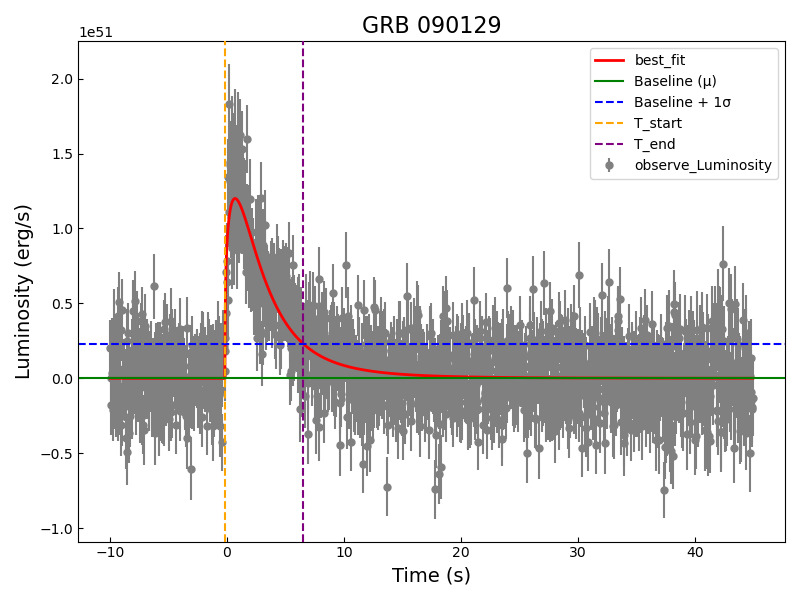}
    \end{subfigure}

    \begin{subfigure}{0.19\textwidth}
        \includegraphics[width=\linewidth]{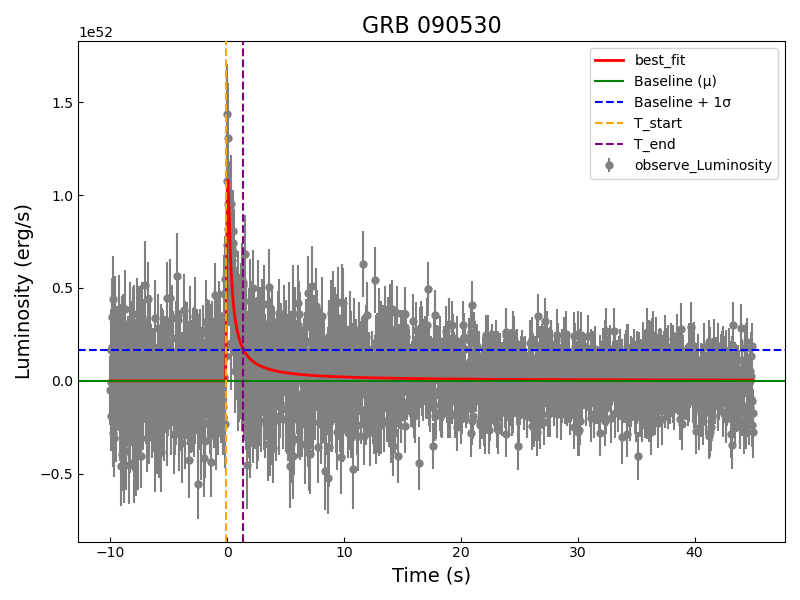}
    \end{subfigure}\hfill
    \begin{subfigure}{0.19\textwidth}
        \includegraphics[width=\linewidth]{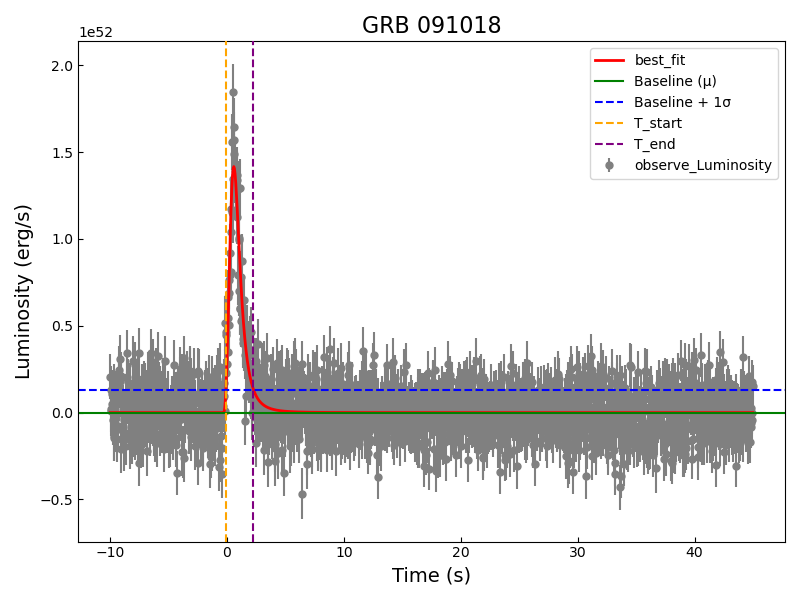}
    \end{subfigure}\hfill
    \begin{subfigure}{0.19\textwidth}
        \includegraphics[width=\linewidth]{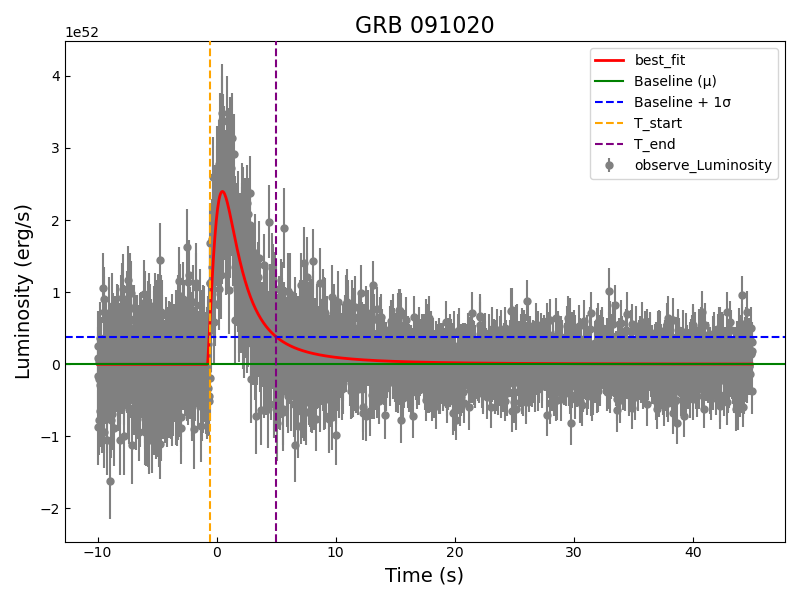}
    \end{subfigure}\hfill
    \begin{subfigure}{0.19\textwidth}
        \includegraphics[width=\linewidth]{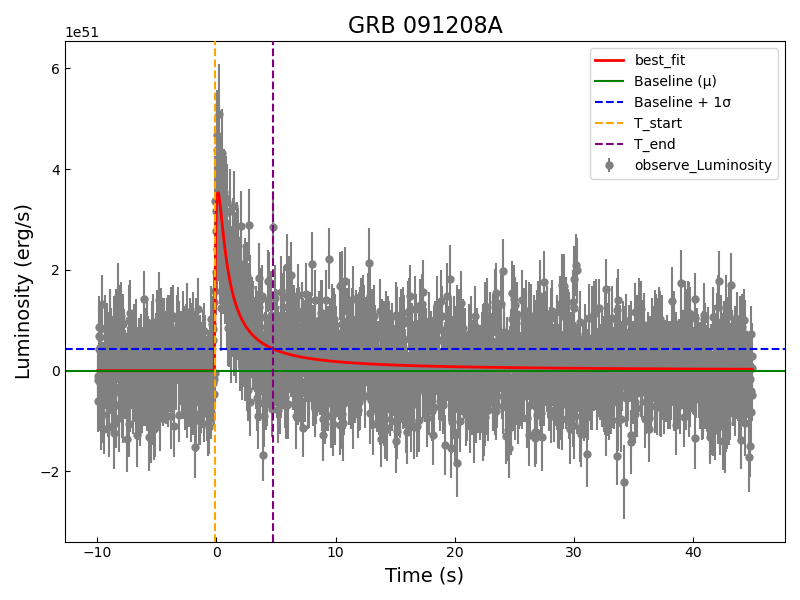}
    \end{subfigure}\hfill
    \begin{subfigure}{0.19\textwidth}
        \includegraphics[width=\linewidth]{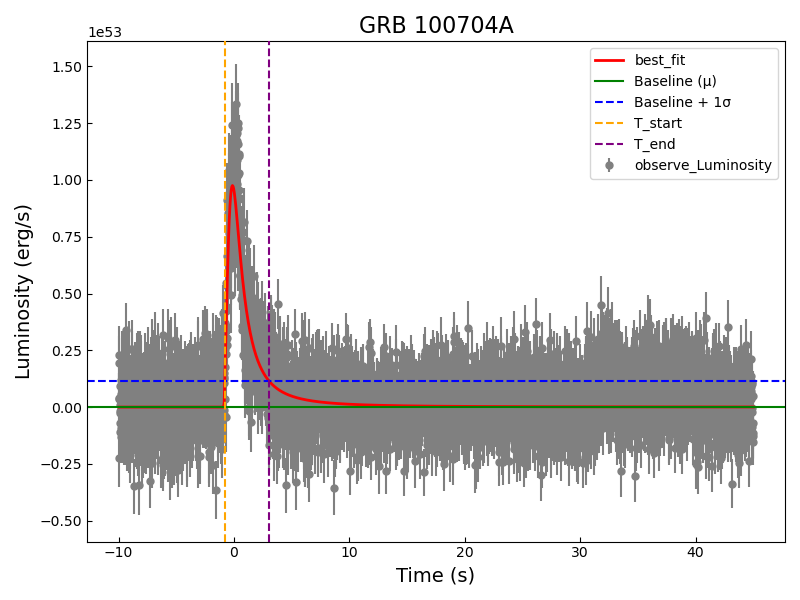}
    \end{subfigure}

    \begin{subfigure}{0.19\textwidth}
        \includegraphics[width=\linewidth]{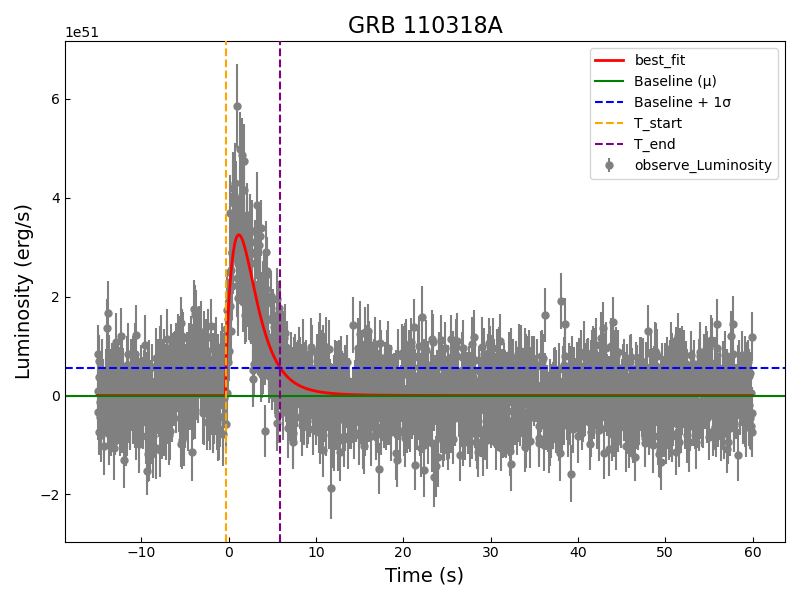}
    \end{subfigure}\hfill
    \begin{subfigure}{0.19\textwidth}
        \includegraphics[width=\linewidth]{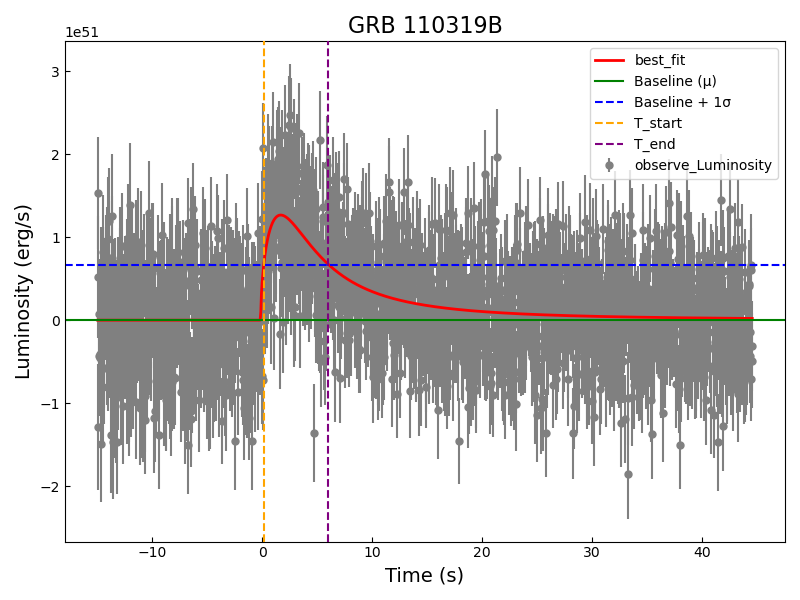}
    \end{subfigure}\hfill
    \begin{subfigure}{0.19\textwidth}
        \includegraphics[width=\linewidth]{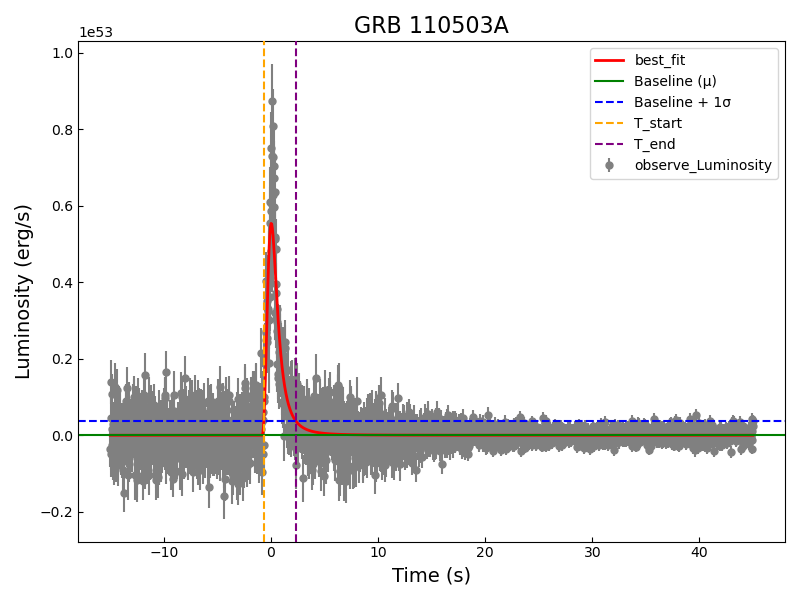}
    \end{subfigure}\hfill
    \begin{subfigure}{0.19\textwidth}
        \includegraphics[width=\linewidth]{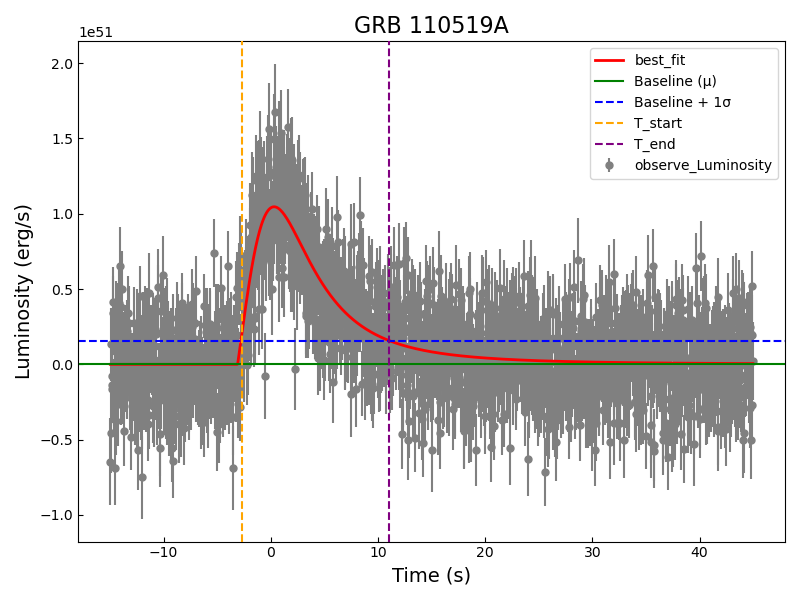}
    \end{subfigure}\hfill
    \begin{subfigure}{0.19\textwidth}
        \includegraphics[width=\linewidth]{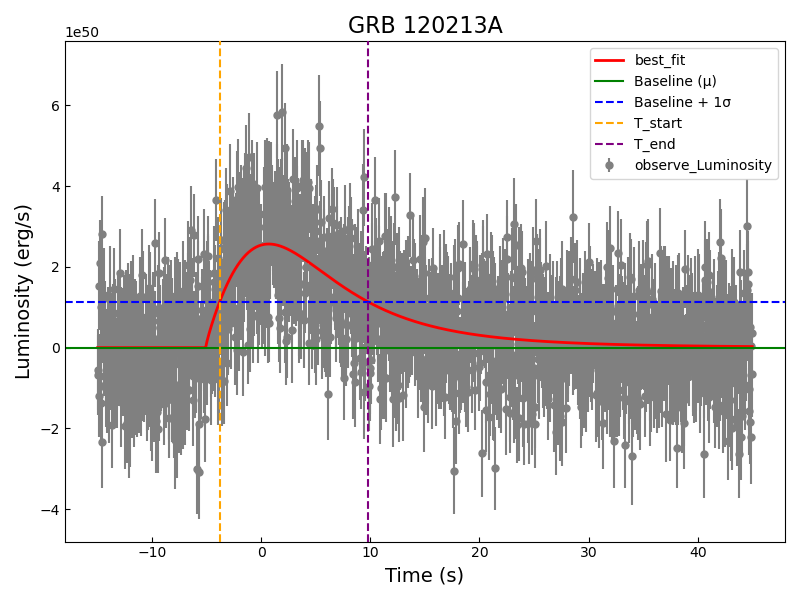}
    \end{subfigure}

    \begin{subfigure}{0.19\textwidth}
        \includegraphics[width=\linewidth]{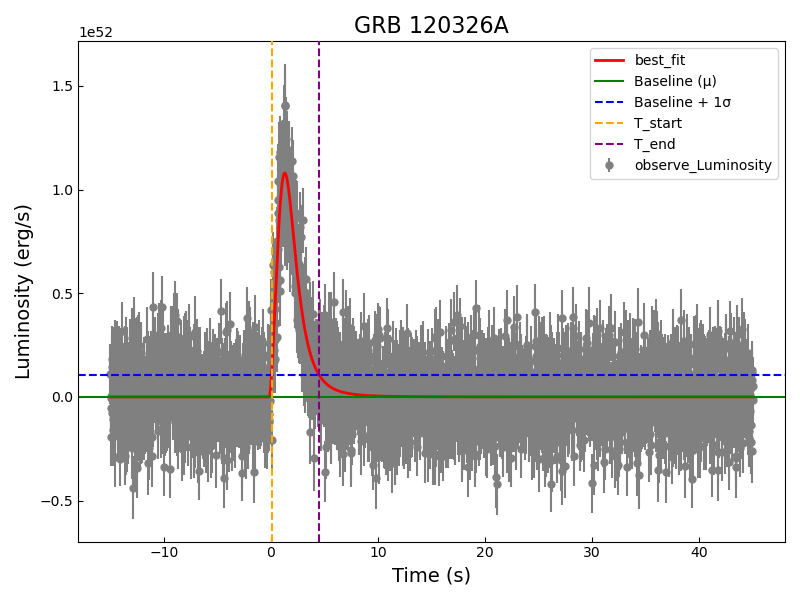}
    \end{subfigure}\hfill
    \begin{subfigure}{0.19\textwidth}
        \includegraphics[width=\linewidth]{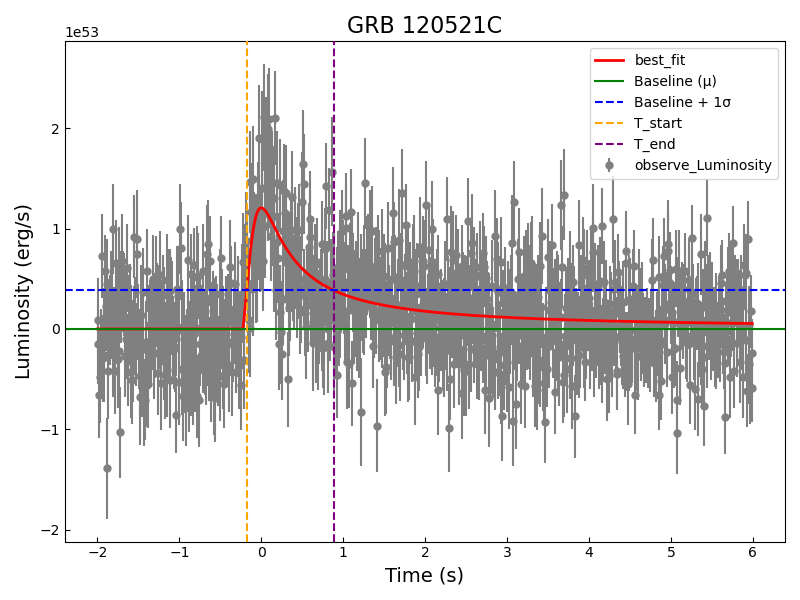}
    \end{subfigure}\hfill
    \begin{subfigure}{0.19\textwidth}
        \includegraphics[width=\linewidth]{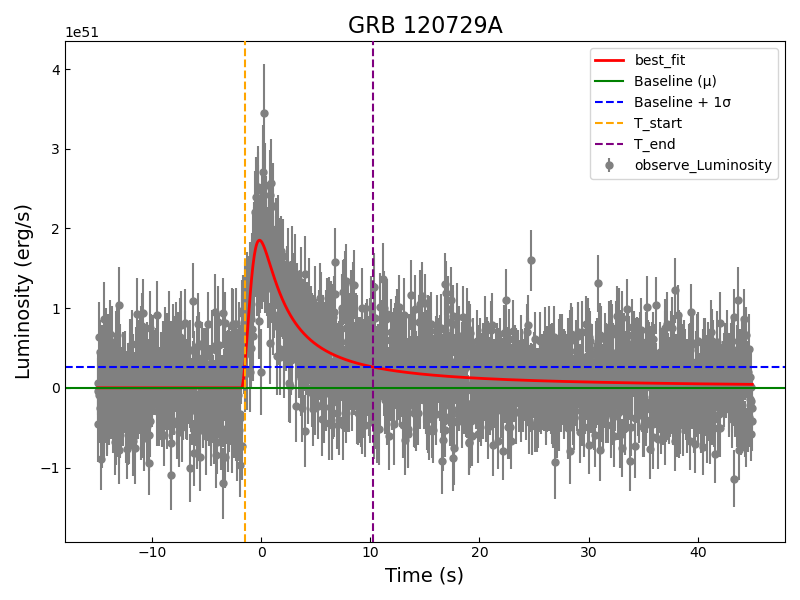}
    \end{subfigure}\hfill
    \begin{subfigure}{0.19\textwidth}
        \includegraphics[width=\linewidth]{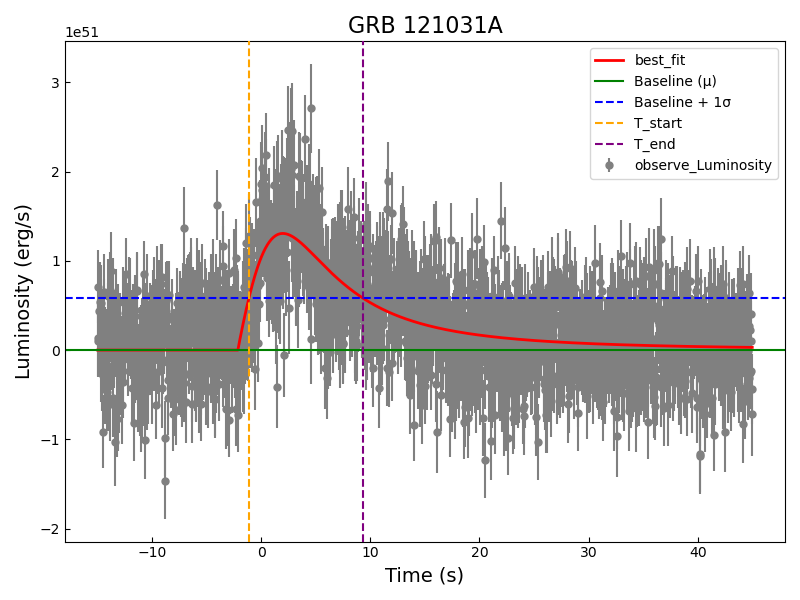}
    \end{subfigure}\hfill
    \begin{subfigure}{0.19\textwidth}
        \includegraphics[width=\linewidth]{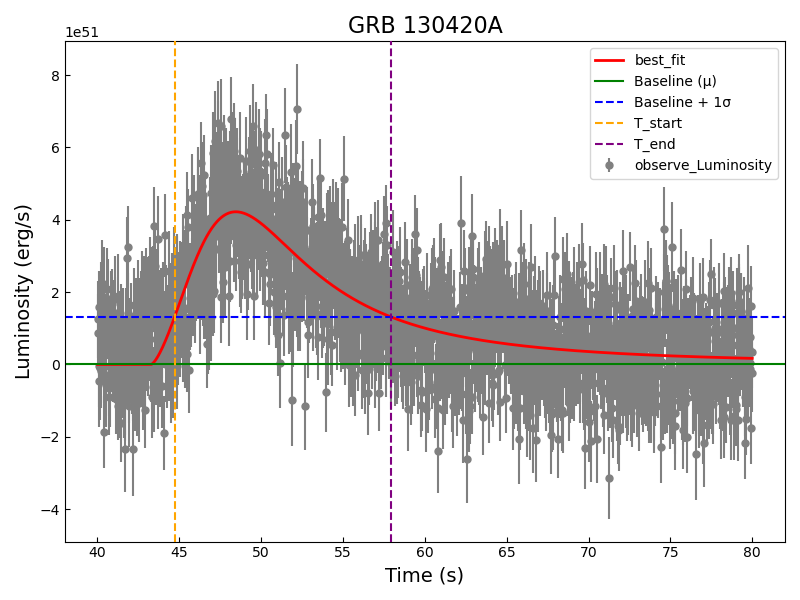}
    \end{subfigure}

    \begin{subfigure}{0.19\textwidth}
        \includegraphics[width=\linewidth]{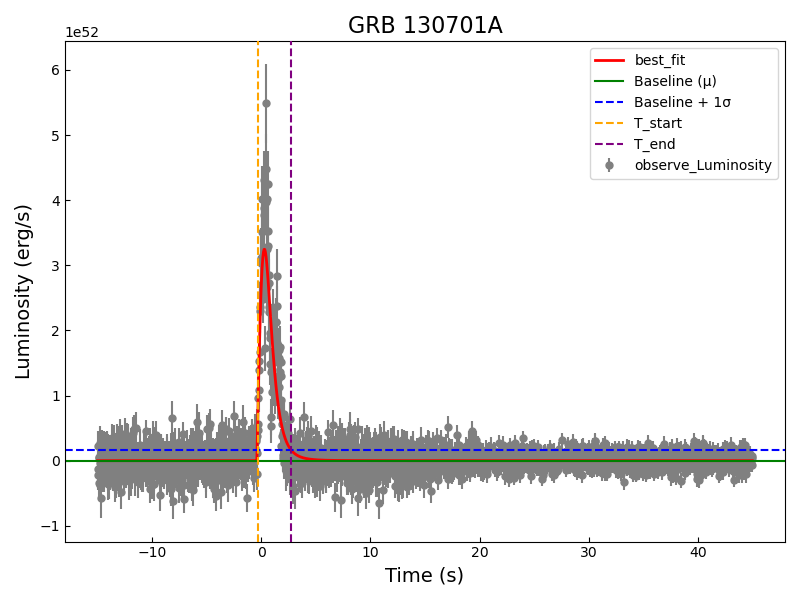}
    \end{subfigure}\hfill
    \begin{subfigure}{0.19\textwidth}
        \includegraphics[width=\linewidth]{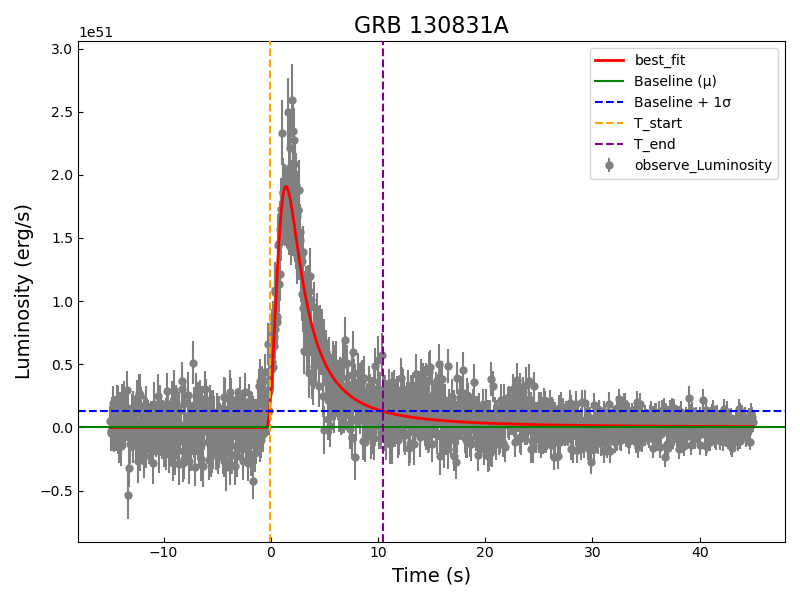}
    \end{subfigure}\hfill
    \begin{subfigure}{0.19\textwidth}
        \includegraphics[width=\linewidth]{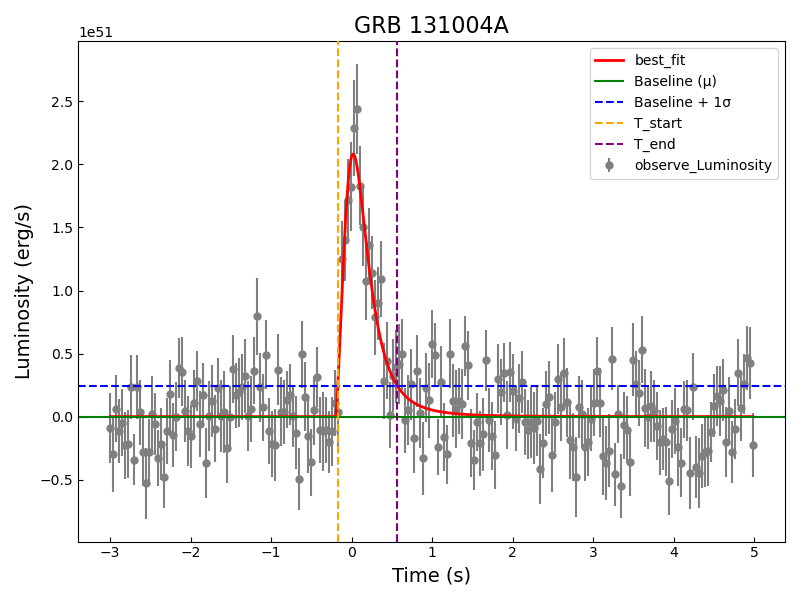}
    \end{subfigure}\hfill
    \begin{subfigure}{0.19\textwidth}
        \includegraphics[width=\linewidth]{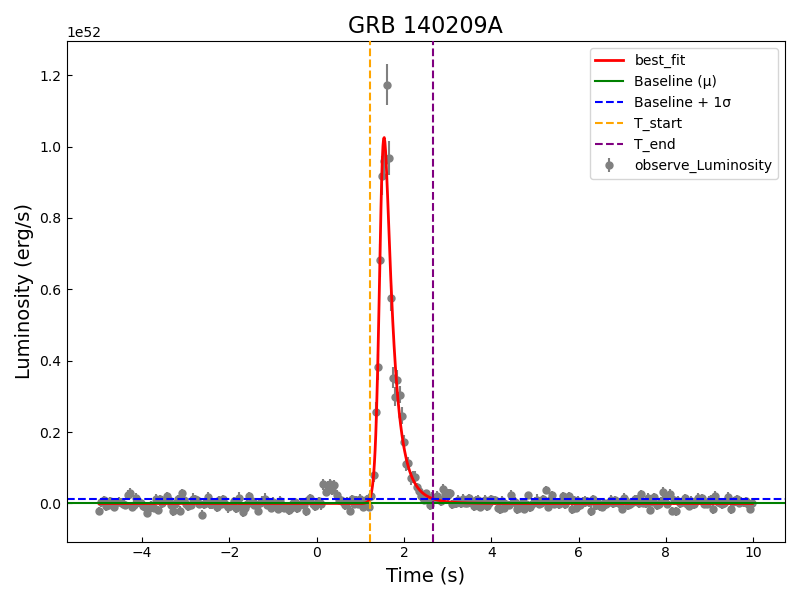}
    \end{subfigure}\hfill
    \begin{subfigure}{0.19\textwidth}
        \includegraphics[width=\linewidth]{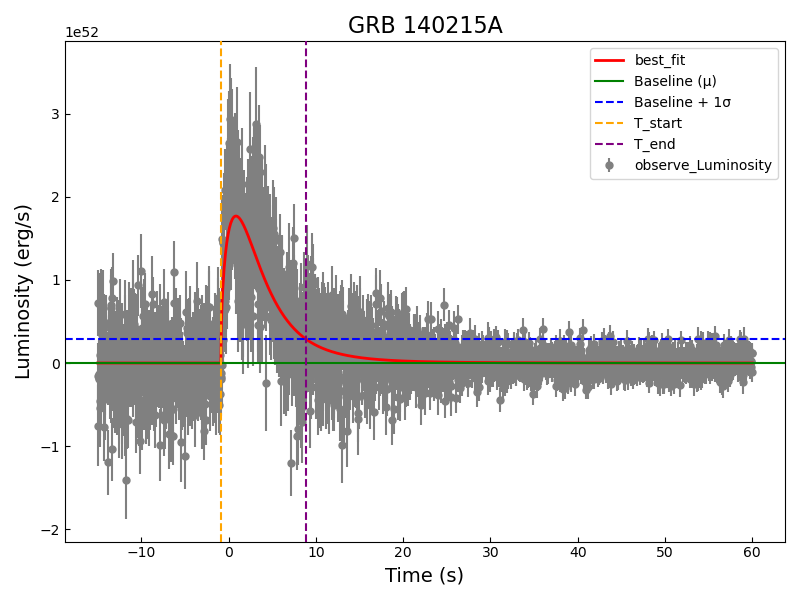}
    \end{subfigure}

    \begin{subfigure}{0.19\textwidth}
        \includegraphics[width=\linewidth]{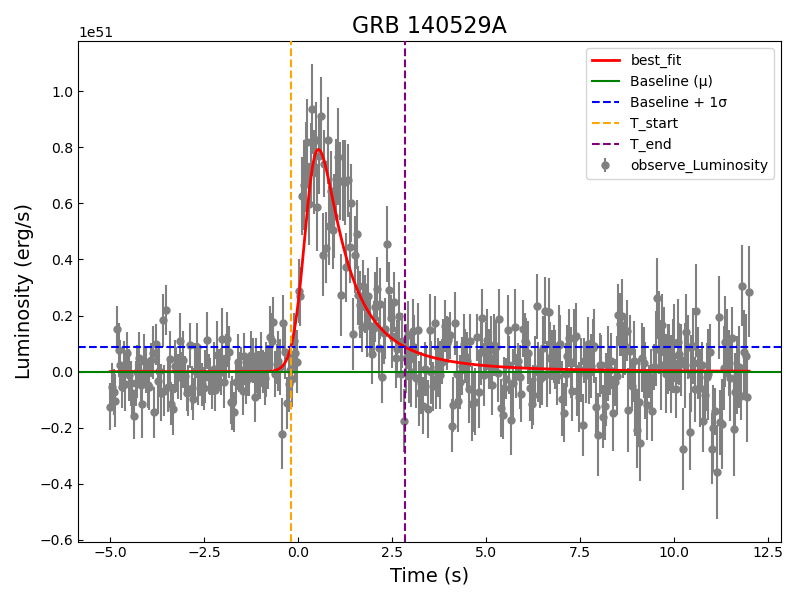}
    \end{subfigure}\hfill
    \begin{subfigure}{0.19\textwidth}
        \includegraphics[width=\linewidth]{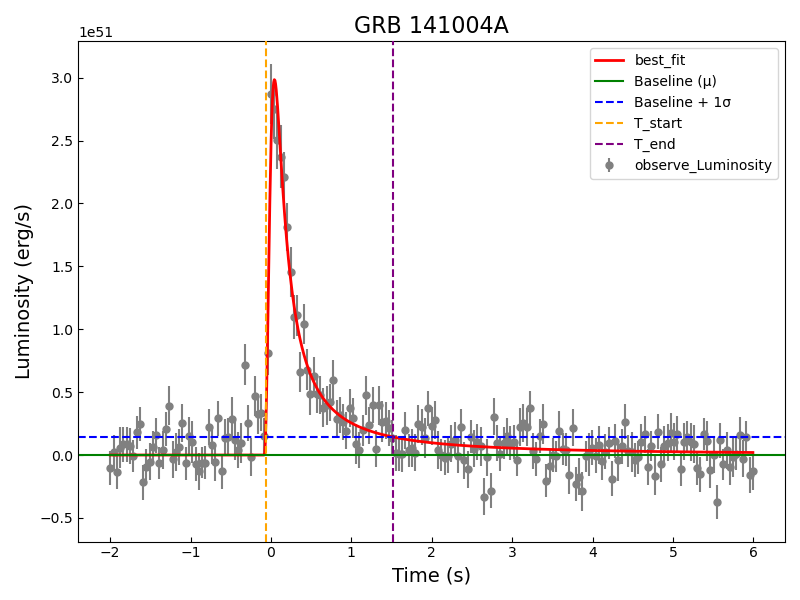}
    \end{subfigure}\hfill
    \begin{subfigure}{0.19\textwidth}
        \includegraphics[width=\linewidth]{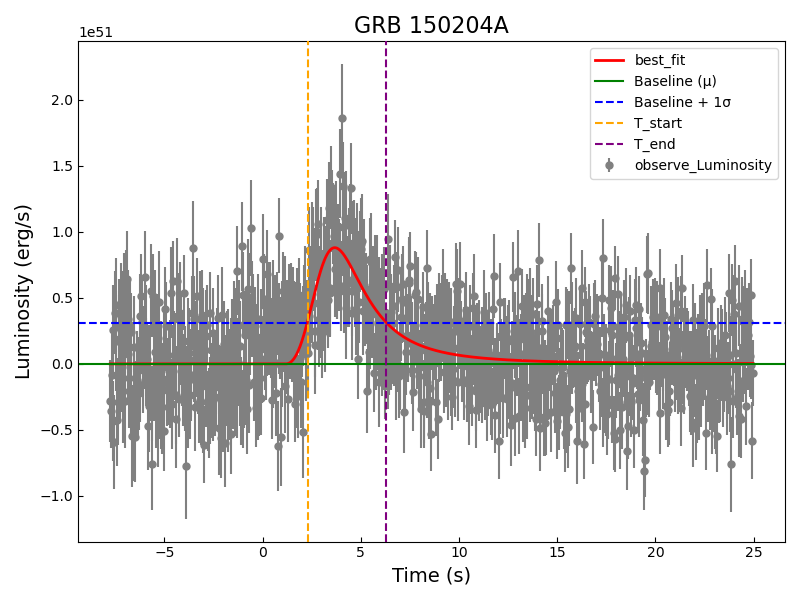}
    \end{subfigure}\hfill
    \begin{subfigure}{0.19\textwidth}
        \includegraphics[width=\linewidth]{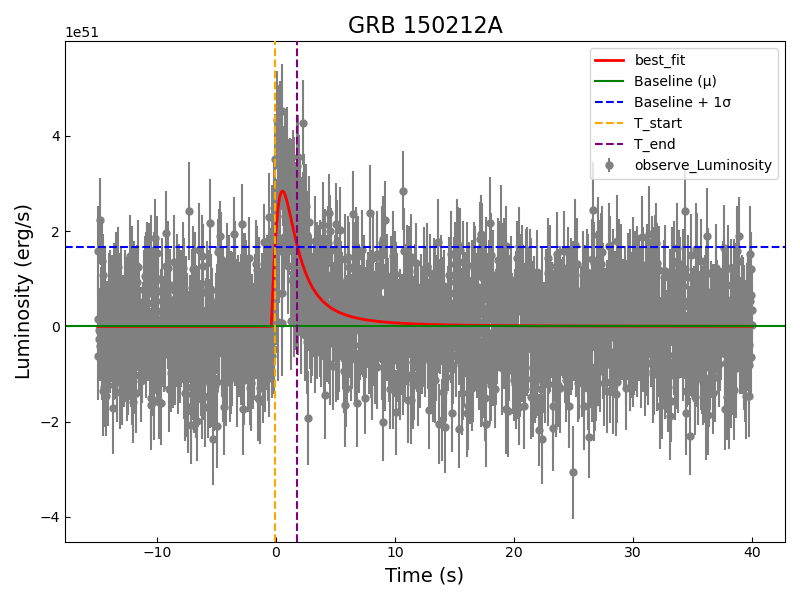}
    \end{subfigure}\hfill
    \begin{subfigure}{0.19\textwidth}
        \includegraphics[width=\linewidth]{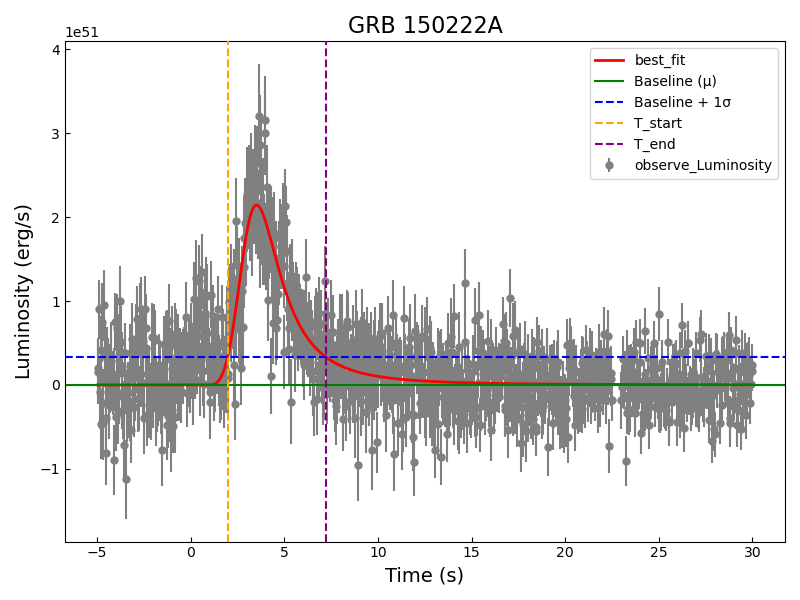}
    \end{subfigure}
    
    \label{fig6_part1}
\end{figure*}

\begin{figure*}[ht]
    \centering
    \captionsetup[subfigure]{aboveskip=-8pt, belowskip=-1pt, margin=0pt}
    \setlength{\belowcaptionskip}{-10pt}
    \setlength{\tabcolsep}{-2pt}
    \ContinuedFloat 
    
    \begin{subfigure}{0.19\textwidth}
        \includegraphics[width=\linewidth]{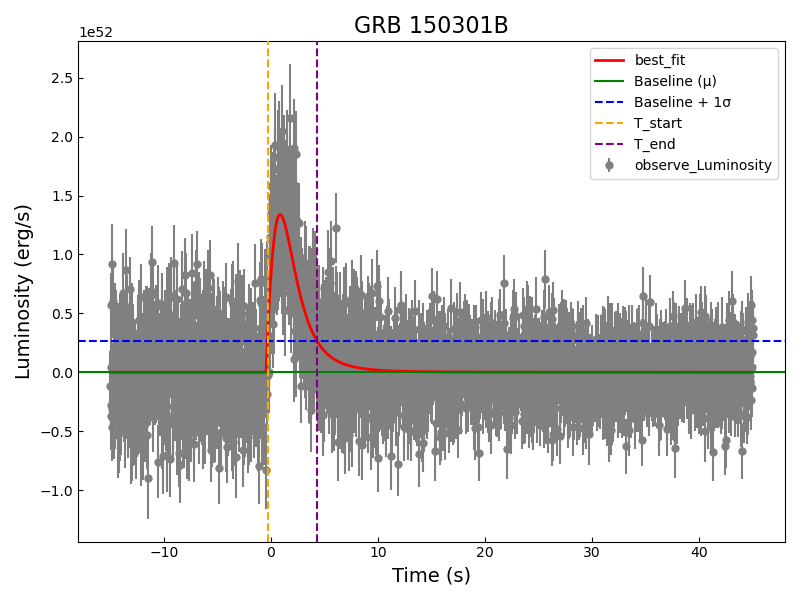}
    \end{subfigure}\hfill
    \begin{subfigure}{0.19\textwidth}
        \includegraphics[width=\linewidth]{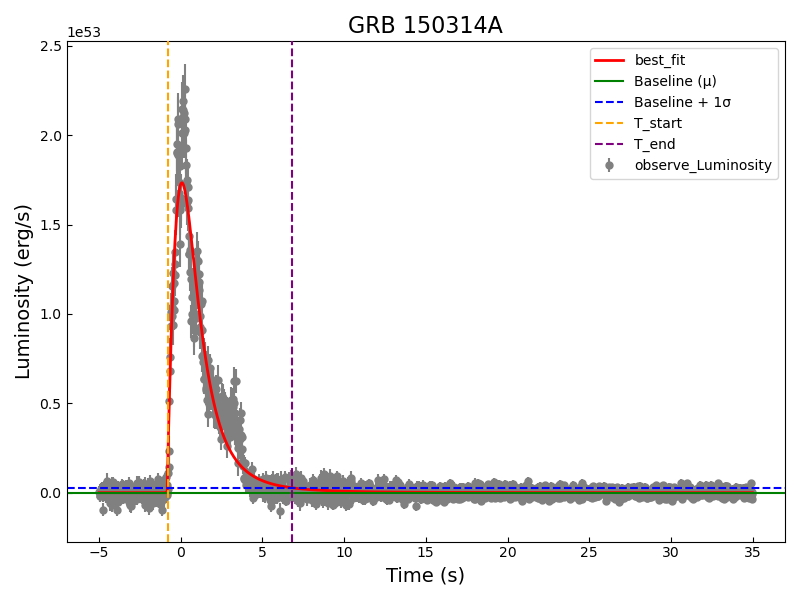}
    \end{subfigure}\hfill
    \begin{subfigure}{0.19\textwidth}
        \includegraphics[width=\linewidth]{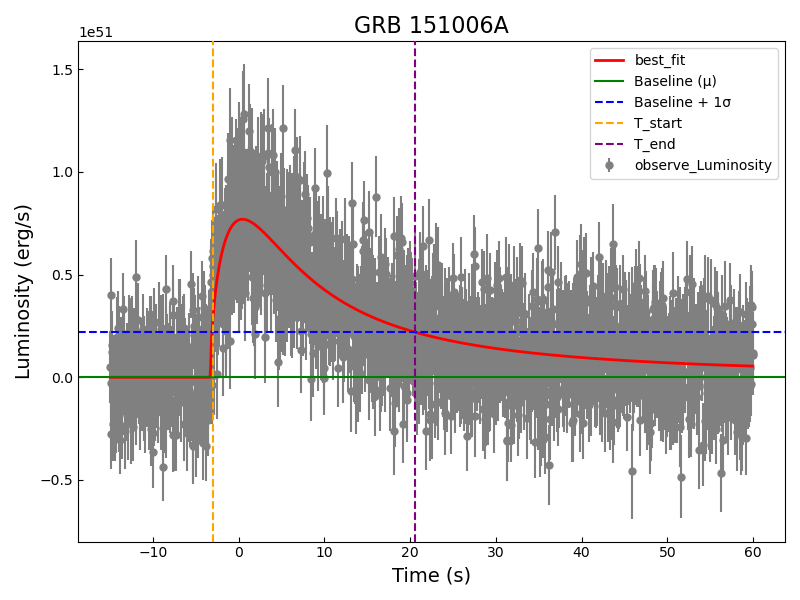}
    \end{subfigure}\hfill
    \begin{subfigure}{0.19\textwidth}
        \includegraphics[width=\linewidth]{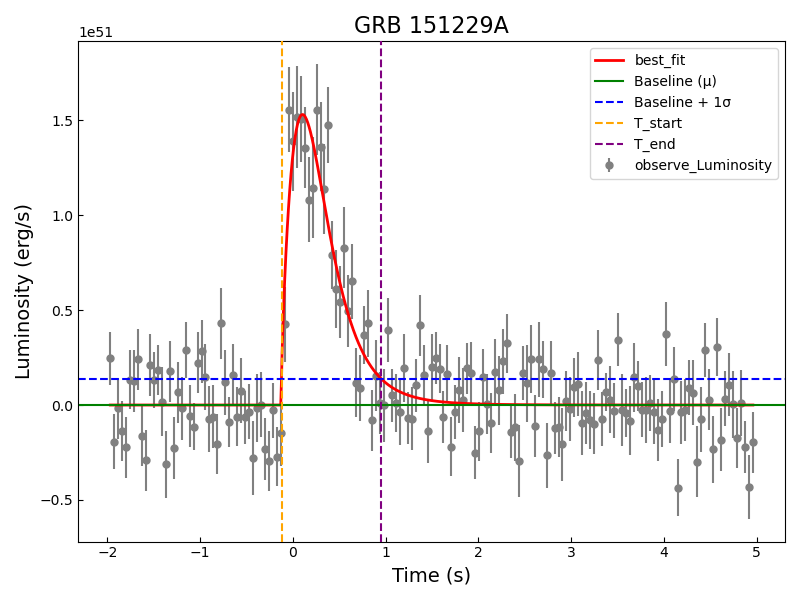}
    \end{subfigure}\hfill
    \begin{subfigure}{0.19\textwidth}
        \includegraphics[width=\linewidth]{GRB160131A_luminosity_fit.png}
    \end{subfigure}

    \begin{subfigure}{0.19\textwidth}
        \includegraphics[width=\linewidth]{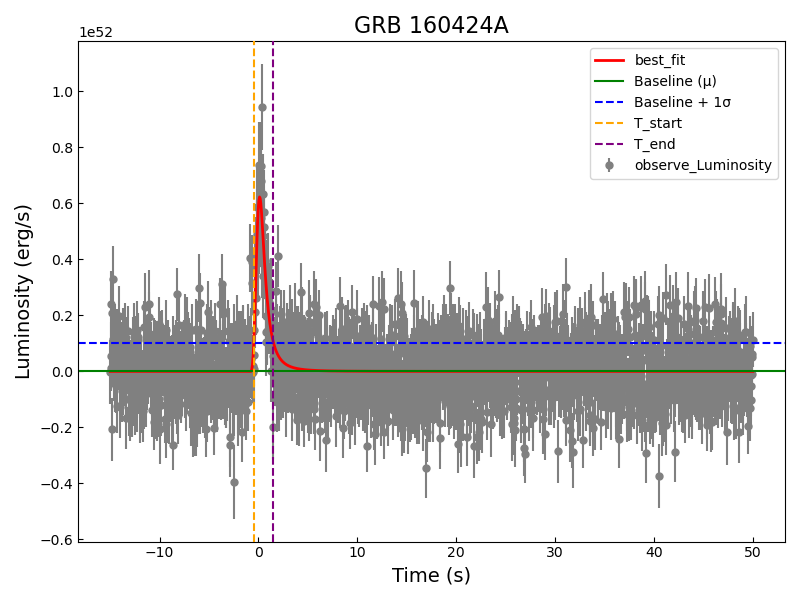}
    \end{subfigure}\hfill
    \begin{subfigure}{0.19\textwidth}
        \includegraphics[width=\linewidth]{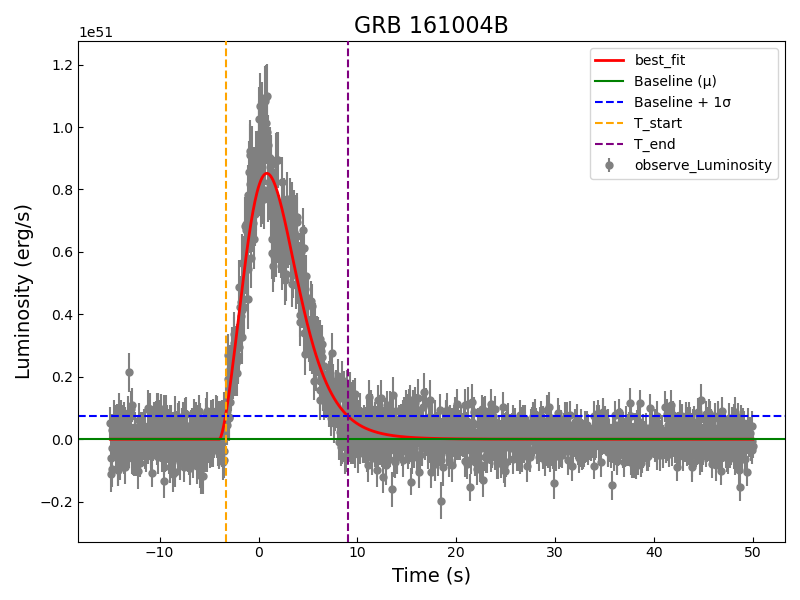}
    \end{subfigure}\hfill
    \begin{subfigure}{0.19\textwidth}
        \includegraphics[width=\linewidth]{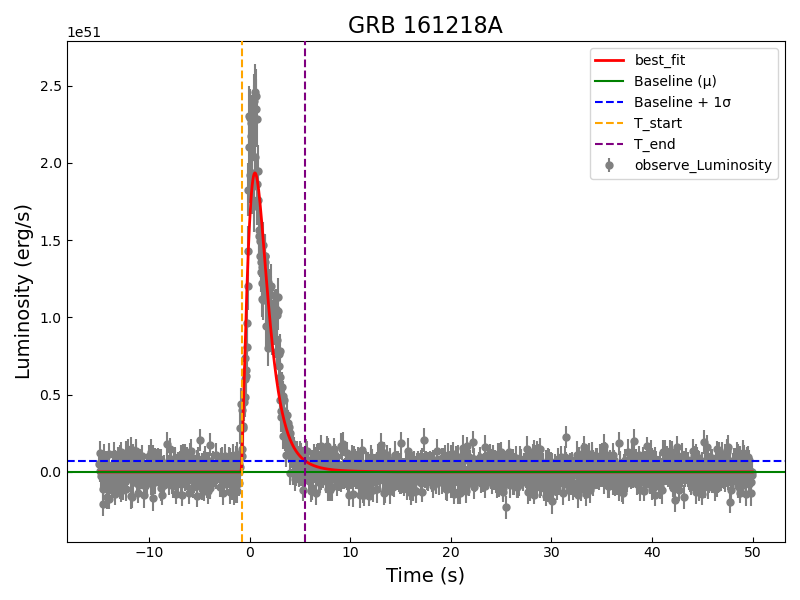}
    \end{subfigure}\hfill
    \begin{subfigure}{0.19\textwidth}
        \includegraphics[width=\linewidth]{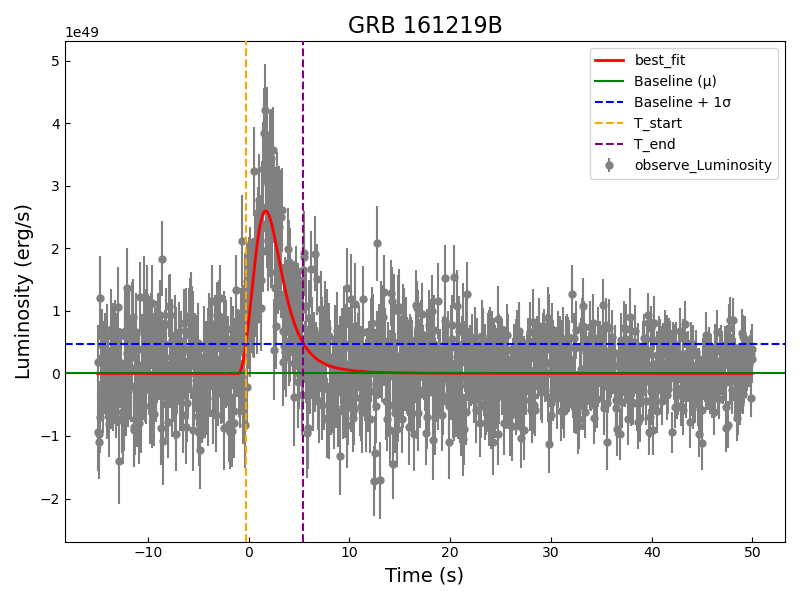}
    \end{subfigure}\hfill
    \begin{subfigure}{0.19\textwidth}
        \includegraphics[width=\linewidth]{GRB170101A_luminosity_fit.png}
    \end{subfigure}

    \begin{subfigure}{0.19\textwidth}
        \includegraphics[width=\linewidth]{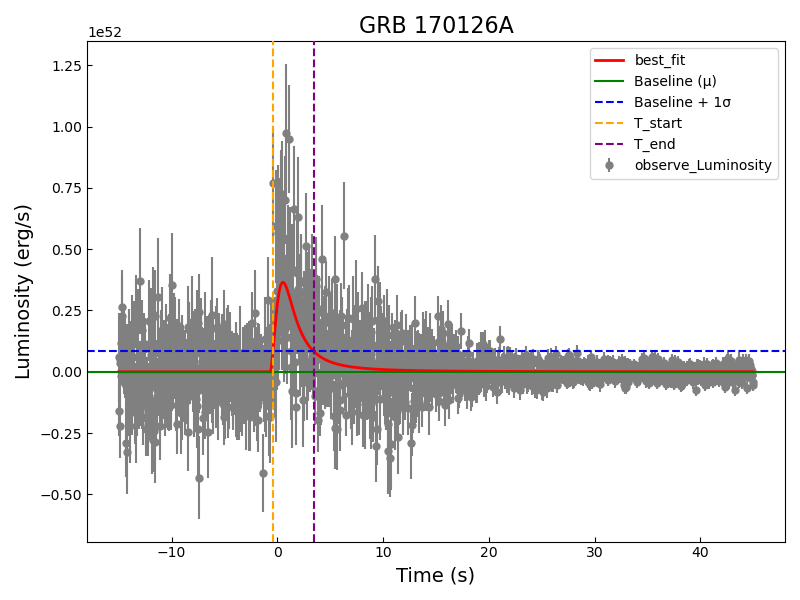}
    \end{subfigure}\hfill
    \begin{subfigure}{0.19\textwidth}
        \includegraphics[width=\linewidth]{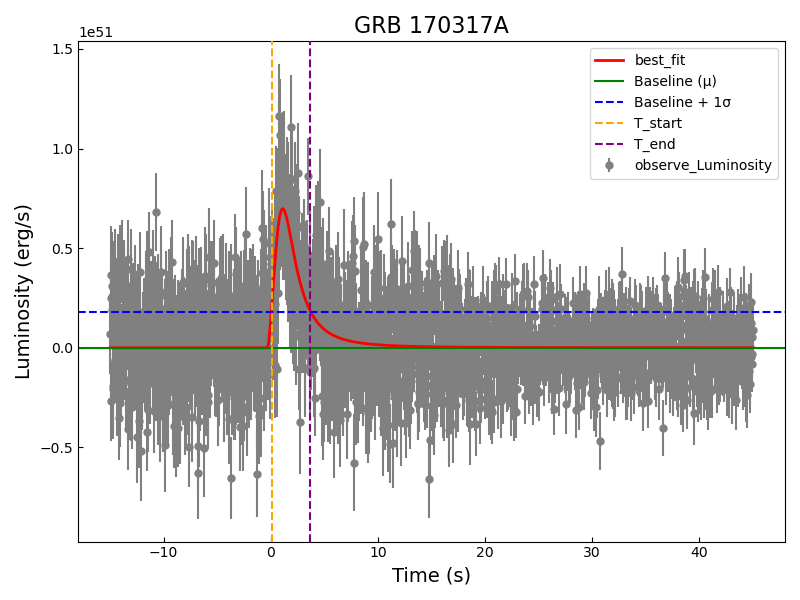}
    \end{subfigure}\hfill
    \begin{subfigure}{0.19\textwidth}
        \includegraphics[width=\linewidth]{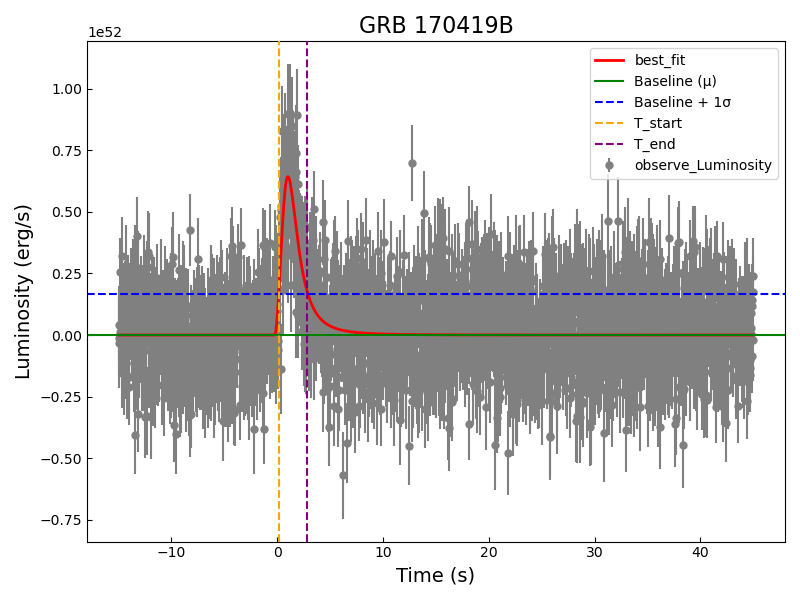}
    \end{subfigure}\hfill
    \begin{subfigure}{0.19\textwidth}
        \includegraphics[width=\linewidth]{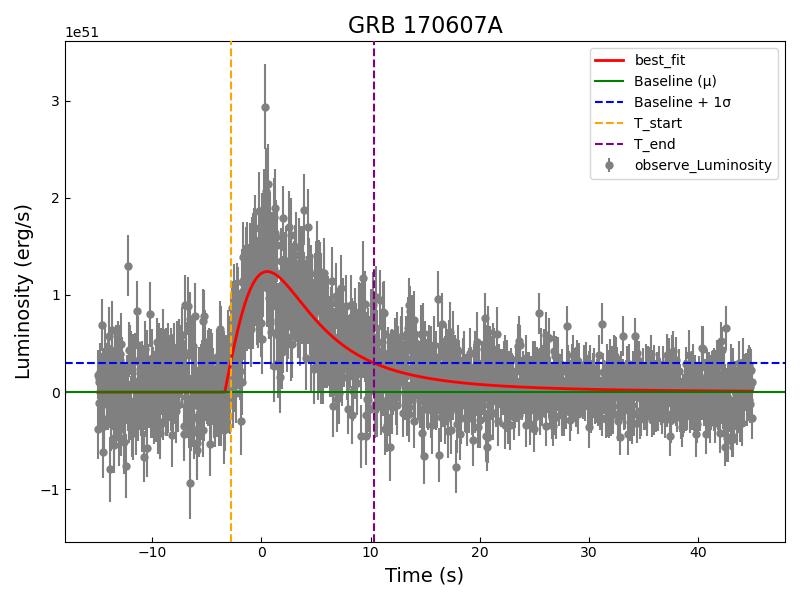}
    \end{subfigure}\hfill
    \begin{subfigure}{0.19\textwidth}
        \includegraphics[width=\linewidth]{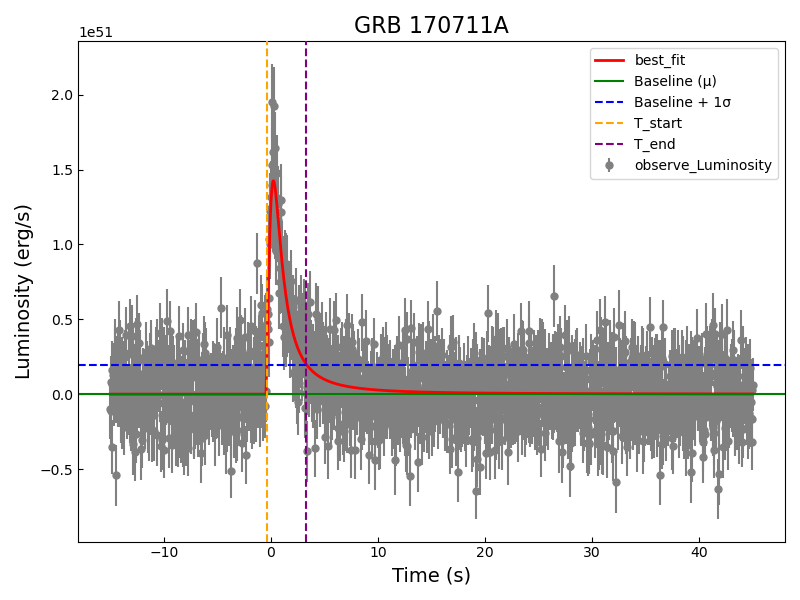}
    \end{subfigure}

    \begin{subfigure}{0.19\textwidth}
        \includegraphics[width=\linewidth]{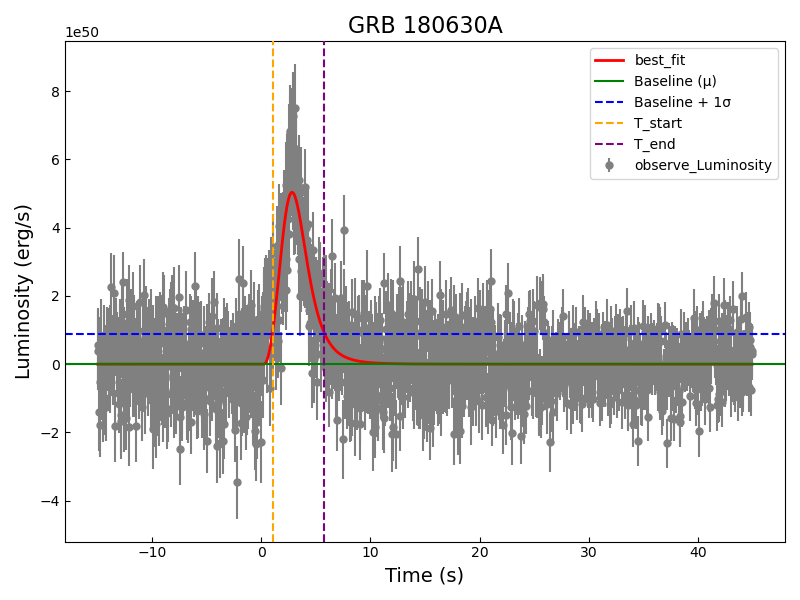}
    \end{subfigure}\hfill
    \begin{subfigure}{0.19\textwidth}
        \includegraphics[width=\linewidth]{GRB180728A_luminosity_fit.png}
    \end{subfigure}\hfill
    \begin{subfigure}{0.19\textwidth}
        \includegraphics[width=\linewidth]{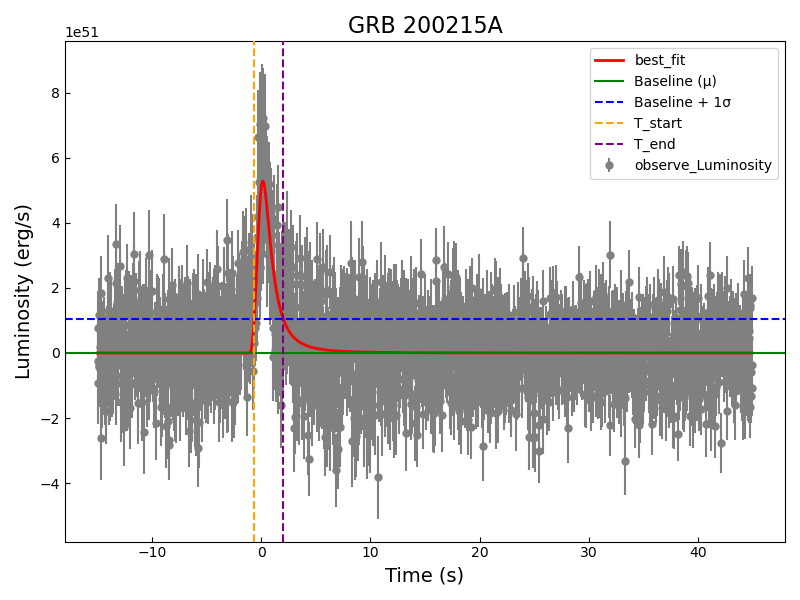}
    \end{subfigure}\hfill
    \begin{subfigure}{0.19\textwidth}
        \includegraphics[width=\linewidth]{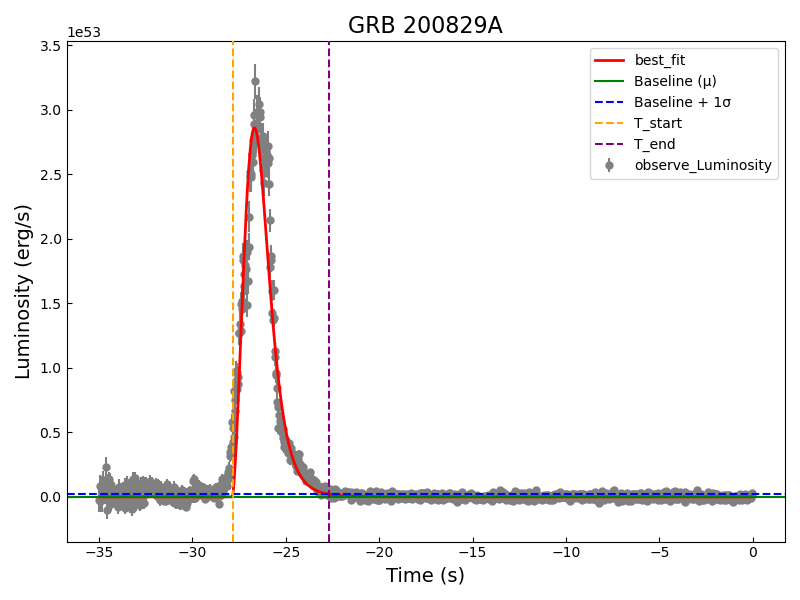}
    \end{subfigure}\hfill
    \begin{subfigure}{0.19\textwidth}
        \includegraphics[width=\linewidth]{GRB200922A_luminosity_fit.png}
    \end{subfigure}

    \begin{subfigure}{0.19\textwidth}
        \includegraphics[width=\linewidth]{GRB201105A_luminosity_fit.png}
    \end{subfigure}\hfill
    \begin{subfigure}{0.19\textwidth}
        \includegraphics[width=\linewidth]{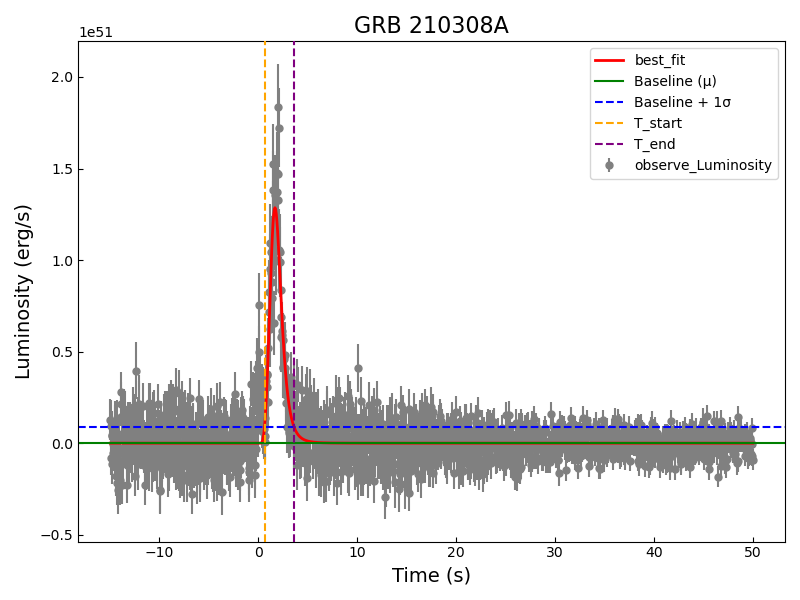}
    \end{subfigure}\hfill
    \begin{subfigure}{0.19\textwidth}
        \includegraphics[width=\linewidth]{GRB210410A_luminosity_fit.png}
    \end{subfigure}\hfill
    \begin{subfigure}{0.19\textwidth}
        \includegraphics[width=\linewidth]{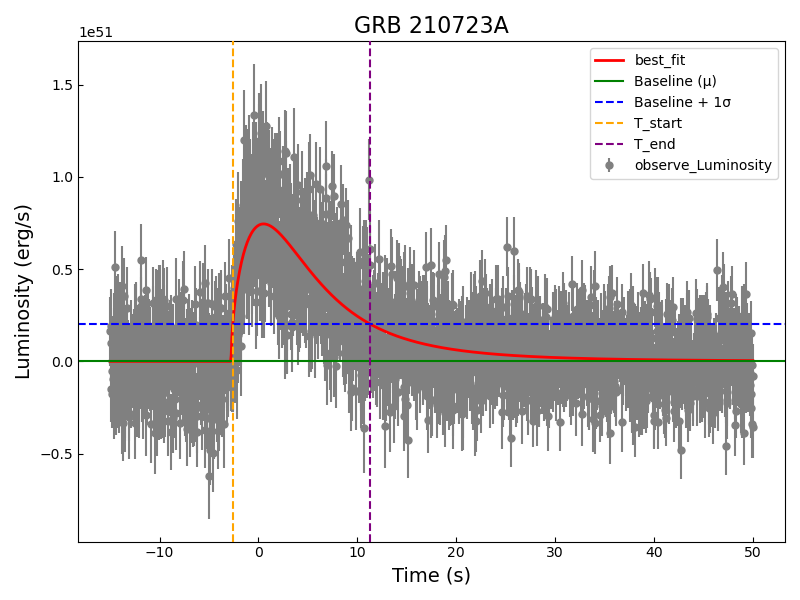}
    \end{subfigure}\hfill
    \begin{subfigure}{0.19\textwidth}
        \includegraphics[width=\linewidth]{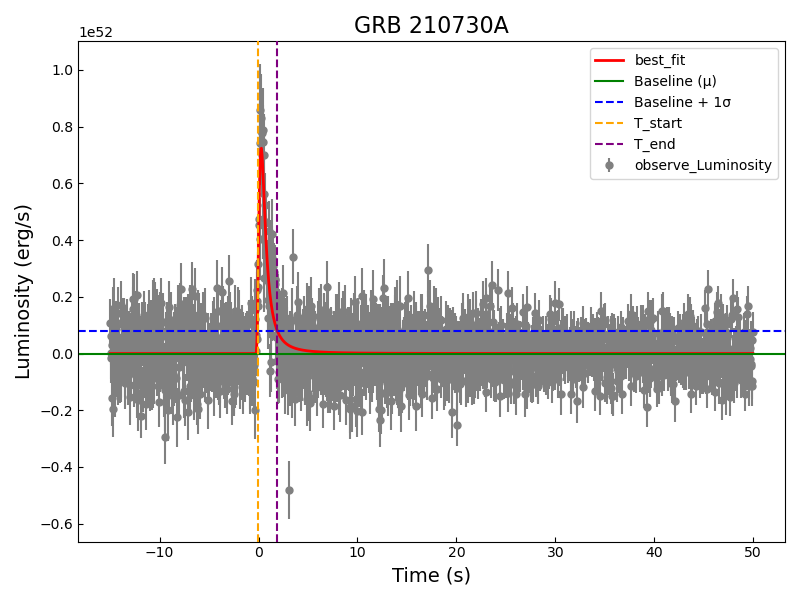}
    \end{subfigure}
    
    \begin{subfigure}{0.19\textwidth}
        \includegraphics[width=\linewidth]{GRB211225B_luminosity_fit.png}
    \end{subfigure}\hfill
    \begin{subfigure}{0.19\textwidth}
        \includegraphics[width=\linewidth]{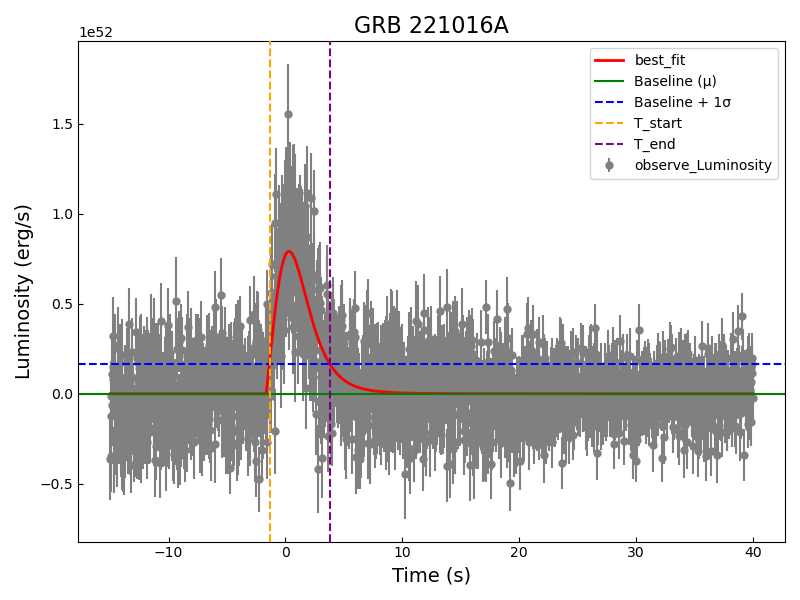}
    \end{subfigure}\hfill
    \begin{subfigure}{0.19\textwidth}
        \includegraphics[width=\linewidth]{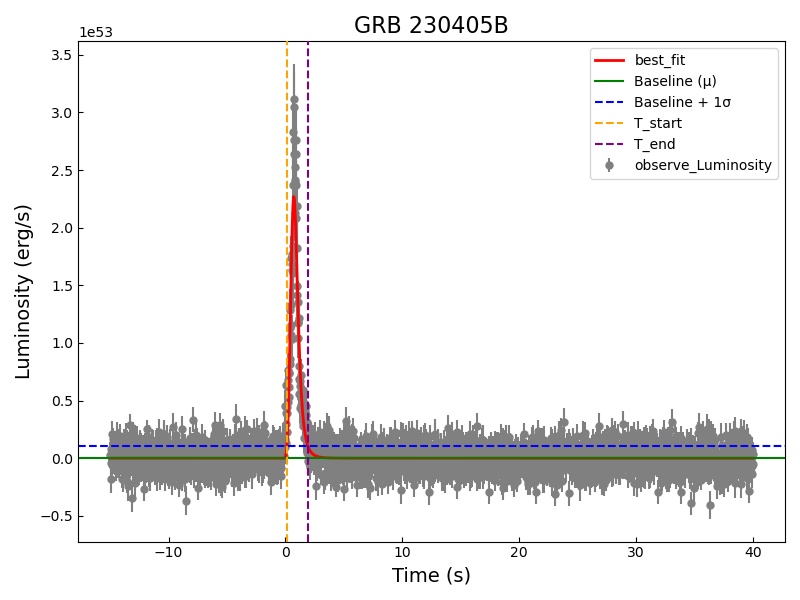}
    \end{subfigure}\hfill
    \begin{subfigure}{0.19\textwidth}
        \includegraphics[width=\linewidth]{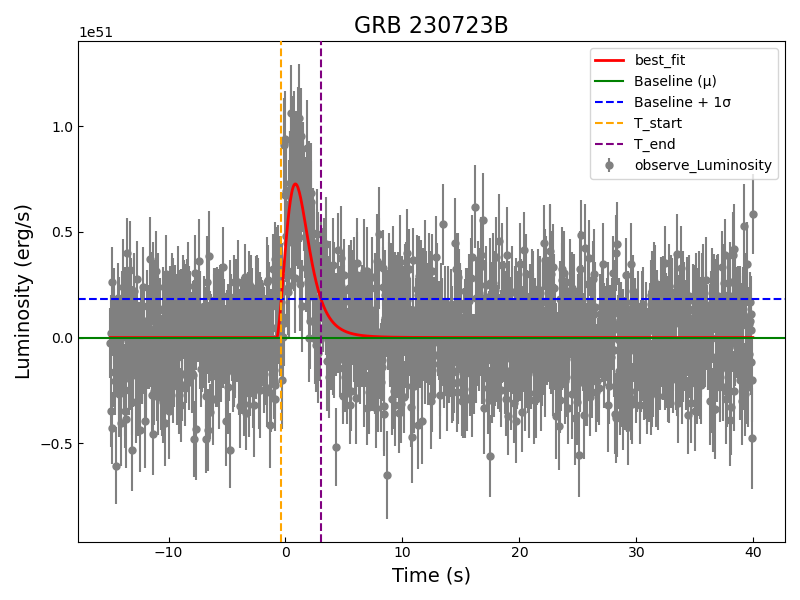}
    \end{subfigure}\hfill
    \begin{subfigure}{0.19\textwidth}
        \includegraphics[width=\linewidth]{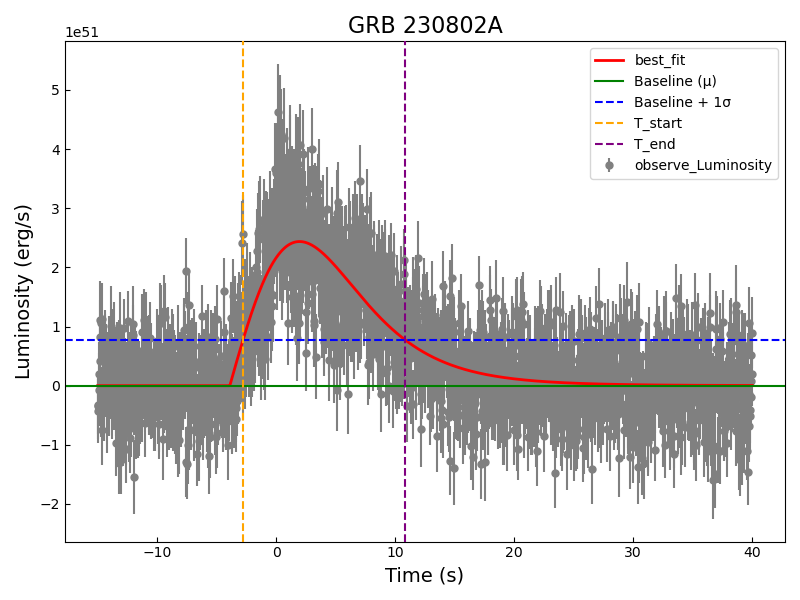}
    \end{subfigure}

    \begin{subfigure}{0.19\textwidth}
        \includegraphics[width=\linewidth]{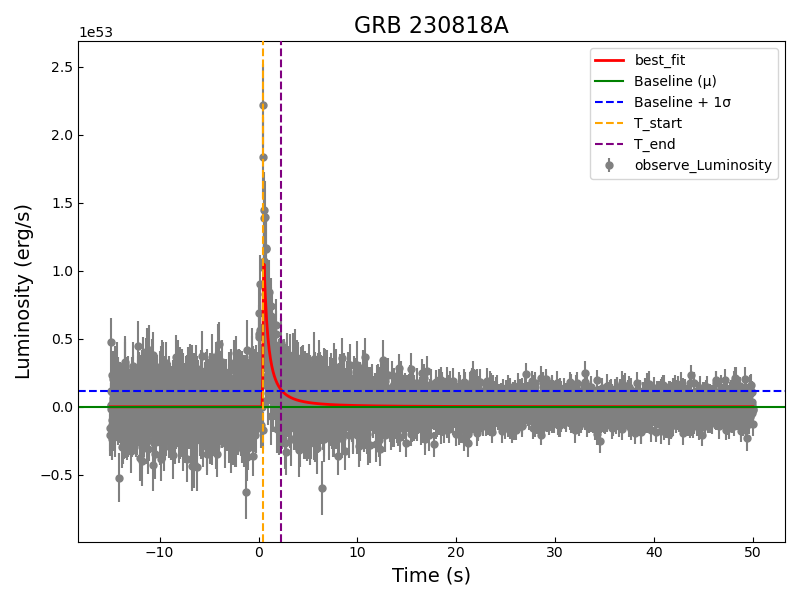}
    \end{subfigure}\hfill
    \begin{subfigure}{0.19\textwidth}
        \includegraphics[width=\linewidth]{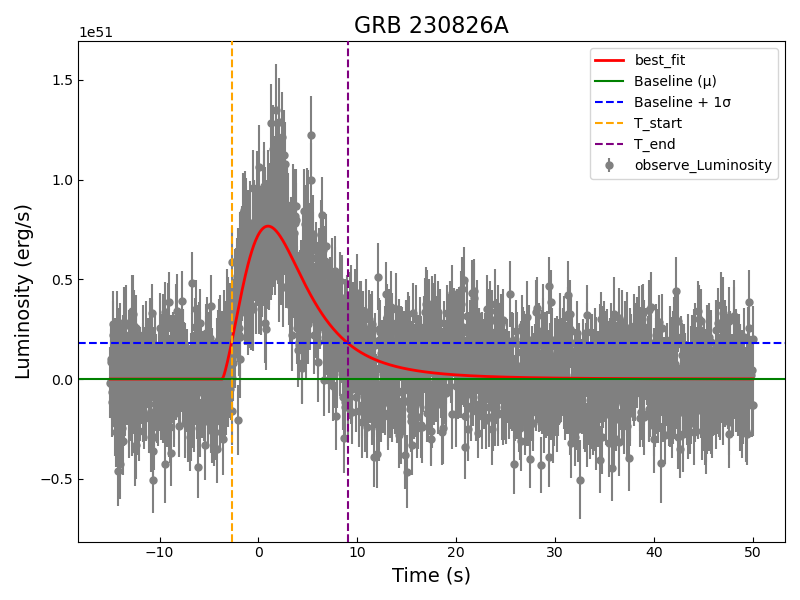}
    \end{subfigure}\hfill
    \begin{subfigure}{0.19\textwidth}
        \includegraphics[width=\linewidth]{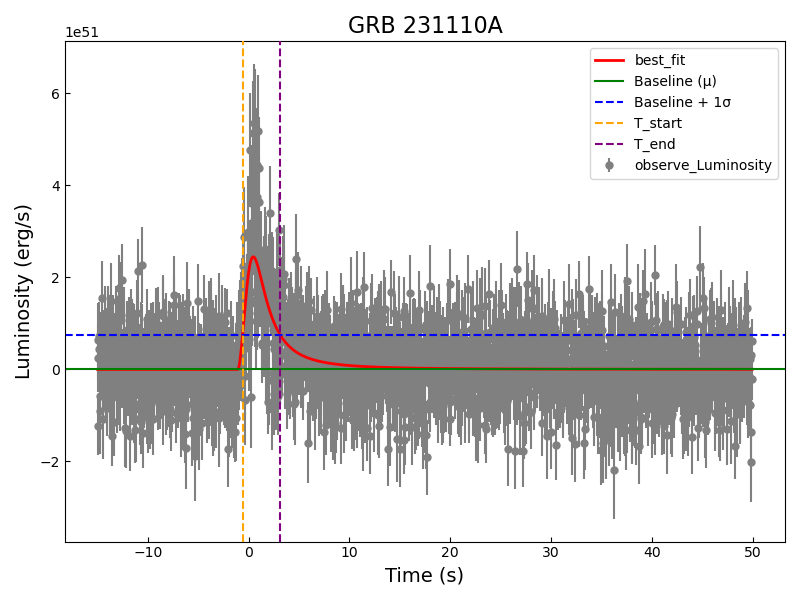}
    \end{subfigure}\hfill
    \begin{subfigure}{0.19\textwidth}
        \includegraphics[width=\linewidth]{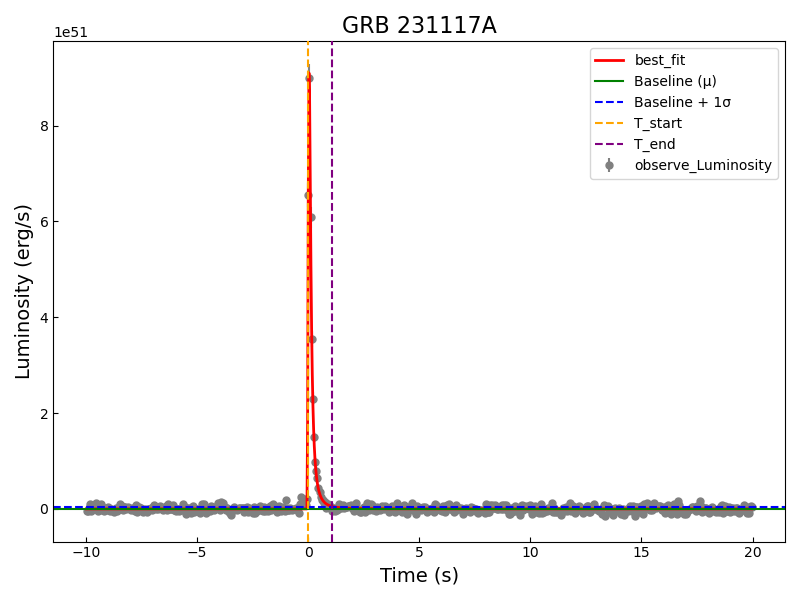}
    \end{subfigure}\hfill
    \begin{subfigure}{0.19\textwidth}
        \includegraphics[width=\linewidth]{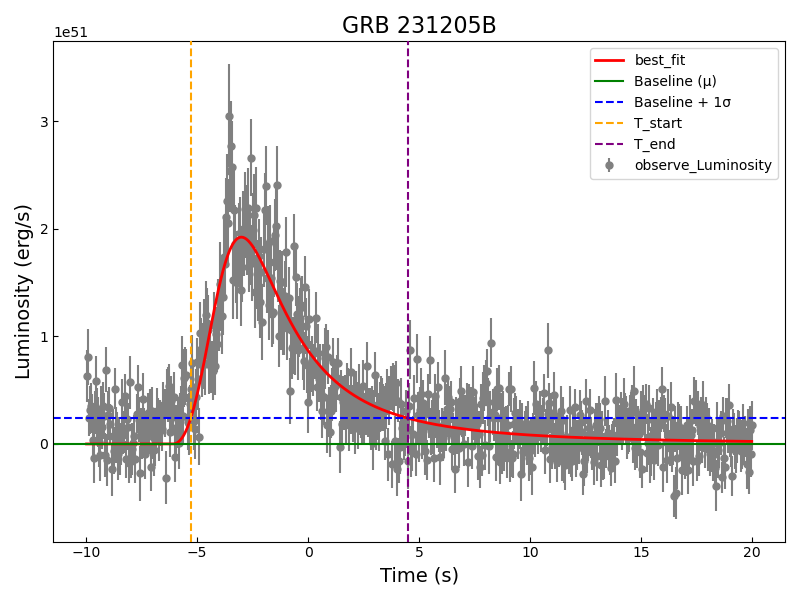}
    \end{subfigure}

    \begin{subfigure}{0.19\textwidth}
        \includegraphics[width=\linewidth]{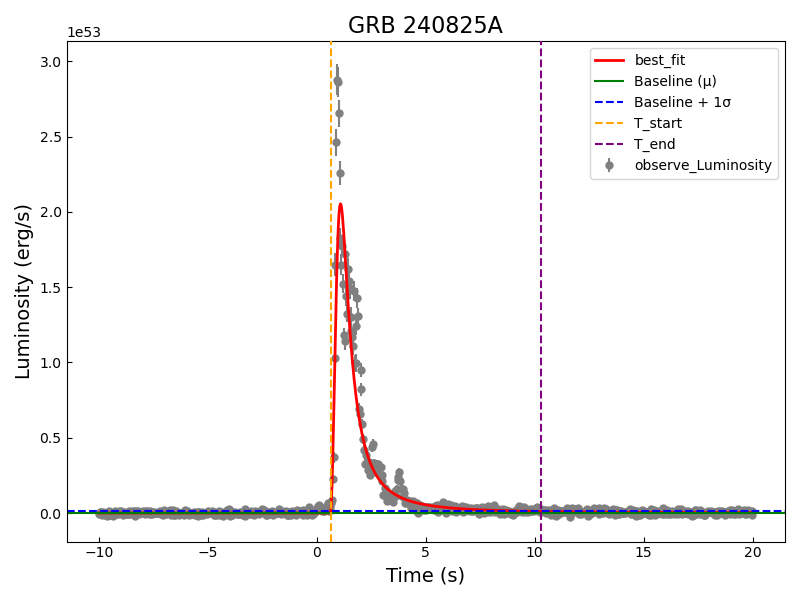}
    \end{subfigure}\hfill
    \begin{subfigure}{0.19\textwidth}
        \includegraphics[width=\linewidth]{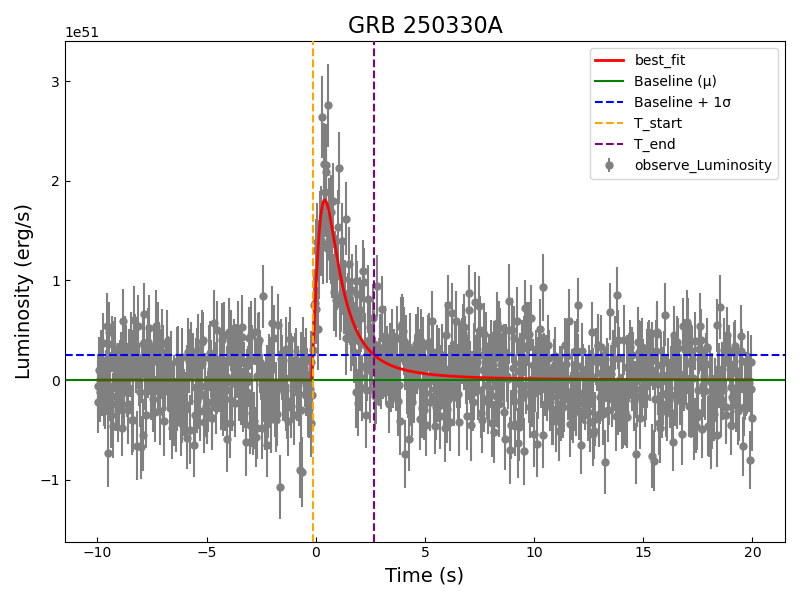}
    \end{subfigure}\hfill
    \begin{subfigure}{0.19\textwidth}
        \includegraphics[width=\linewidth]{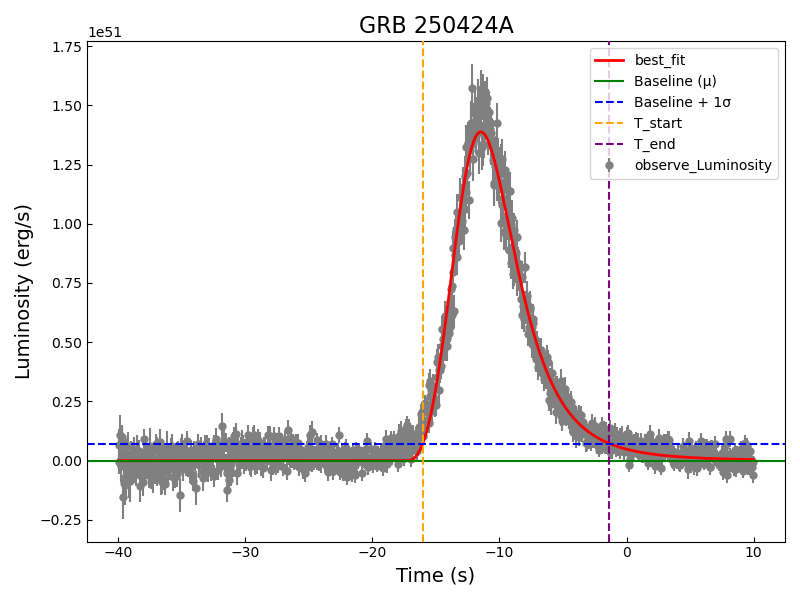}
    \end{subfigure}\hfill
    \begin{subfigure}{0.19\textwidth}
        \includegraphics[width=\linewidth]{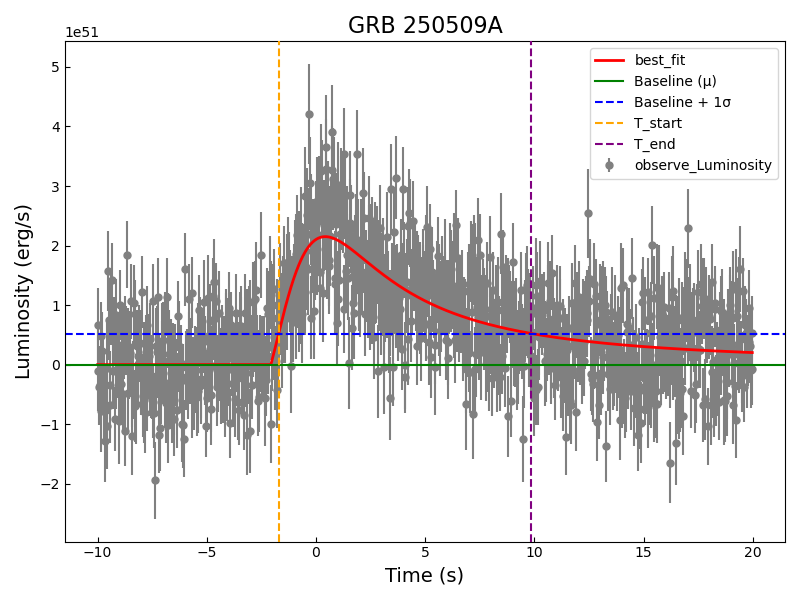}
    \end{subfigure}\hfill
    \begin{subfigure}{0.19\textwidth}
        \includegraphics[width=\linewidth]{GRB250605A_luminosity_fit.png}
    \end{subfigure} 
    \caption[]{Light-curve fitting of 85 GRBs using the KRL function (red lines). The dotted lines indicate the pulse start and end times. The data have been corrected for redshift, and the observed data are shown in grey.}
    \label{fig6}
\end{figure*}

\end{appendix}

\end{document}